\documentclass[useAMS,usegraphicx,usenatbib]{mn2e}
\usepackage{color} 

\usepackage{multirow}
\usepackage{amssymb}

\usepackage{amsmath}

\newcommand{\kms}{\mbox{$\>{\rm km\, s^{-1}}$}}
\newcommand{\pc}{\>{\rm pc}}
\newcommand{\kpc}{\mbox{$\>{\rm kpc}$}} 
\newcommand{\Gyr}{\mbox{$\>{\rm Gyr}$}}
\newcommand{\Myr}{\mbox{$\>{\rm Myr}$}}

\newcommand{\Msun}{\>{\rm M_{\odot}}}
\newcommand{\Rd}{\mbox{$R_{\rm d}$}} 
\newcommand{\zd}{\mbox{$z_{\rm d}$}} 
\newcommand{\md}{\mbox{$M_{\rm d}$}} 
 
\newcommand\degrees{^\circ}
\newcommand{\avg}[1]{\mbox{$\left<{#1}\right>$}}
\newcommand{\sig}[1]{\mbox{$\sigma_{#1}$}}
\newcommand{\feh}{\mbox{$\rm [Fe/H]$}}
\newcommand{\al}{\mbox{$\rm \alpha$}}
\newcommand{\alfe}{\mbox{$\rm [O/Fe]$}}
\newcommand{\tf}{\mbox{$\tau$}} 
\newcommand{\rs}{\mbox{$R_s$}} 
\newcommand{\abar}{\mbox{$A_{\rm bar}$}}

\def\ie{{\it i.e.}}

\title[Kinematic Fractionation by an Evolving Bar]{Separation of
  Stellar Populations by an Evolving Bar: implications for the Bulge
  of the Milky Way}
 
\author[Debattista et al.]{Victor P. Debattista$^{1}$\thanks{E-mail:
    vpdebattista@gmail.com}, Melissa Ness$^2$, Oscar A. Gonzalez$^3$,
  K. Freeman$^4$, 
  \newauthor Manuela Zoccali$^{5,6}$, Dante Minniti$^{6,7,8}$ \\
  $^1$ Jeremiah Horrocks Institute, University of Central Lancashire,
  Preston PR1 2HE, UK \\
  $^2$ Max-Planck-Institut f\"ur Astronomie, K\"onigstuhl 17, D-69117 Heidelberg, Germany \\
  $^3$ UK Astronomy Technology Centre, Royal Observatory, Blackford Hill, Edinburgh EH9 3HJ, UK \\
  $^4$ Research School of Astronomy and Astrophysics, Mount Stromlo Observatory, Cotter Road, Weston Creek ACT 2611, Australia \\
  $^5$ Instituto de Astrof\'isica, Facultad de F\'isica, Pontificia Universidad Cat\'olica de Chile, Av. Vicu\~na Mackenna 4860, 782-0436 Macul, \\ Santiago, Chile \\
  $^6$ The Millennium Institute of Astrophysics (MAS), Av. Vicu\~na
  Mackenna 4860, 782-0436 Macul,  Santiago, Chile \\
  $^7$ Departamento de Ciencias Fisicas, Universidad Andres Bello,
  Republica 220, Santiago, Chile \\
  $^8$ Vatican Observatory, V-00120 Vatican City State, Italy
}

\begin{document}   

\date{{\it Draft version on \today}}
\pagerange{\pageref{firstpage}--\pageref{lastpage}} \pubyear{----}
\maketitle

\label{firstpage}

\begin{abstract} 

We present a novel interpretation of the previously puzzling different
behaviours of stellar populations of the Milky Way's bulge.  We first
show, by means of pure $N$-body simulations, that initially {\it
  co-spatial} stellar populations with different in-plane random
motions separate when a bar forms.  The radially cooler populations
form a strong bar, and are vertically thin and peanut-shaped, while
the hotter populations form a weaker bar and become a vertically
thicker box.  We demonstrate that it is the radial, not the vertical,
velocity dispersion that dominates this evolution.  Assuming
that early stellar discs heat rapidly as they form, then both the
in-plane and vertical random motions correlate with stellar age and
chemistry, leading to different density distributions for metal-rich
and metal-poor stars.  We then use a high-resolution simulation, in
which all stars form out of gas, to demonstrate that this is what
happens.  When we apply these results to the Milky Way we show that a
very broad range of observed trends for ages, densities, kinematics
and chemistries, that have been presented as evidence for
contradictory paths to the formation of the bulge, are in fact
consistent with a bulge which formed from a continuum of disc stellar
populations which were kinematically separated by the bar.  For the
first time we are able to account for the bulge's main trends via a
model in which the bulge formed largely in situ.  Since the model is
generic, we also predict the general appearance of stellar population
maps of external edge-on galaxies.

\end{abstract}


%%%%%%%%%%%%%%%%%%%%%%%%%%%%%%%%%%%%%%%%%%%%%%%%%%%%%%%%%%%%%%%%%%%%%%%%%%%%%%

\section{Introduction}
\label{sec:intro}

The origin of the bulge of the Milky Way has been the subject of
considerable discussion.  On the one hand, many lines of evidence now
point to the fact that the Milky Way hosts a bar
\citep{devaucouleurs64, peters75, cohen_few76, liszt_burton80,
  gerhard_vietri86, mulder_liem86, binney+91, nakada+91,
  whitelock_catchpole92, weiland+94, paczynski+94, dwek+95, zhao+96,
  sevenster96, binney+97, nikolaev_weinberg97, stanek+97,
  hammersley+00}.  This leads to the opportunity for secular evolution
to have sculpted the centre of the Milky Way into the bulge we see
today.  Evidence of this comes from the morphology and kinematics of
the bulge.  The X-shape of the red clump distribution in the bulge is
manifested by the bimodal distribution of their distances
\citep{mcwilliam_zoccali10, nataf+10, saito+11, wegg_gerhard13,
  gonzalez+15}.  This distribution is produced by looking through the
near- and far-side corners of a peanut-shaped bulge
\citep[e.g.][]{li_shen12}.  Such box-/peanut-shaped (B-/P-shaped)
bulges are produced by the buckling instability of bars
\citep{raha+91, merritt_sellwood94, bureau_athanassoula05,
  debattista+06} or by orbit trapping \citep{combes_sanders81,
  combes+90, quillen02, quillen+14} and is observed in many external
galaxies.  \citet{li_shen12} showed that a buckled bar model is fully
able to account for this morphology in the Milky Way.  Meanwhile data
from the BRAVA, ARGOS and APOGEE surveys have shown that the bulge is
cylindrically rotating \citep{howard+08, ness+13b, ness+16b}, which is
the typical velocity field produced by bars.  A more detailed
comparison of a barred simulation with kinematics from the BRAVA
survey \citep{howard+08, kunder+12} by \citet{jshen+10} noted that the
velocity and velocity dispersion profiles can be reproduced provided a
slowly rotating bulge component constitutes less than $8\%$ of the
disc mass.  Thus, these properties all favour the presence of a bulge
formed by the vertical thickening of the bar (see the reviews of
\citet{kormendy_kennicutt04} and \citet{fisher_drory16}).

The alternative path to the formation of bulges is their accretion as
part of the hierarchical growth of galaxies \citep{kauffmann+93,
  guedes+13}.  Various lines of evidence favouring the existence of
such a component in the Galactic bulge have been put forth.
Photometric studies of resolved stellar populations have generally
found {\it uniformly} old stars across the bulge, typically older than
$10\Gyr$ \citep{ortolani+95, kuijken_rich02, zoccali+03, ferreras+03,
  sahu+06, clarkson+08, clarkson+11, brown+10, valenti+13,
  calamida+14}.  For instance, \citet{clarkson+11} estimate that any
component younger than $5\Gyr$ must account for less than $3.4\%$ of
the bulge.  The Milky Way bulge also exhibits a metallicity gradient
in the vertical direction \citep{zoccali+08, gonzalez+11, johnson+11,
  johnson+13}.  This vertical gradient was thought to be impossible if
the bulge formed via the buckling instability.  Evidence for an
accreted bulge was also inferred from the old and metal-poor RR~Lyrae
at the centre of the Galaxy, which \citet{dekany+13} showed traced a
more axisymmetric, weakly barred shape than the red clump stars.
Lastly, the velocity dispersion of metal-poor stars on the minor-axis
above $|b| \simeq 4\degrees$ becomes flat, while that of metal-rich
stars continues declining \citep{babusiaux16}, suggesting the
existence of a second component.

One interpretation of these seemingly conflicting results is that the
Milky Way has a compound bulge \citep{athanassoula05, debattista+05},
with both a secular B/P bulge and an accreted classical bulge present
\citep{babusiaux+10, hill+11, zoccali+14}.  Many external galaxies
have been found to host compound bulges.  For instance,
\citet{mendez-abreu+14} found that $\sim 30\%$ of their sample of
face-on barred galaxies showed signs of multiple bulge components,
while \citet{erwin+15} presented a sample of nine galaxies with composite
bulges \citep[see also][]{erwin+03}.  Thus, composite bulges are not
rare, and it is conceivable that the Milky Way has one too.  A
compound bulge can easily produce the vertical metallicity gradient
observed via the variation of the relative importance of two
components \citep{grieco+12}.

One of the challenges that pure secular evolution models have faced is
why the properties of the bulge should be different for different
populations (usually defined via the metallicity of the stars).  Since
gravity cannot act differently on stars of different metallicities, a
number of studies have postulated that the different stellar
populations represent different structural components, or perhaps
different regions of the initial disc.
\citet{martinez-valpuesta_gerhard13} demonstrate that, contrary to
naive expectations, buckling can generate a vertical metallicity
gradient, consistent with that observed, out of steep {\it radial}
gradients.  \citet{pdimatteo+14} instead consider pure $N$-body
simulations including a classical bulge and find that stars initially
in the disc as far out as the outer Lindblad resonance of the bar
contribute to the boxy bulge, comprising up to $30\%$ of stars at high
latitudes.  In these scenarios, assuming that the initial disc has a
declining metallicity profile, a puzzling observation is that the
X-shape is weak or absent in the metal-poor stars \citep{ness+12,
  uttenthaler+12, rojas-arriagada+14}.  \citet{bekki_tsujimoto11}
instead considered the evolution of a thin$+$thick disc model, where
the thick disc was metal-poor and the thin disc was metal-rich.  In
this way, they were able to build a vertical metallicity gradient in
the bulge.  \citet{pdimatteo+15} argued that a single thin disc
model is unable to produce the observed split red clump in the
metal-rich stars and no split in the metal-poor red clump stars.

As argued by \citet{pdimatteo+15}, we do not expect that a single thin
stellar disc is able to explain the full range of data uncovered for
stars in the bulge because there is nothing to distinguish different
chemical populations.  Like those authors, we find that {\it multiple}
discs are needed to explain the observational data.  However, we show
that it is not primarily the disc thickness that drives the
differences between stellar populations in the bulge, but rather the
initial random motion in the disc plane, a situation which arises
naturally in a disc during its early rapid growth phase.  In this
paper, we consider both purely collisionless $N$-body simulations
comprised of multiple stellar populations with different kinematics
and a simulation with star formation, in order to understand how the
structure of the Milky Way's bulge arises.  We discover that stellar
populations separate (further) in the presence of a bar, a process we
refer to as {\it kinematic fractionation}. We explore the implications
of this separation for the morphology, kinematics, ages, chemistry and
microlensing of the bulge.

This paper is organized as follows.  In Section \ref{sec:theory}, we
present a simple dynamical interpretation for why kinematic
fractionation occurs.  Section \ref{sec:nbodysims} presents a study
based on pure $N$-body simulations of kinematic fractionation, showing
that the radial velocity dispersion is the main driving factor
in the separation and that the vertical velocity dispersion plays a
smaller role.  We then demonstrate the same behaviour in a simulation
with gas and star formation in Section \ref{sec:run708}.  Section
\ref{sec:milkyway} compares the trends in the resultant bulge of this
simulation with those in the Milky Way's bulge, showing that all the
observed trends can be reproduced, with only an additional $5\%$ hot
component (a stellar halo) needed to match the kinematics at low
metallicity.  We also predict new trends for testing for kinematic
fractionation in the Milky Way.  Section \ref{sec:discussion}
discusses these results including making predictions for the
metallicity and age distributions of B/P bulges in edge-on galaxies,
and ends with a summary of our conclusions.

%%%%%%%%%%%%%%%%%%%%%%%%%%%%%%%%%%%%%%%%%%%%%%%%%%%%%%%%%%%%%%%%%%%%%%%%%%%%%

\section{Theoretical considerations}
\label{sec:theory}

Below we present models in which stellar populations with different
initial in-plane kinematics, but identical density distribution,
separate as the bar forms and grows.  We refer to this behaviour as
{\it kinematic fractionation}, in analogy with chemical fractionation
produced by phase transitions.  The key result of these simulations is
that the populations with larger initial in-plane random motion go on
to become thicker and less peanut-shaped, and to host weaker bars.

Before presenting these simulations, we seek to understand
qualitatively why radially hotter discs become vertically thicker.
\citet{merritt_sellwood94} showed that bending instabilities in discs
occur when the vertical frequency of stars, $\nu$, exceeds the
frequency with which they encounter a vertical perturbation, allowing
them to respond in phase with a perturbation, thereby enhancing it.
Bars are the most important source of bending instabilities for two
reasons.  First, because of their intrinsic geometry, bars enforce
$m=2$ (saddle-shaped) bends, which generally are very vigorous.
Secondly, bars raise the in-plane velocity dispersion, but do not
substantially alter the vertical dispersion unless they suffer bending
instabilities.  Why does this matter?  Consider the case of a vertical
$m=2$ perturbation rotating at frequency $\Omega_p$, the pattern speed
of the bar.  The frequency at which stars encounter this perturbation
is $2(\Omega - \Omega_p)$, where $\Omega$ is the mean angular
frequency of the stars.  Thus, stars will support the perturbation as
long as
\begin{equation}
\nu > 2(\Omega - \Omega_p),
\label{eqn:freqs}
\end{equation}
enabling it to grow inside corotation.  If we separate disc stars into
two populations based on their radial random motion, we can label
population `h' (hot) as the one with the larger radial velocity
dispersion and population `c' (cool) the one with the smaller
dispersion.  It is well known that the asymmetric drift, $\Delta V
\equiv V_{circ} - \avg{V_\phi} \simeq x \sig{R}^2/V_{circ}$,
\citep[e.g.][section 4.2.1(a)]{bt87}, where $V_{circ}$ is the circular
velocity and $\avg{V_\phi}$ is the mean streaming velocity of the
stars and $x$ is a factor that depends on the density distribution and
orientation of the velocity ellipsoid; $-1 < x < 1$ at one disc
scalelength.  Thus
\begin{equation}
\Omega \simeq \Omega_{circ} - \frac{x}{RV_{circ}}\sig{R}^2,
\end{equation}
where $\Omega_{circ}$ is the frequency of circular orbits at radius
$R$.  Therefore, $\Omega_h < \Omega_c$: raising \sig{R}\ lowers
$\Omega$.  While we have assumed near axisymmetry to make this
argument, the underlying principle, that hotter populations rotate
more slowly, holds also when the perturbation is stronger since the
mean streaming motion is just one component supporting a population
against collapse.  Stars in population `h' satisfy inequality
\ref{eqn:freqs} at a smaller $\nu$ than those in population `c'.  A
smaller $\nu$ is achieved when the maximum height, $z_{max}$, that the
star reaches is larger.  Since, as a population thickens, $\nu$
declines faster than $\Omega$ (Appendix \ref{app:orbits} presents
orbit integrations which demonstrate this result), a bend in population
`h' will saturate at a larger $z_{max}$ than in population `c'.  We
conclude that in general radially hotter populations can thicken more
than radially cooler ones.

This separation of different kinematic populations does not end at the
buckling instability.  Bars transfer angular momentum to the halo
slowing down in the process \citep{weinberg85, debattista_sellwood98,
  debattista_sellwood00, athanassoula02, oneill_dubinski03}.  With
$\Omega_p$ declining and $\sigma_R$ rising, the right-hand side of
equation \ref{eqn:freqs} drops, allowing more stars to respond in
phase with further perturbations.  As long as the bar is slowing,
therefore, the disc will continue to thicken at different rates for
different radial dispersion populations, allowing the separation of
populations to persist in subsequent evolution.  As examples of this
behaviour, \citet{martinez-valpuesta+06} found that slowing bars that
had buckled already can buckle again, while \citet{debattista+06} was
able to induce buckling in an otherwise stable bar by slowing it down
with an impulsively imposed torque.

%%%%%%%%%%%%%%%%%%%%%%%%%%%%%%%%%%%%%%%%%%%%%%%%%%%%%%%%%%%%%%%%%%%%%%%%%%%%%

\section{Demonstration using pure $N$-body models}
\label{sec:nbodysims}

The initial conditions usually employed in pure $N$-body simulations
studying disc galaxy evolution have generally set up a disc composed
of a single distribution function, equivalent to a single stellar
population.  In reality, discs are composed of multiple populations of
different ages with different dispersions; even in the Solar
neighbourhood, the radial, tangential and vertical velocity
dispersions are all a function of age \citep[e.g.][]{wielen77,
  nordstrom+04, holmberg+09, aumer_binney09, casagrande+11}.  In the
early Universe, disc galaxies were subject to significant internal
turmoil \citep[e.g.][]{glazebrook13}, and had lower mass, which
resulted in rapid heating.  As a consequence, in young galaxies, the
stellar random motion increases rapidly with age over a relatively
short period of time.  Alternatively, early discs may form directly
hot, with subsequent populations forming in cooler, thinner
distributions \citep{kassin+12, bird+13, stinson+13, wisnioski+15,
  grand+16}.  The outcome of this upside-down evolution is similar:
discs with random motions increasing as a function of age.

In order to explore the effect of the age dependence of random motions
on the evolution of different populations, we first study the
behaviour of multipopulation discs by means of carefully controlled
pure $N$-body (\ie\ gasless) simulations where the initial conditions
are comprised of multiple superposed stellar discs with different
initial kinematics.  We run two types of simulations: one in which
multiple stellar discs are exactly co-spatial but have different
in-plane kinematics (described in Section \ref{ssec:run741}), and
another set with only two discs, with different initial heights
(described in Section \ref{ssec:run742}).

\subsection{Co-spatial discs}
\label{ssec:run741}

In the first set of simulations, we build systems in which the initial
disc is comprised of five separate populations with different in-plane
velocity dispersions but identical density distribution.  We set up
initial conditions using {\sc GalactICS} \citep{widrow+08,
  widrow_dubinski05, kuijken_dubinski95}.  We construct a model
consisting of an exponential disc and a Navarro-Frenk-White (NFW) halo
\citep{nfw96}, with no bulge component.  We truncate this NFW halo as
follows:
\begin{equation}
\rho(r) = \frac{2^{2-\gamma} \sigma_h^2}{4\pi a_h^2}
\frac{1}{(r/a_h)^\gamma (1+r/a_h)^{3-\gamma}}C(r;r_h,\delta r_h),
\end{equation}
\citep{widrow+08}.  Here, $C(r;r_h,\delta r_h)$ is a cutoff function to
smoothly truncate the model at a finite radius, and is given by:
\begin{equation}
C(r;r_h,\delta r_h) =
\frac{1}{2}\mathrm{erfc}\left(\frac{r-r_h}{\sqrt{2}\delta r_h}\right).
\end{equation}
Our choice of parameters is $\sigma_h = 400\kms$, $a_h =
16.7\kpc$, $\gamma = 0.873$, $r_h = 100\kpc$ and $\delta r_h =
25\kpc$.

\begin{figure}
  \includegraphics[angle=-90.,width=\hsize]{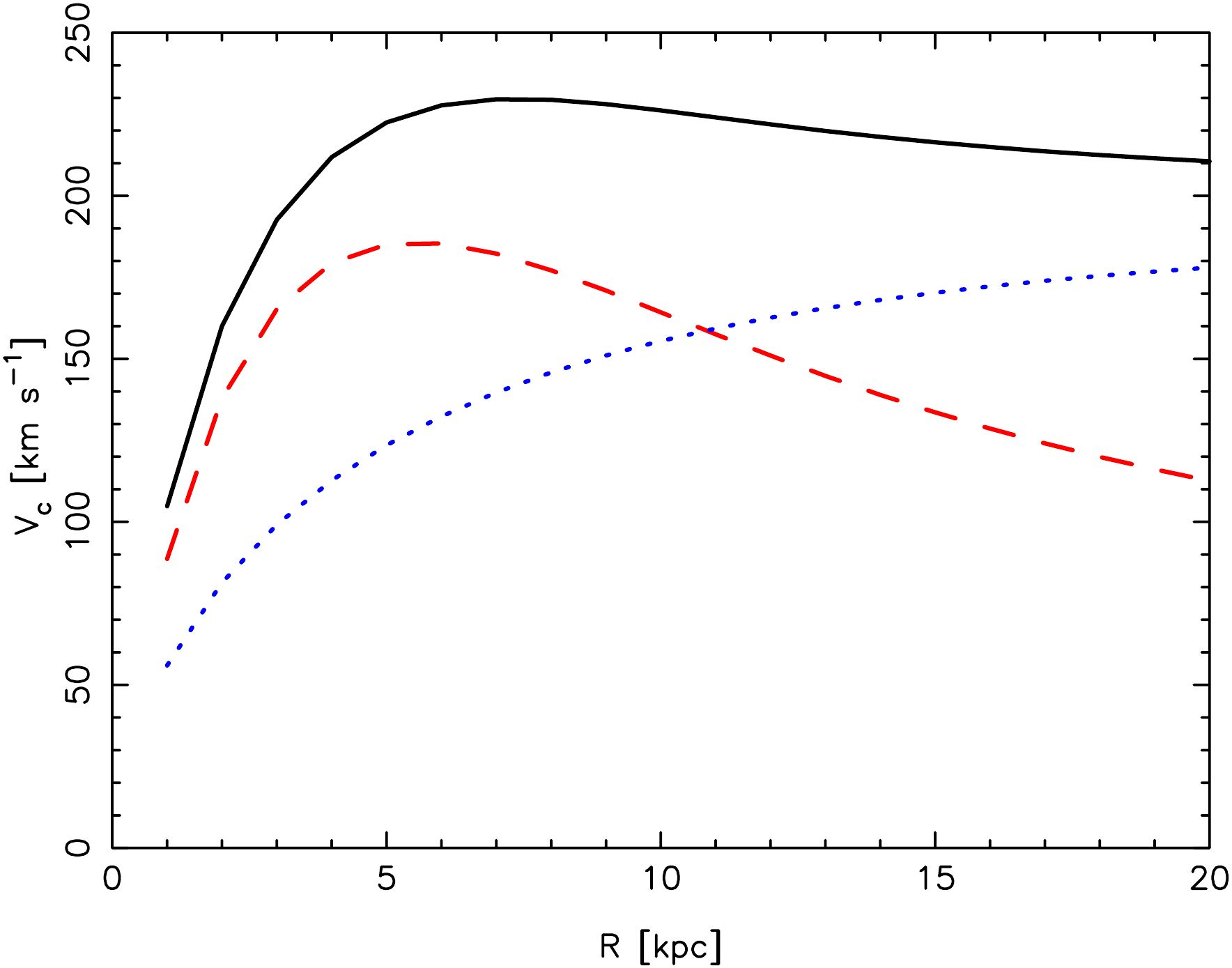}
  \includegraphics[angle=-90.,width=\hsize]{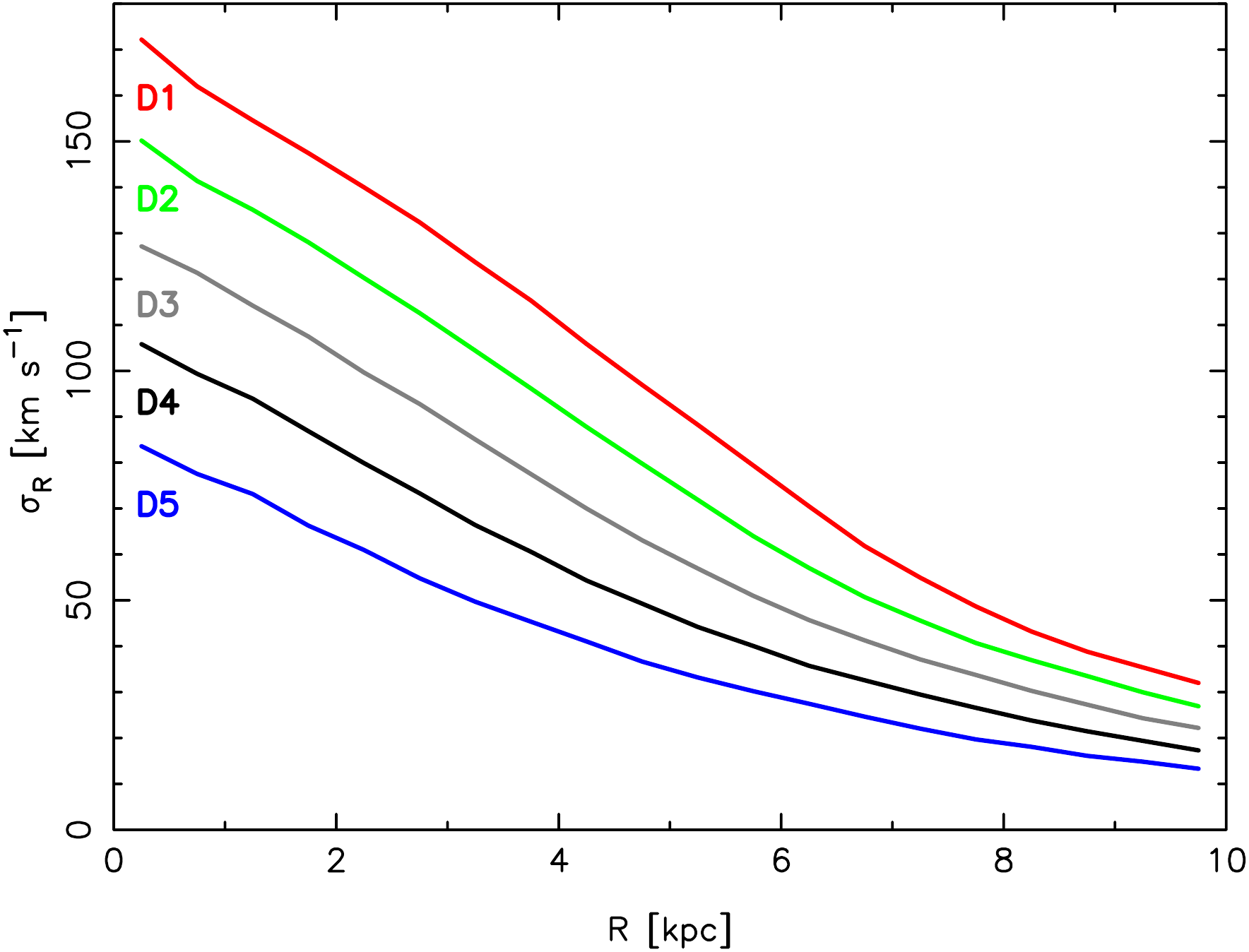}
  \includegraphics[angle=-90.,width=\hsize]{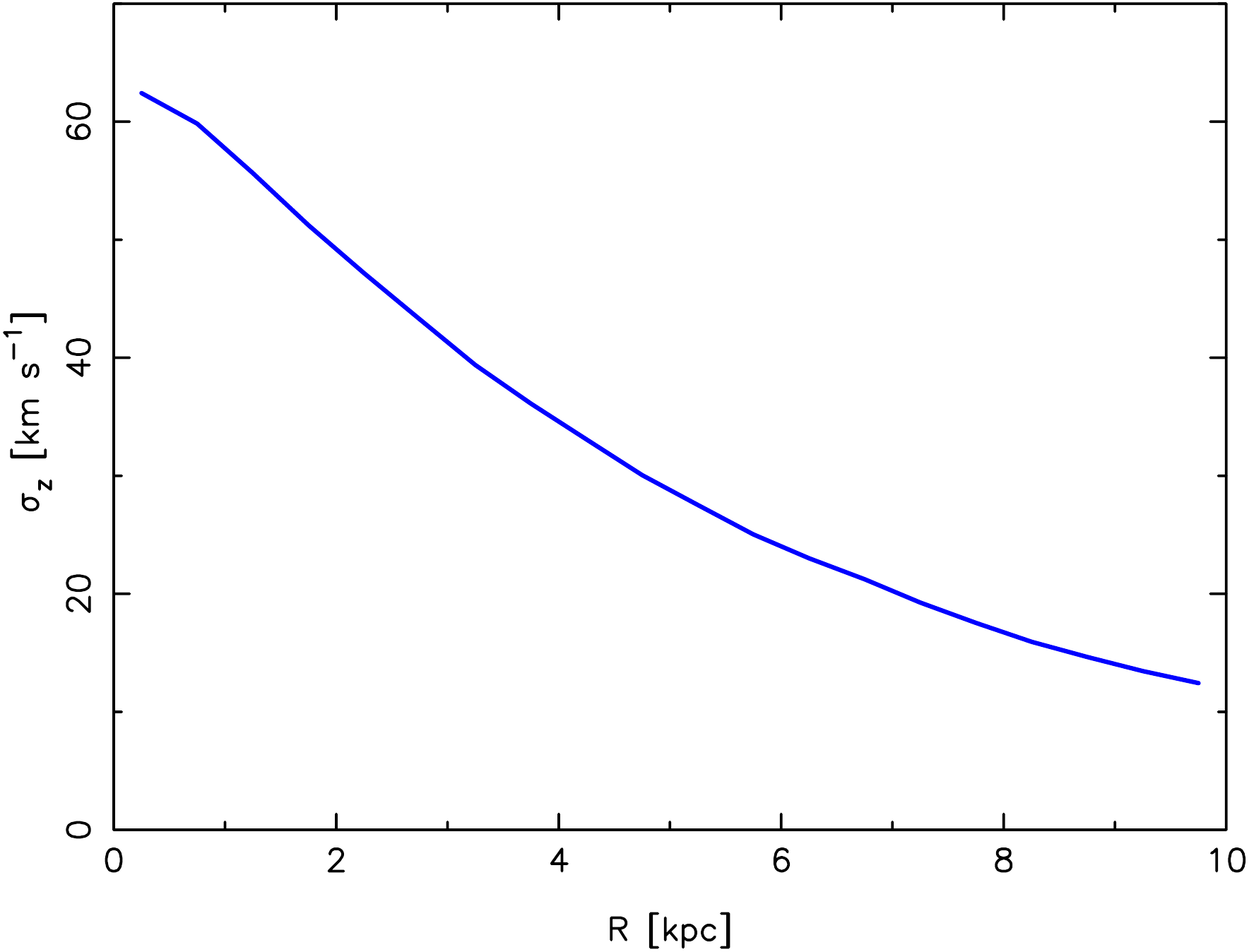}
\caption{Top: the rotation curve of the pure $N$-body models D1-D5.
  The dotted (blue) curve shows the halo contribution, the dashed
  (red) curve shows that of the stars, and the solid (black) curve
  shows the full rotation curve.  Middle: radial velocity dispersions,
  \sig{R}, for discs D1 (red) to D5 (blue).  Bottom: vertical
  velocity dispersion, \sig{z}, of the disc in model D5.  Models D1-D4
  have \sig{z}\ almost identical to this.
  \label{fig:run741ICs}}
\end{figure}

We build model discs with an exponential profile:
\begin{equation}
\Sigma(R,z) = \Sigma_{0} \exp(-R/\Rd)~\mathrm{sech}^2 (z/\zd),
\end{equation}
where \Rd\ is the disc scalelength and \zd\ is the scaleheight.  Our
discs have $\md = 2 \pi \Sigma_0 \Rd^2 = 5.2\times10^{10} \Msun$, $\Rd
= 2.4\kpc$ and $\zd = 250\pc$.  The rotation curve resulting from this
set of parameters is shown in the top panel of Fig.
\ref{fig:run741ICs}.  We construct five versions of this disc density,
with different radial velocity dispersion profiles:
\begin{equation}
\sigma_R^2(R) = \sigma_{R0}^2 \exp(-R/R_\sigma).
\end{equation}
We vary the central velocity dispersion, \sig{R0}, fixing $R_\sigma$
to $2.5\kpc$.  We select five values of \sig{R0}, $190\kms$ (disc D1),
$165\kms$ (disc D2), $140\kms$ (disc D3), $115\kms$ (disc D4) and
$90\kms$ (disc D5).  This large range of \sig{R0}\ was chosen to
enhance the effect of kinematic fractionation; the evolution of
further models with a narrower range of \sig{R0}\ is qualitatively
similar to what we will show here. The middle panel of Fig.
\ref{fig:run741ICs} shows the different radial velocity dispersion
profiles of the five discs.  Although we vary the in-plane kinematics
of the initial discs, the vertical density profile of each disc is
identical and therefore the vertical velocity dispersion, \sig{z},
shown in the bottom panel of Fig. \ref{fig:run741ICs}, is too.
We use $6\times10^6$ particles in each disc and $4\times10^6$
particles in the halo, giving particles of mass $\simeq 1.1 \times
10^4 \Msun$ (disc) and $\simeq 1.7\times 10^5 \Msun$ (halo).

Once we build the five discs, we construct a further two compound
discs which are superpositions of different subsamples of D1-D5.  In
model CU (`compounded uniformly'), we sample each of the discs equally.
Thus, we extract from each one-fifth of their particles and join them
together to form a new composite disc with ratio D1:D2:D3:D4:D5 set to
1:1:1:1:1.  The second disc, model CL (`compounded linearly'), samples
the discs assuming a linearly decreasing contribution from D1 to D5 in
the ratio 5:4:3:2:1.  These composite systems mimic a system with a
continuum of stellar populations, in the case of CU with a constant
star formation rate, while CL corresponds to a declining star
formation rate (assuming that the kinematically hotter populations are
older).  We therefore refer to each of disc D1-D5 in models CU and CL
as different populations.

Because the overall mass distribution is unchanged by the resampling,
by the linearity of the Boltzmann and Poisson equations, the composite
discs CL and CU are also in equilibrium.  We evolve these models using
{\sc Pkdgrav} \citep{pkdgrav}, with a particle softening of $\epsilon
= 50$ and $100\pc$ for star and halo particles, respectively.  Our
base timestep is $\Delta t = 5\Myr$ and we refine timesteps such that
each particle's timestep satisfies $\delta t = \Delta t/2^n < \eta
\sqrt{\epsilon/a_g}$, where $a_g$ is the acceleration at the
particle's current position; we set $\eta = 0.2$.  Our gravity
calculation uses an opening angle of the tree code $\theta = 0.7$.

\subsubsection{Evolution of single-population discs}

\begin{figure}
\centerline{\includegraphics[angle=0.,width=\hsize]{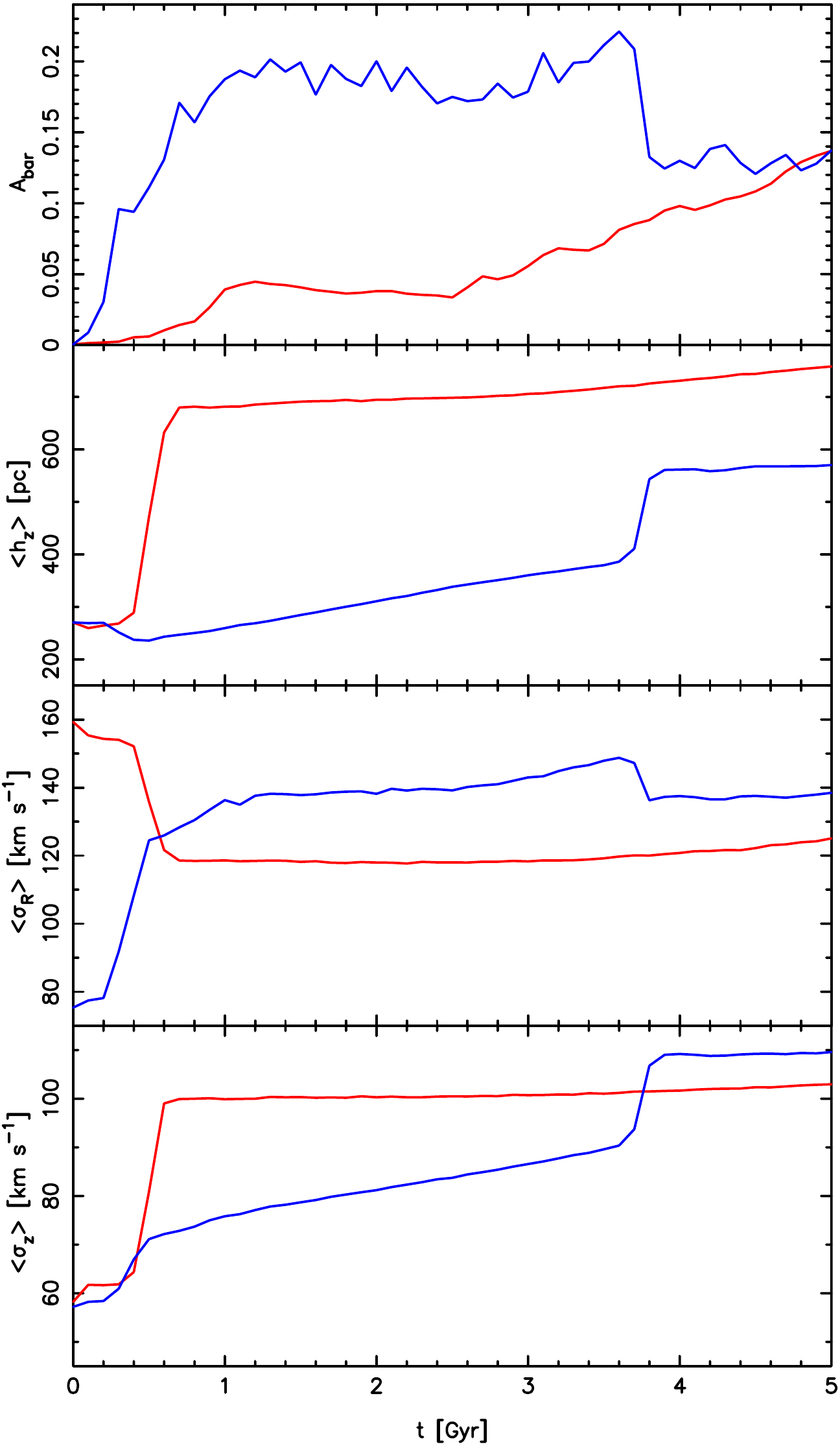}}
\caption{Evolution of the pure $N$-body single-population models with
  discs D1 (red lines) and D5 (blue lines) when evolved as
  single-population systems.  From top to bottom, we plot the bar
  amplitude, \abar, the average height, $\avg{h_z}$, the average
  radial velocity dispersions $\avg{\sig{R}}$ and the average vertical
  velocity dispersion $\avg{\sig{z}}$.  Profiles are averaged in the
  radial range $R \leq 2\kpc$ to match the region where $h_z$ peaks.
  The two discs evolve very differently, with disc D1 (the hotter
  disc) heating vertically at the expense of in-plane random motions.
  \label{fig:d1d5evol}}
\end{figure}

Before considering the multi-population discs, we demonstrate that the
single-population discs evolve very differently from each other using
the two most different models, disc D1 and D5.
Fig.~\ref{fig:d1d5evol} shows their evolution.  The top panel of
Fig. \ref{fig:d1d5evol} shows the bar amplitude, defined as
\begin{equation}
\abar = \left|\frac{\sum_i m_i e^{2 i \phi_i}}{\sum_i m_i}\right|
\end{equation}
where the sums extend over all star particles in a given disc
population, $m_i$ is the mass of the $i$th particle (which is the
same for all star particles in these pure $N$-body simulations), and
$\phi_i$ is its cylindrical angle.
In both models the bar forms by 1 Gyr, although the bar in model D1 is
best characterized as a very weak oval at this point.  Later, the two
bars reach a semi-major axis of $\sim 6.5\kpc$.  The bar in model D5
buckles at $\sim 3.7 \Gyr$, at which point its amplitude decreases.
In model D1 instead an axisymmetric buckling occurs at 0.4 \Gyr,
before the bar forms.  We measure the radial profile of the
root-mean-square height of star particles, $h_z$, and average this
profile to a radius of $2\kpc$, \avg{h_z}; the second panel from the
top shows its time evolution.  Whereas \avg{h_z} evolves mildly in
model D5 until the bar buckles, in model D1 \avg{h_z}\ rises very
sharply from $250$ to $\sim 700\pc$, while the bar is still forming;
thereafter it barely evolves any further.  In model D5 instead the
rise in \avg{h_z} is initially driven by slow heating by the bar and
abruptly by the buckling.  The third and fourth panels show the
evolution of the radial and vertical velocity dispersions averaged
over the same radial range.  In model D5 \avg{\sig{R}} rises rapidly
once the bar forms, while \avg{\sig{z}} increases only slightly.  The
buckling in model D5 increases \avg{\sig{z}} at the expense of
\avg{\sig{R}}.  Instead, in model D1 a large fraction of the random
motion in the plane is transformed into vertical random motion.  At
later times, both \avg{\sig{R}}\ and \avg{\sig{z}}\ evolve very
little.

The vertical evolution of a system composed of a single population is
therefore strongly determined by the radial velocity dispersion before
bar formation.  A radially hot disc efficiently transforms a large
fraction of its in-plane random motion into vertical random motion.
If a hot disc and a cold disc are coincident, a reasonable expectation
is that the cold disc drives the formation of a strong bar, which
transforms a large part of the hot disc's radial random motion into
vertical random motion, thickening the system overall.

\subsubsection{Evolution of multi-population co-spatial discs}

\begin{figure}
\centerline{\includegraphics[angle=0.,width=\hsize]{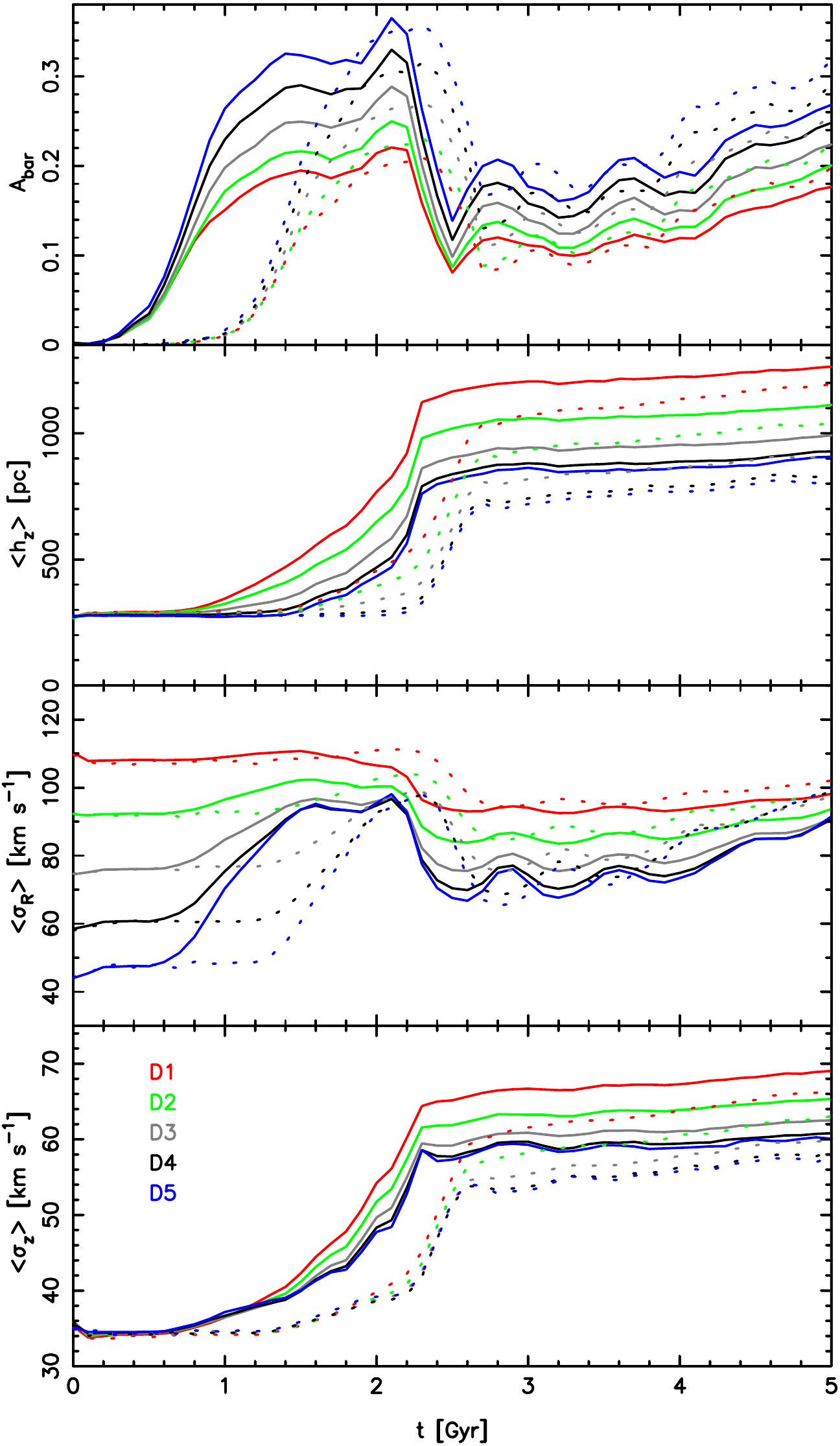}}
\caption{Evolution of the different populations in model CL.  Top: bar
  amplitudes.  Upper middle: average heights, \avg{h_z}.  Lower
  middle: average radial velocity dispersions, \avg{\sig{R}}.  Bottom:
  average vertical velocity dispersions, $\avg{\sig{z}}$.  Populations
  D1-D5 are indicated by red to blue solid lines (respectively).
  The dotted lines show the evolution of the same system when we
  employ a quiet start, as described in the text.  Profiles are
  averaged in the radial range $2 \leq R/\kpc \leq 6$, to match the
  region of peak $h_z$.  Note the separation of the different
  populations, both in the plane (in bar strength) and vertically.
  \label{fig:run741CLevol}}
\end{figure}

Fig. \ref{fig:run741CLevol} presents the evolution of model CL
(compounded linearly), separately for each disc population.  The top
panel shows the bar amplitude; the bar is strongest in the coolest
population (D5) and weakest in the hottest (D1).  This is the case at
bar formation ($t\simeq 1.5\Gyr$) and continues to be so after the
strong buckling starting at around $t=2.1\Gyr$.  The average height,
\avg{h_z}, shown in the second panel of Fig.~\ref{fig:run741CLevol},
increases the least in population D5, while D1 starts thickening once
the bar starts forming, when the disc starts developing bends in the
vertical density distribution.  Even after buckling saturates,
population D1 continues to be the thickest population, as anticipated
by the discussion in Section \ref{sec:theory}.  The third row of
Fig.~\ref{fig:run741CLevol} shows the average radial velocity
dispersion, \avg{\sig{R}}; this rises rapidly for population D5 at bar
formation and then drops at buckling as the bar is weakened, whereas
for population D1 bar formation does not change \avg{\sig{R}} while
buckling radially {\it cools} it slightly.  The bottom row of
Fig.~\ref{fig:run741CLevol} shows the vertical velocity dispersion,
\avg{\sig{z}}, which rises for all populations.  While the evolution
tends to drive \avg{\sig{R}}\ towards convergence, that of
\avg{\sig{z}}\ tends to be parallel for the different populations,
even though they are evolving in the same potential.  The bar
amplitude, height and random motions all vary nearly monotonically
with increasing initial in-plane random motions.

We have verified that the evolution, and especially the rapid vertical
thickening of the different populations, is indeed driven by the
growing bar (rather than being a numerical artefact or imperfect
initial conditions) by employing a quiet start \citep{sellwood83}, in
which $2.5\%$ of the particles from each initial disc population, and
from the halo, are drawn from the full system.  Each chosen particle,
with phase space coordinates $(x,y,z,v_x,v_y,v_z)$, is then replaced
by 40 particles as follows:
\begin{equation}
\left(
\begin{array}{c}
  x\\
  y\\
  z\\
  v_x\\
  v_y\\
  v_z\\
  \end{array}
  \right)
\rightarrow \sum_{n=0}^{19} {\cal
  R}\left(\frac{n\pi}{10}\right)
\left(
\begin{array}{c}
  x\\
  y\\
  z\\
  v_x\\
  v_y\\
  v_z\\
 \end{array}
 \right)
 + \sum_{n=0}^{19} {\cal
   R}\left(\frac{n\pi}{10}\right)
\left(
\begin{array}{c}
  x\\
  y\\
  -z\\
  v_x\\
  v_y\\
  -v_z\\
 \end{array}
 \right)
\end{equation}
where ${\cal R}(\phi)$ is the rotation matrix for a rotation by an
angle $\phi$ about the $z$-axis.  This setup leads to a very low seed
$m=2$ perturbation out of which the bar must grow, delaying its
formation.  The dotted lines in Fig. \ref{fig:run741CLevol} show that
the bar forms about $0.5\Gyr$ later with this quiet start.  The
thickening and heating of the different populations is then also
delayed, which would not occur if thickening arises for artificial
numerical reasons, and must therefore be driven by the bar.

\begin{figure*}
\centerline{\includegraphics[angle=0.,width=\hsize]{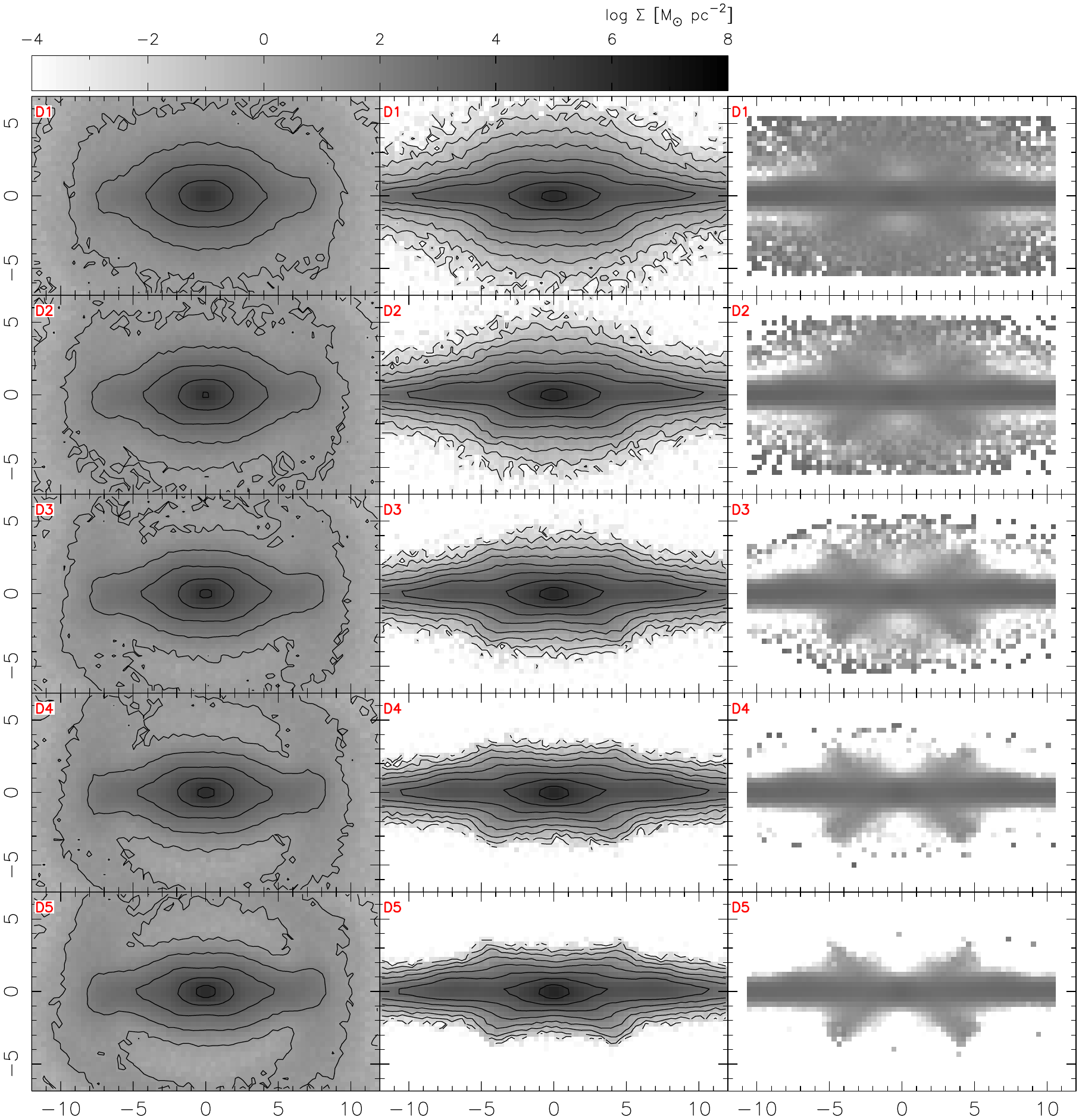}}
\caption{The face-on (left-hand panels) and edge-on (middle panels) surface
  density of stars in each disc population of model CU at $5\Gyr$.
  These two columns share a common density scale indicated by the
  wedge at top.  The right-hand panels show an unsharp mask of the edge-on
  image using a square kernel of width $1.4\kpc$.  From top to bottom
  are shown populations D1 (initially radially hottest) to D5
  (initially radially coolest).  Decreasing initial \sig{R}\ leads to
  a stronger quadrupole, a thinner disc and a more prominent peanut
  shape.
  \label{fig:run741CUdens}}
\end{figure*}

The evolution of model CU (compounded uniformly) is qualitatively
similar to that of CL.  Because the number of particles in each
population is the same in model CU, it makes visual comparison between
different populations easier.  Fig. \ref{fig:run741CUdens} presents
the final density distribution of each population in model CU.
In the face-on view, population D1 has a significantly rounder bar
with a weaker quadrupole moment, while the bar in population D5 has a
stronger quadrupole moment. None the less, the bar has the same size in
all five populations.  The difference in the vertical structure is even
more striking; while population D1 has a thick centre and a somewhat
boxy shape, population D5 is thinner at the centre and thus exhibits a
peanut shape.  The edge-on unsharp masks in the right column reveal a
photometric X-shape in populations D3-D5 which is smeared out in
populations D1 and D2.

We have verified that the separation of populations is not dependent
on the presence of very hot populations by running a further model
comprised of an equal mix of discs D4 and D5, the two coolest discs.
The two populations separate to a degree comparable to what they did
in model CL.  Therefore kinematic fractionation does not require very
hot populations to be present.

\subsection{Evolution of discs with different heights}
\label{ssec:run742}

In this section, we present four different simulations each containing
two discs of different heights.  We use the same version of the
thicker disc in all four simulations.  The thinner discs have the same
density distribution and therefore also the same vertical random
motions in all four cases, but differ in their radial random motions.
In this way, we are able to study the relative importance of vertical
and radial random motions to the final B/P bulge morphology of the two
different populations.

\begin{figure}
\centerline{\includegraphics[angle=0.,width=\hsize]{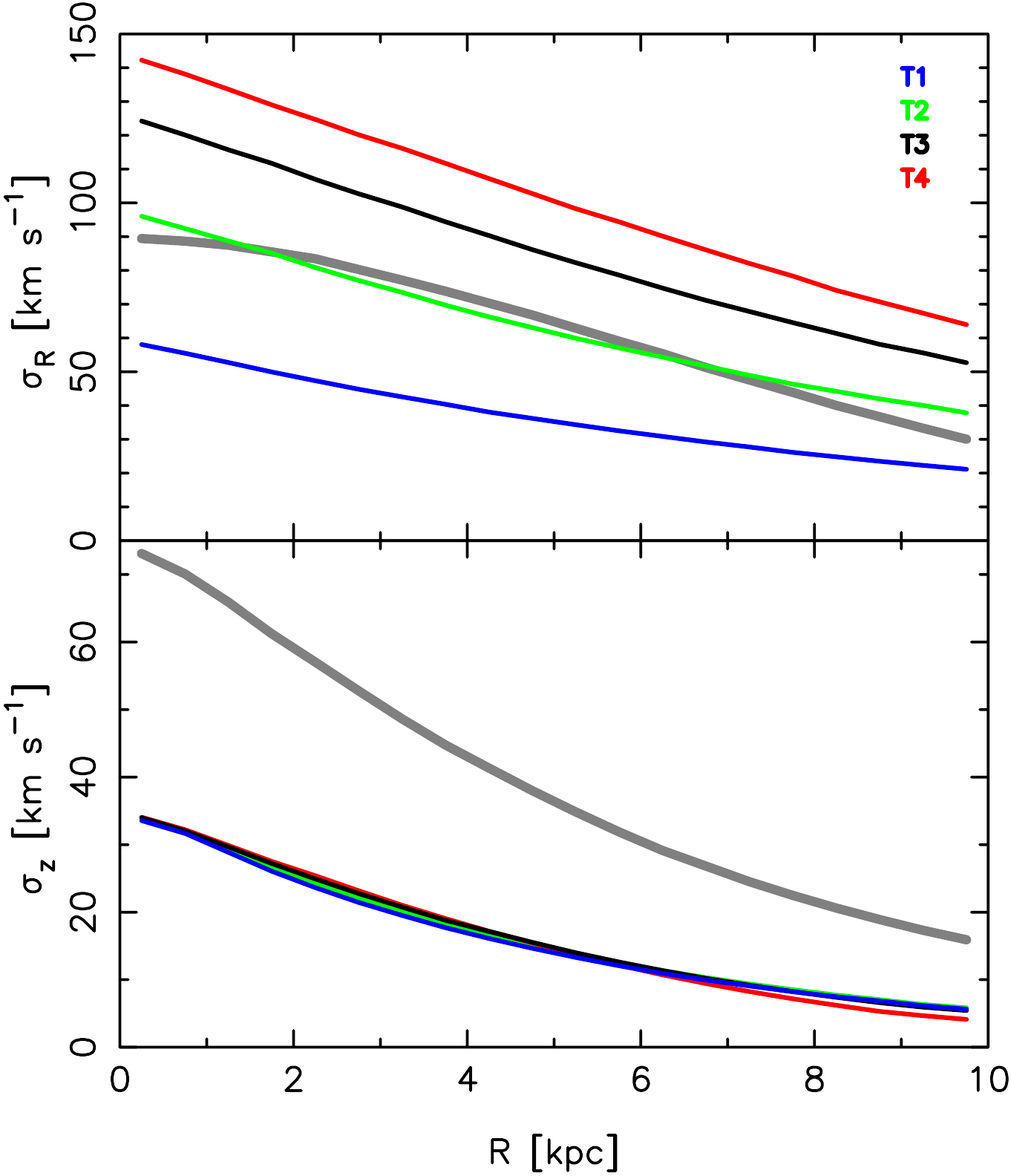}}
\caption{The initial conditions for the thinner$+$thicker discs
  simulations.  The top panel shows \sig{R}, while the bottom panel
  shows \sig{z}.  In both panels, the thick grey line shows the thicker
  disc, which is common to all four models, while the different thinner
  discs, T1-T4, are shown by the coloured lines as indicated in the
  top panel.  In model T2, the \sig{R}\ profile of the thinner disc
  (green line) is set to approximately match that of the thicker disc
  (grey line).
  \label{fig:742ICs}}
\end{figure}

In order to set up initial conditions with discs of different
thickness, we again use {\sc GalactICS}, which allows two discs with
different density distributions to be set up.  We choose identically
the same halo parameters as in models CL and CU, and again include no
bulges in the models.  Instead of a pseudo-continuum of populations as
in models CL and CU, we use only two discs, one thinner and the other
thicker.  Both discs have a scale-length $\Rd = 2.4\kpc$.  Each disc
has a mass $\md = 2.6\times 10^{10}\Msun$ (\ie\ half that of the total
disc in models CL and CU) and is represented by $3 \times 10^6$
particles.  One disc has a scaleheight $\zd = 100\pc$ (thinner disc),
while the other has $\zd = 400\pc$ (thicker disc).  These two discs
are therefore not to be thought of as the thin and thick discs of a
galaxy such as the Milky Way.  Indeed both discs are considerably
thinner than the Milky Way's thick disc and closer to the thin disc
\citep{juric+08}, which accounts for our terminology `thinner' and
`thicker' rather than `thin' and `thick'.  As before, $\sig{R}^2$
declines exponentially, from $\sig{R0} = 90\kms$ for the thicker disc.
We produce four thinner disc models.  In T1, the thinner disc is
radially cooler than the thicker disc, while in models T3 and T4, the
thinner disc is radially hotter.  In model T2, we set up the thinner
disc such that it has almost the same \sig{R}\ profile as the thicker
disc.  Fig. \ref{fig:742ICs} shows the initial conditions of these
systems.

\begin{figure}
\centerline{\includegraphics[angle=-90.,width=\hsize]{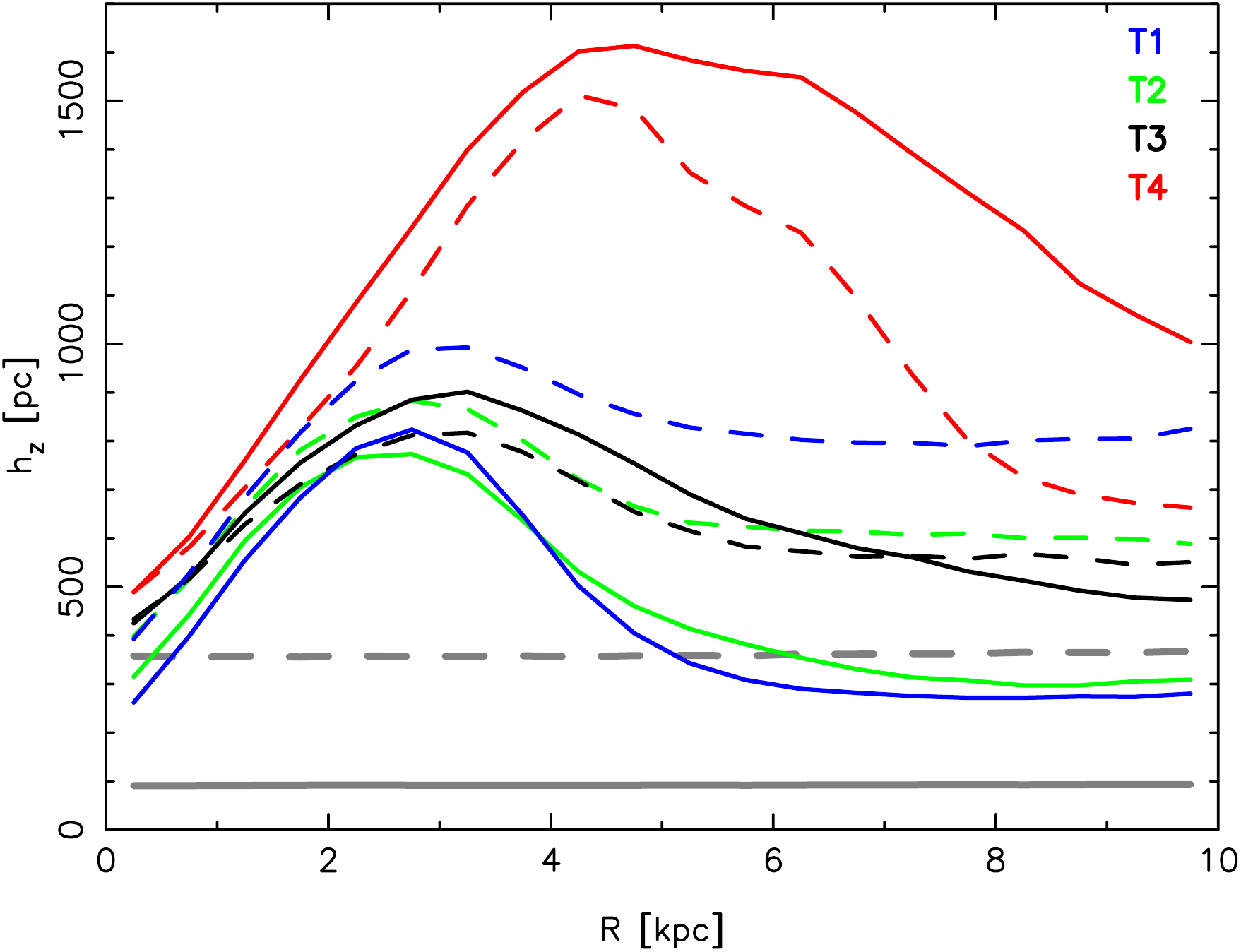}}
\caption{Initial and final height profiles for the thinner$+$thicker
  discs simulations.  The horizontal grey lines correspond to the
  initial conditions, with the solid line showing the thinner disc and
  the dashed line showing the thicker disc.  The remaining lines show
  the height profiles after evolving for $5\Gyr$.  Note the inversion
  in final thickness, with the thinner discs attaining larger heights
  than the thicker discs, in models T3 and, especially, T4.
  \label{fig:742heights}}
\end{figure}

\begin{figure}
\centerline{\includegraphics[angle=0.,width=\hsize]{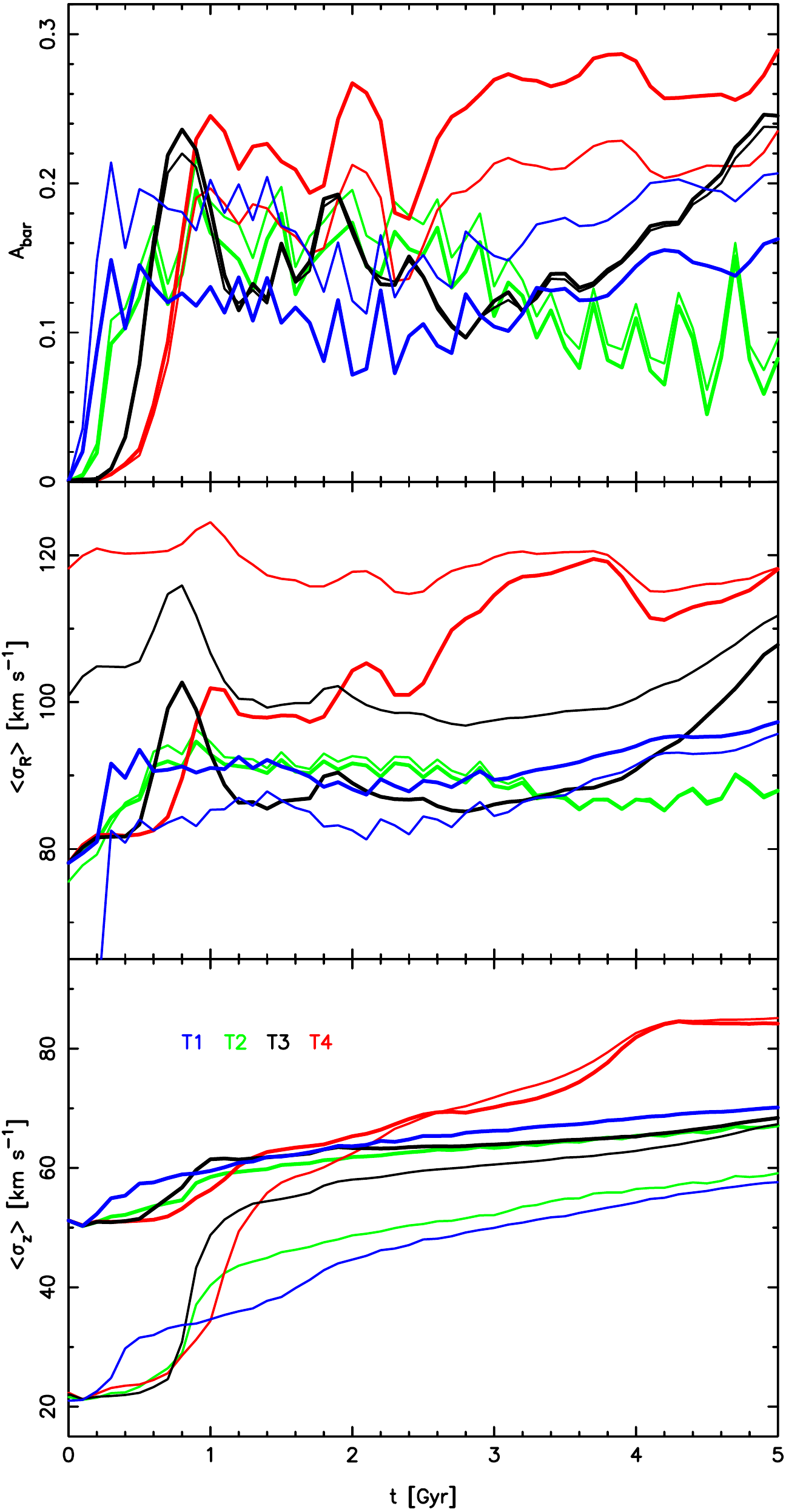}}
\caption{Evolution of the thinner$+$thicker discs simulations.  From
  top to bottom we plot the bar amplitude, \avg{\sig{R}}, and
  \avg{\sig{z}}.  The thin lines show the thinner discs with the thick
  lines showing the thicker discs.  Based on
  Fig. \ref{fig:742heights}, we average the profiles in the radial
  range $1 \leq R/\kpc < 5$, to match the region of peak $h_z$.  Note
  the change in the relative strength of the bars in the thinner and
  thicker discs between models T1 and T4, and the strong vertical
  heating in model T4.
  \label{fig:742evolution}}
\end{figure}

As with models CU and CL, we evolve these models using {\sc Pkdgrav}
with the same numerical parameters.  Fig. \ref{fig:742heights}
compares the final $h_z$ profiles of these models.  All four thinner
disc models start out with the same density distribution.  The radial
extent and degree to which both discs heat vertically increases from
model T1 to T4, \ie\ with increasing in-plane random motion of the
thinner disc.  In models T1 and T2, the thinner disc remains thinner
than the thicker disc, but in models T3 and T4, the thinner disc heats
to the extent that their height becomes larger than that of the
thicker disc at the end of the simulation.
Fig. \ref{fig:742evolution} shows the overall evolution of the
systems.  In models T2 and T3, the bar has equal amplitude in both
discs, while the bar is stronger in the thinner disc in T1 and in the
thicker (but radially cooler) disc in model T4.  The velocity
dispersions increase in all models, with \avg{\sig{z}}\ increasing
rapidly in the thinner disc around the time of bar formation.
\avg{\sig{z}} increases the most in model T4, for which the thinner
disc become vertically hotter than the thicker disc.  Buckling occurs
in these models at times ranging from 0.3 \Gyr\ (model T1) to 1.1
\Gyr\ (model T4).

\begin{figure*}
\centerline{\includegraphics[angle=0.,width=\hsize]{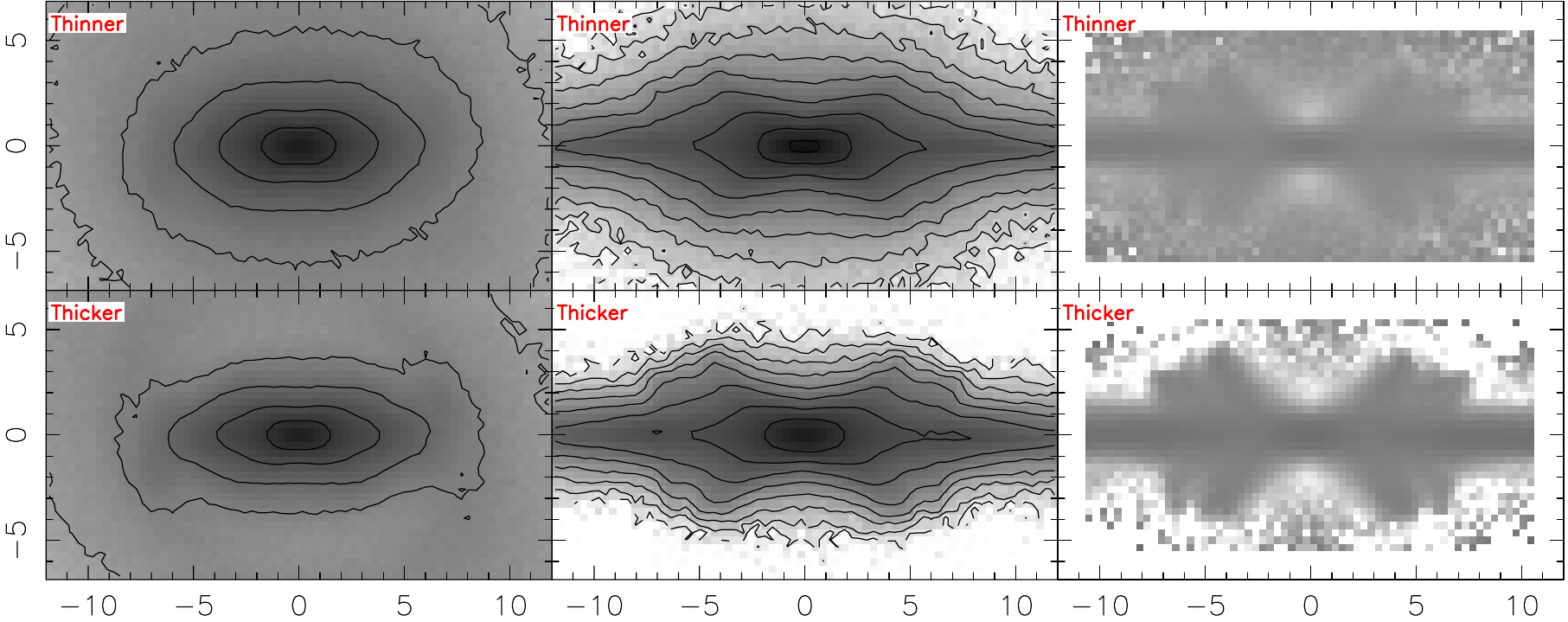}}
\caption{The face-on (left-hand panels) and edge-on (middle panels) surface
  density of stars in the thinner (top row) and thicker (bottom row)
  disc populations of model T4 at $5\Gyr$.  These two columns share a
  common density scale with the first two columns of
  Fig.~\ref{fig:run741CUdens}.  The right-hand panels show an unsharp mask
  of the edge-on image using a square kernel of width $1.4\kpc$.  The
  initially thinner disc ends thicker, with a less peanut-shaped
  bulge, than the initially thicker disc, demonstrating that the main
  driver of the final morphology is {\it not} the thickness of the
  initial disc, but its in-plane random motion.
  \label{fig:742Cdensity}}
\end{figure*}

Fig. \ref{fig:742Cdensity} shows face-on and edge-on density maps as
well as edge-on unsharp mask images of the thinner and thicker
populations in model T4.  The thicker population, which, initially,
has lower radial velocity dispersion than the thinner population, ends
up vertically cooler with a more readily apparent peanut shape, which
stands out clearly in the unsharp mask.  The thinner disc instead ends
up more vertically extended and hosting a weaker bar with a weaker
peanut shape than the thicker disc.  Thus, it is not the initial height
of the disc that is the strongest driver of the final bulge morphology
(\ie\ whether it is boxy or peanut-shaped), but the in-plane random
motions.

\subsection{Summary of the pure $N$-body simulations}

The key insights these pure $N$-body experiments provide can be
summarized as follows:
\begin{enumerate}
  
\item The formation and evolution of a bar causes co-spatial stellar
  populations, with different initial radial velocity dispersions, to
  separate.  The initially radially hotter populations are lifted to
  larger heights than the cooler populations, form a weaker bar, and
  give rise to a boxy/spheroidal shape, not a peanut shape.  The
  cooler populations instead form a peanut shape.

\item A thin, radially hot population thickens more than a thick,
  radially cool population.  While it is natural that a radially
  hotter population will also be vertically hotter and thicker, the
  initial thickness of the disc is not the main driver of the final
  morphology.  Instead, the initial in-plane random motions are more
  important in determining the subsequent bulge morphology than is the
  initial vertical dispersion.
  
\end{enumerate}

We stress that the initial conditions we have used in these pure
$N$-body simulations are not always meant to represent realistic
systems; indeed it would be hard to conceive of a formation scenario
which leads to a vertically thicker disc being radially cooler than a
thinner disc.  The goal of these experiments has been to show the
relative importance of the vertical and in-plane random motions to the
final morphology.

%%%%%%%%%%%%%%%%%%%%%%%%%%%%%%%%%%%%%%%%%%%%%%%%%%%%%%%%%%%%%%%%%%%%%%%%%%%%%

\section{Simulation with Star Formation}
\label{sec:run708}

Having explored the effect of different in-plane kinematics with pure
$N$-body simulations, we now examine a simulation in which the stars
form continuously out of gas, allowing the ages and random motions
(in-plane and vertically), as well as the chemical properties, to be
correlated.
We consider the evolution of the model presented by \citet{cole+14}
and \citet{ness+14}.  Briefly, in this simulation, a disc galaxy forms
entirely out of gas cooling from a spherical corona and settling into
a disc, triggering continuous star formation.  This star-forming
simulation, which was evolved with {\sc Gasoline} \citep{gasoline},
the smooth particle hydrodynamics version of {\sc Pkdgrav}, is
extremely useful for deciphering the evolution of the Milky Way's
bulge for a number of reasons.  It has high resolution ($50\pc$
spatially and with a stellar mass of $9.5 \times 10^3\Msun$), forms a
strong bar, and all stars form from gas, rather than put in by hand as
in most previous simulation models of the Milky Way
\citep[e.g.][]{sellwood85, fux97, fux99, jshen+10,
  martinez-valpuesta_gerhard11, saha+12, ness+13b, gardner+14}.  Other
studies \citep[e.g.][]{samland02, brook+04TD, house+11, guedes+13,
  bird+13} have considered the evolution of various aspects of the
Milky Way via simulations in which the disc forms from gas, but the
bar and bulge have received comparatively less attention in such work.
The simulation tracks the dispersal of iron and oxygen, using the
yields of \citet[][SN~II]{raiteri+96}, \citet[][SN~Ia]{thielemann+86},
and \citet[][stellar winds]{weidemann87}, allowing us to dissect the
model not just by age, but also by metallicity, \feh, and
$\alpha$-enhancement, \alfe.

\citet{cole+14} showed that the model forms between 2 and 4 \Gyr,
slowly growing longer after then (their figs 1 and 2).  By 10~\Gyr, a
B/P-shaped bulge has formed as can be seen in fig. 2 of
\citet{ness+14}.  Because we are interested in comparing the model to
the Milky Way, we scale the model in size and velocity to approximate
the size of the bar and the bulge rotational velocity as seen from the
Sun, as described in \citet{ness+14} (multiplying coordinates by 1.2
and velocities by 0.48).  We use the same scale factors at all times
when presenting this simulation.

\begin{figure}
\centerline{
\includegraphics[angle=0.,width=0.9\hsize]{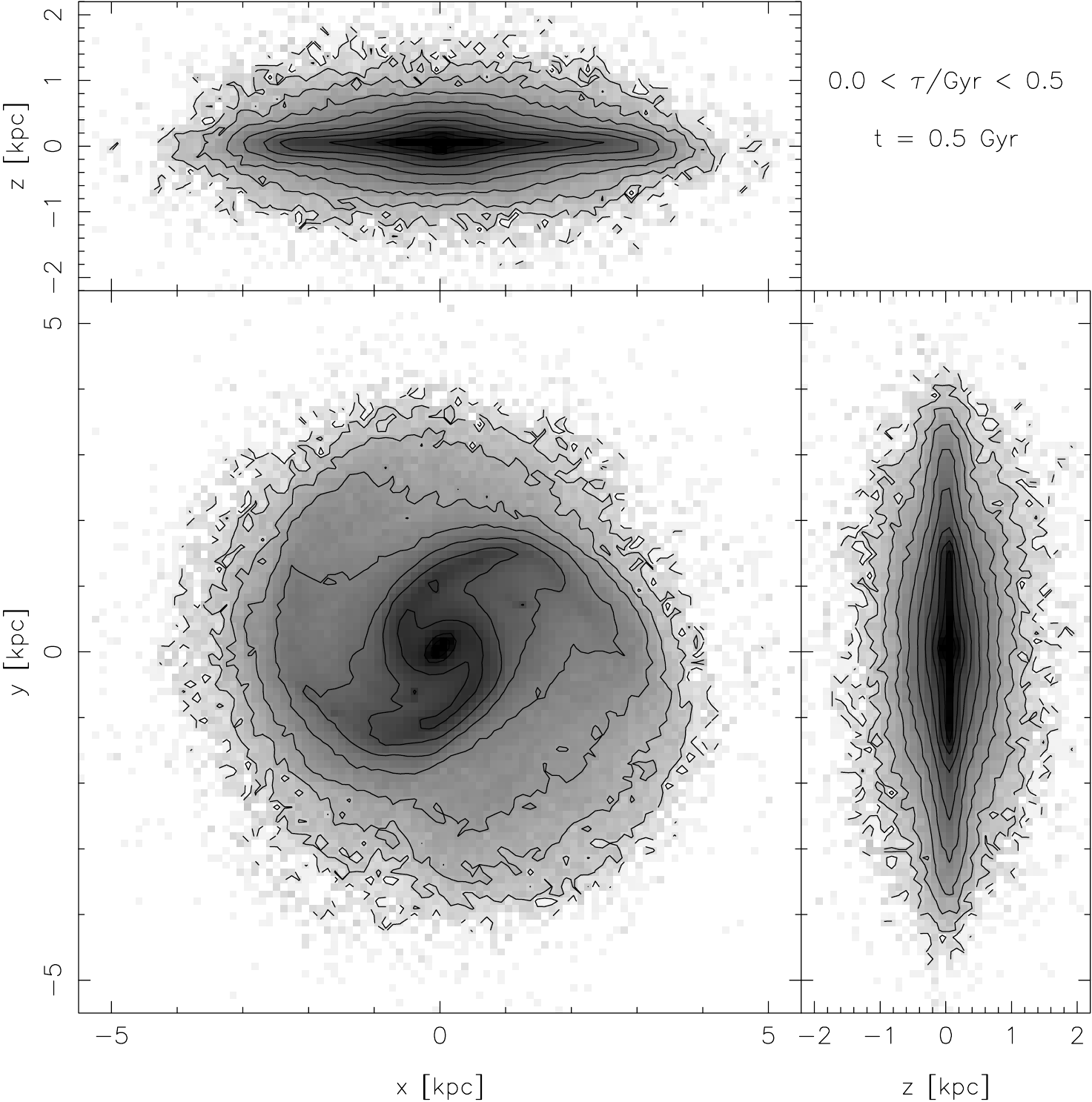}}
  \caption{The oldest stellar population ($0.0 \le \tf \le 0.5 \Gyr$) at
  $t=0.5\Gyr$ in the star-forming simulation. The distribution of
  these stars is discy and thin at formation, not a thick spheroid. }
\label{fig:oldest}
\end{figure}

\subsection{The system before bar formation}
\label{ssec:prebar}

\begin{figure}
\centerline{\includegraphics[angle=0.,width=\hsize]{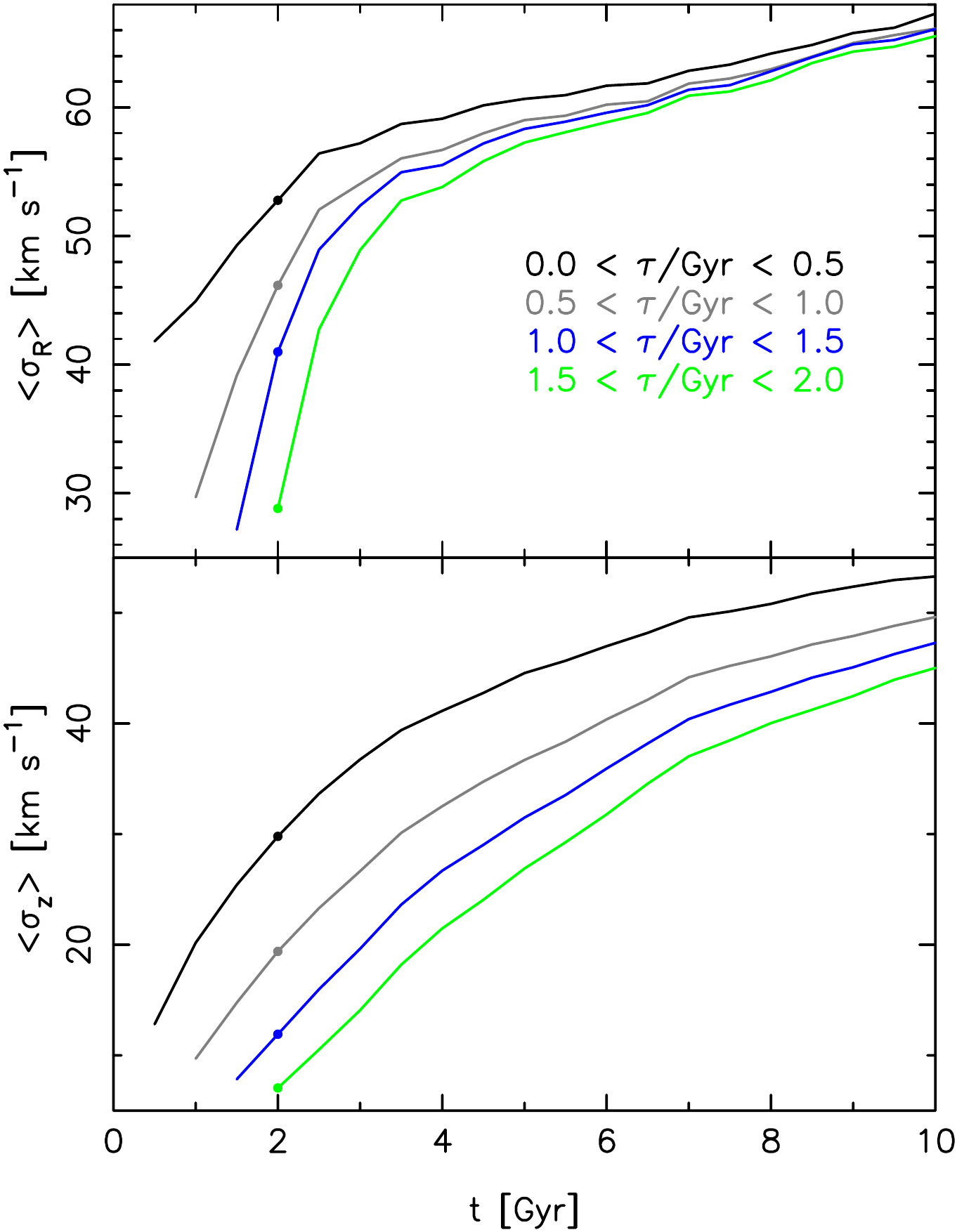}}
\caption{The evolution of \avg{\sig{R}}\ (top), and
  \avg{\sig{z}}\ (bottom) in the star-forming simulation.  The
  different \tf\ populations are indicated in the top panel.  Averages
  are taken over the radial range $1.5 \leq R/\kpc \leq 3$.  Note that
  at $2\Gyr$ (marked by the filled circles), as the bar starts
  forming, the radial and vertical dispersions increase with \tf.}
\label{fig:kinematics}
\end{figure}

At $2\Gyr$, the model has already formed $55\%$ of the total final
stellar mass, which is not uncommon for Sb-type galaxies
\citep[e.g.][]{kennicutt+94, tacchella+15}, although on average Milky
Way mass galaxies reached half their mass at $z \sim 1-1.4$
\citep{patel+13, behroozi+13, vandokkum+13, terrazas+16}.  We measure
the properties of the stellar distribution at this time as a function
of the time of formation, \tf, of the stars.  We separate stars in 0.5
Gyr bins in \tf\ and from here on refer to each such bin as a separate
population.

We emphasize that the stars initially form in a discy distribution.
Fig. \ref{fig:oldest} shows the density distribution of the oldest
stellar population, with $0.0 \le \tf \le 0.5 \Gyr$, at $t=0.5\Gyr$,
\ie\ just after formation.  Strong spirals can be seen, as can a few
very weak star-forming clumps.  At $t=10\Gyr$, this population still
retains a discy character, although the stars have spread out, both
radially and vertically, considerably.

\begin{figure*}
\centerline{
\includegraphics[angle=0.,width=0.25\hsize]{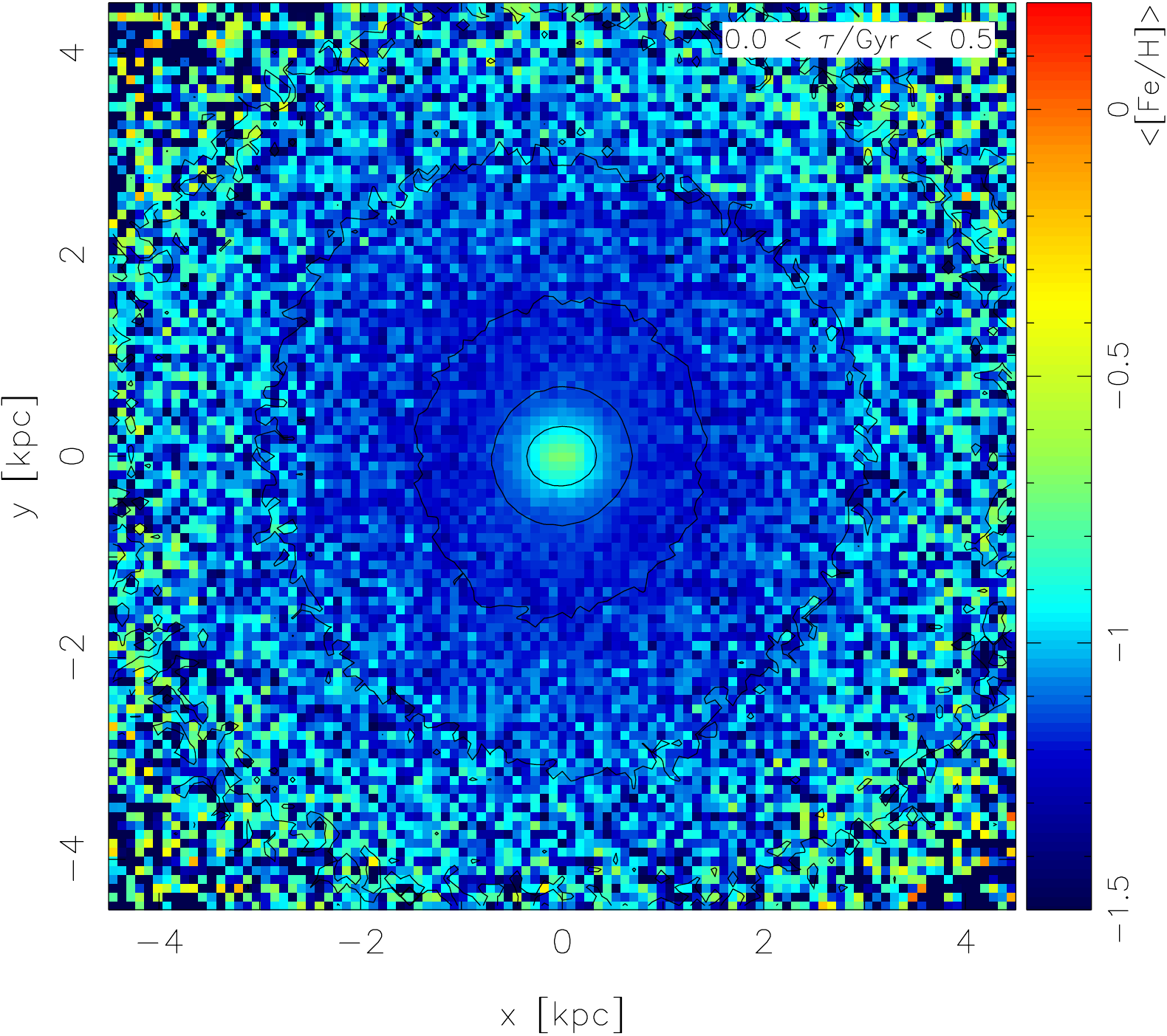}
\includegraphics[angle=0.,width=0.25\hsize]{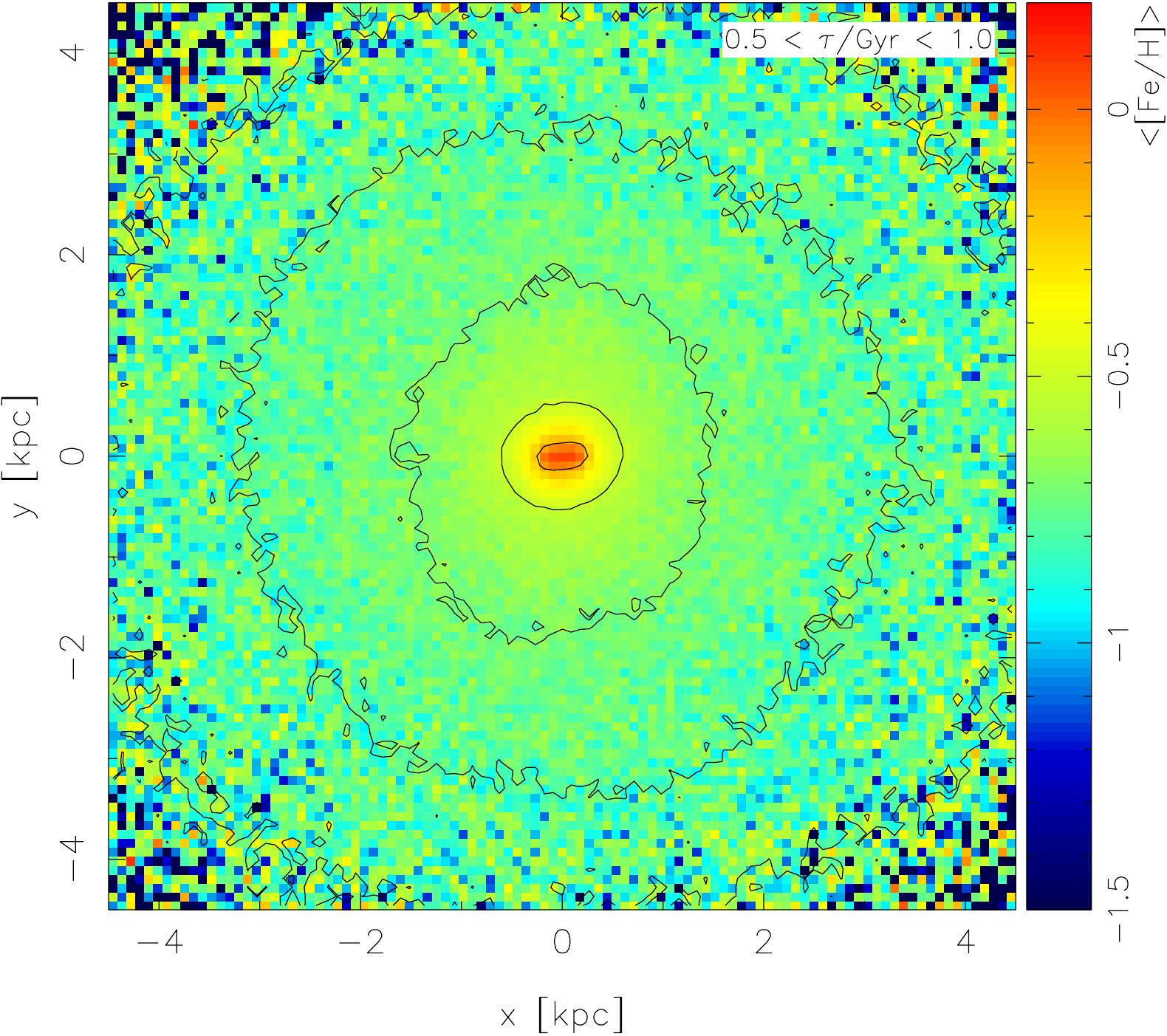}
\includegraphics[angle=0.,width=0.25\hsize]{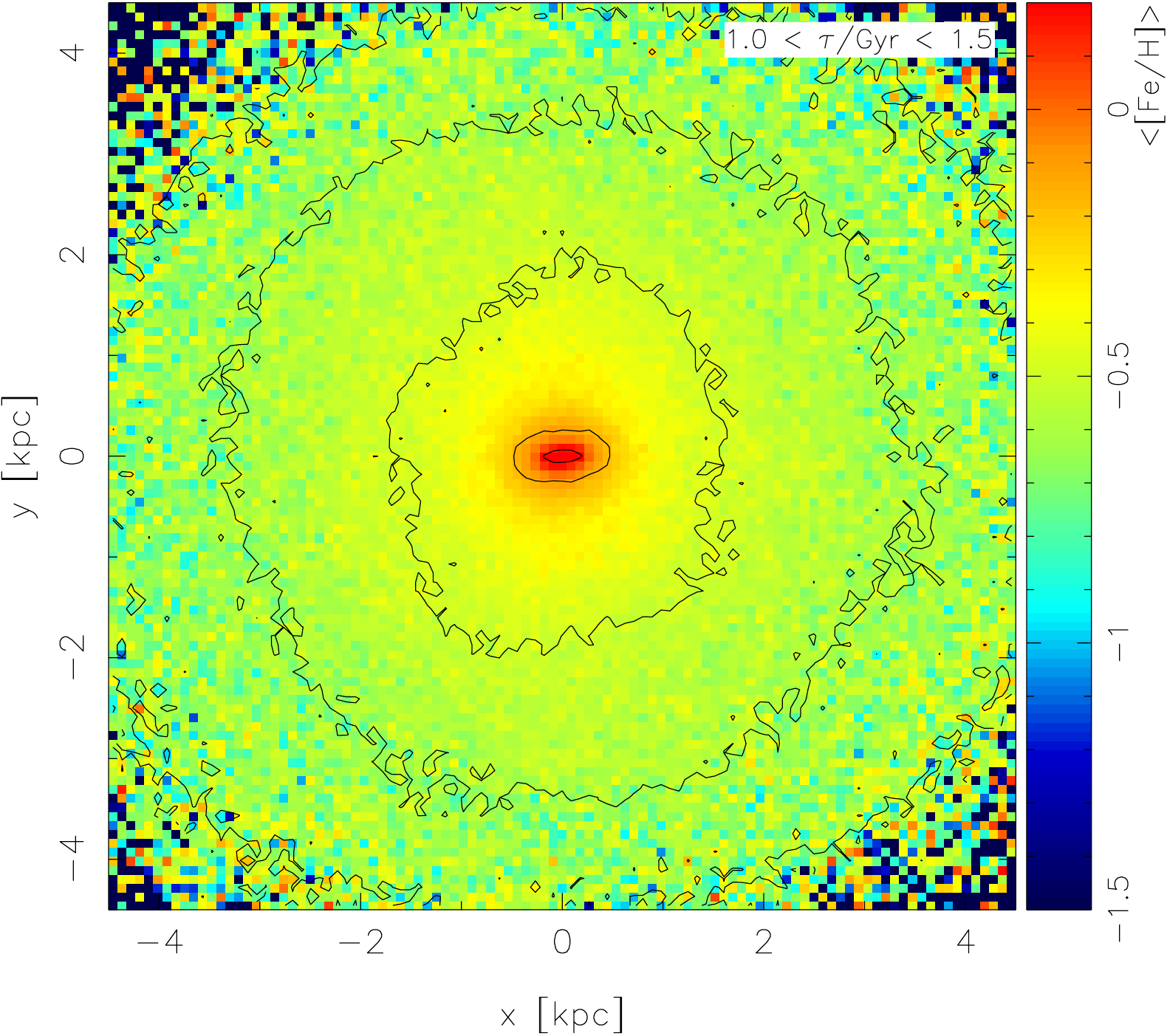}
\includegraphics[angle=0.,width=0.25\hsize]{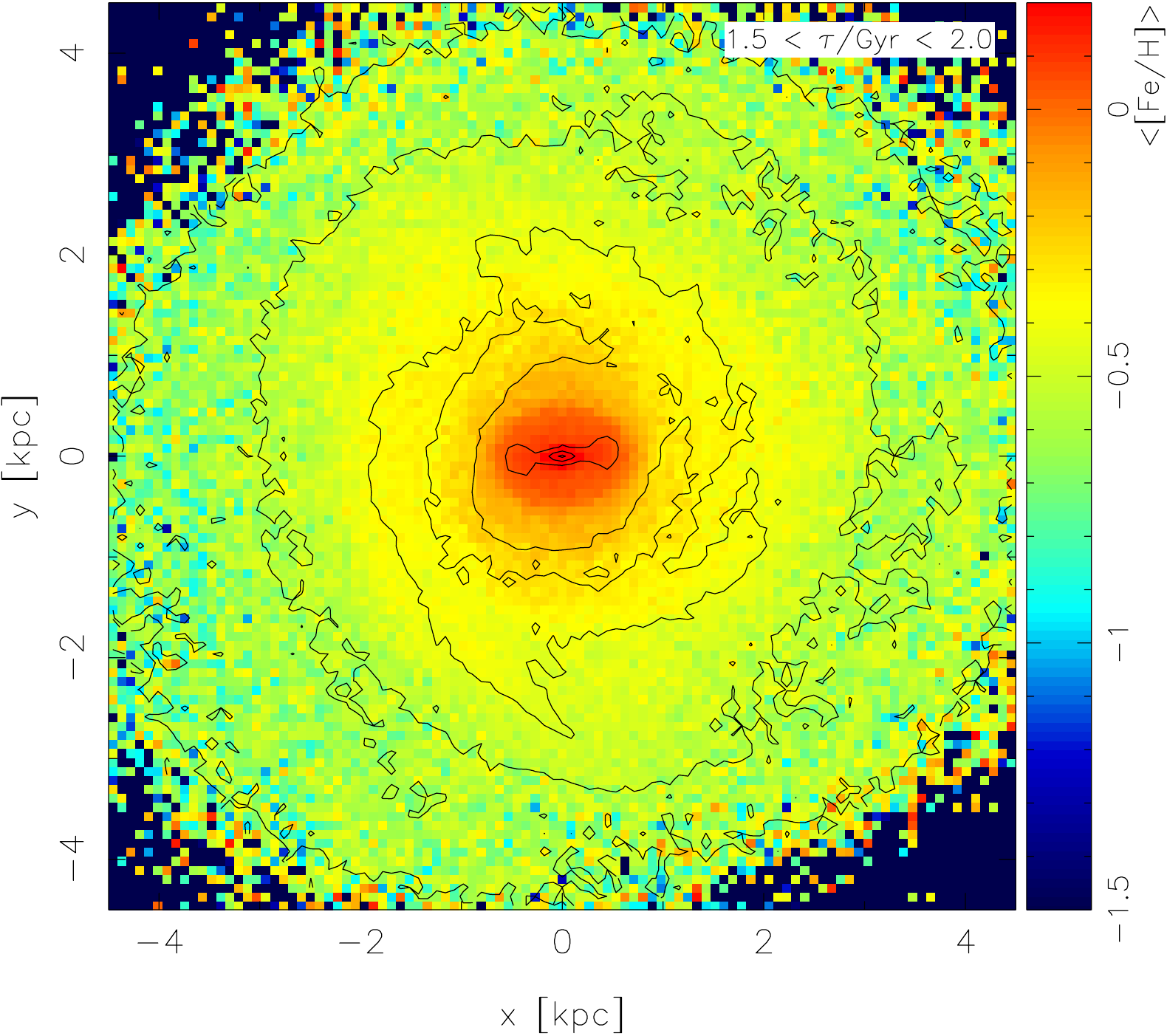}
}
\centerline{
\includegraphics[angle=0.,width=0.25\hsize]{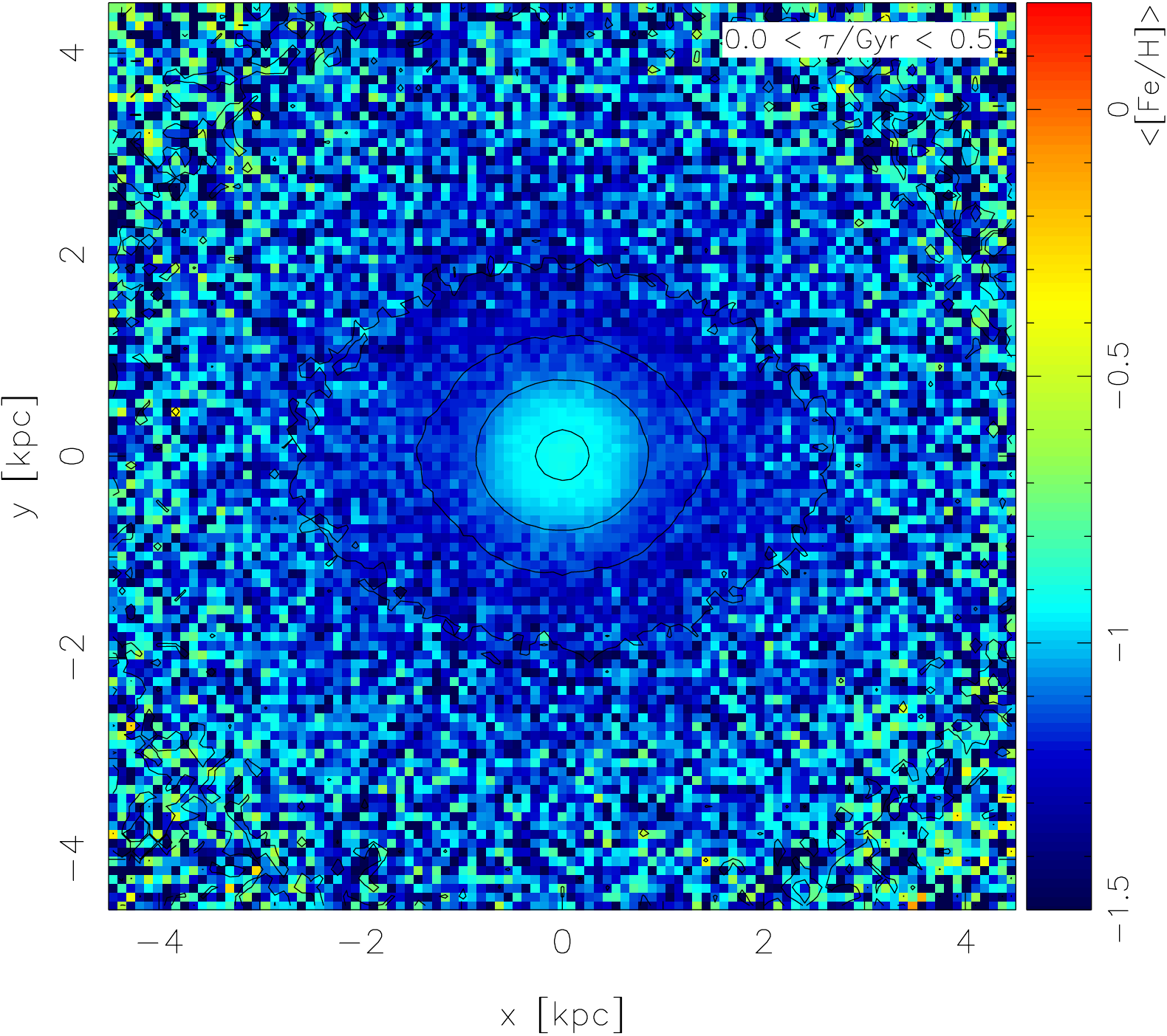}
\includegraphics[angle=0.,width=0.25\hsize]{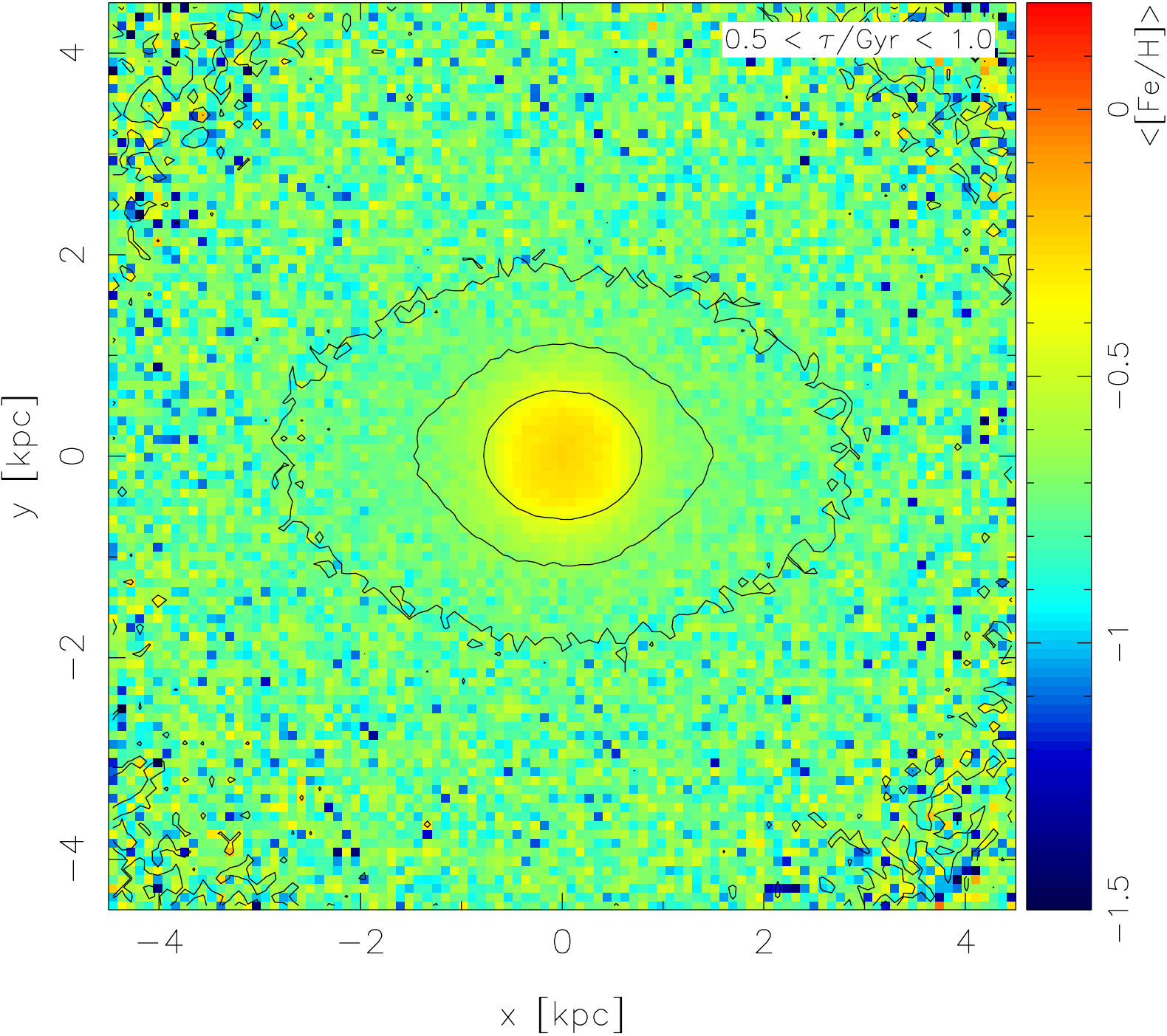}
\includegraphics[angle=0.,width=0.25\hsize]{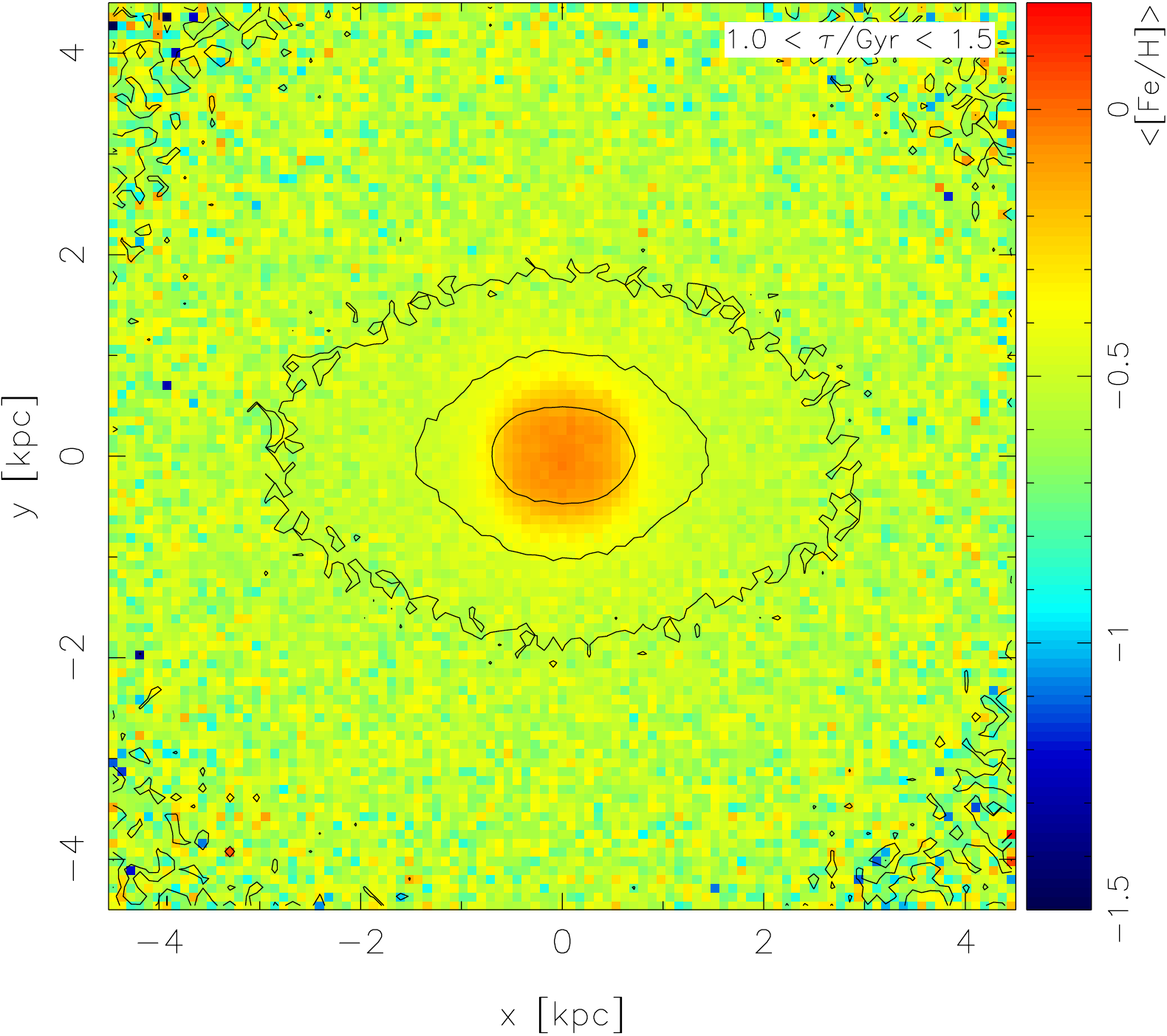}
\includegraphics[angle=0.,width=0.25\hsize]{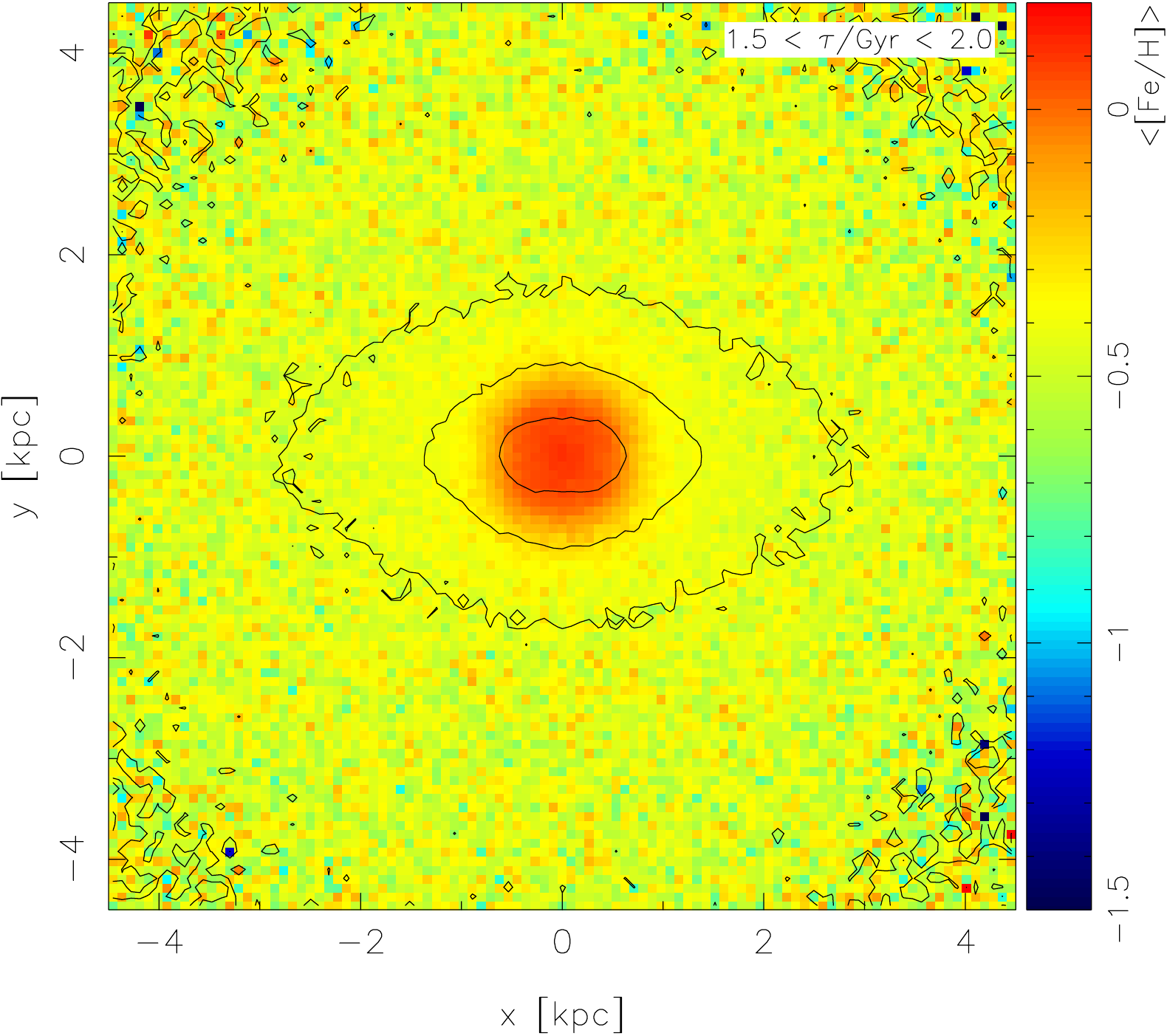}
}
\caption{Maps of \avg{\feh} at $t=2\Gyr$ (top row) and at $t=10\Gyr$
  (bottom row) for different \tf\ populations as indicated at the
  top right of each map (from left to right, $0.0 \leq \tf/\Gyr \leq
  0.5$, $0.5 \leq \tf/\Gyr \leq 1.0$, $1.0 \leq \tf/\Gyr \leq 1.5$ and
  $1.5 \leq \tf/\Gyr \leq 2.0$), in the star-forming simulation.  The
  colour scale, indicated by the wedge beside each map, is common to
  all panels and spans $-1.5 \leq \feh \leq 0.2$.  Contours indicate
  the surface density of the different populations.  The metallicity
  of populations rises rapidly with \tf.}
\label{fig:feh200and1000}
\end{figure*}

Fig. \ref{fig:kinematics} shows the average radial, \avg{\sig{R}}, and
average vertical, \avg{\sig{z}}, velocity dispersions, for the stars
born before $2\Gyr$.  At $2\Gyr$, the older stellar populations are
radially and vertically hotter than the younger populations.
The top row of Fig. \ref{fig:feh200and1000} maps the average
metallicity, \avg{\feh}, at $t=2\Gyr$.  \avg{\feh}\ rises rapidly in
the first $2 \Gyr$.  For all populations, the central $\sim 500 \pc$ is
more metal-rich with a small \feh\ gradient outside this region.  A
small-scale ($\sim 0.5\kpc$) bar is evident in the mass distribution,
which provides the seed around which the larger bar grows.
We also examined maps of the average $\alpha$-abundance, \avg{\alfe}.
These show that \avg{\alfe}\ drops rapidly with increasing \tf.

\begin{figure}
\centerline{
\includegraphics[angle=0.,width=\hsize]{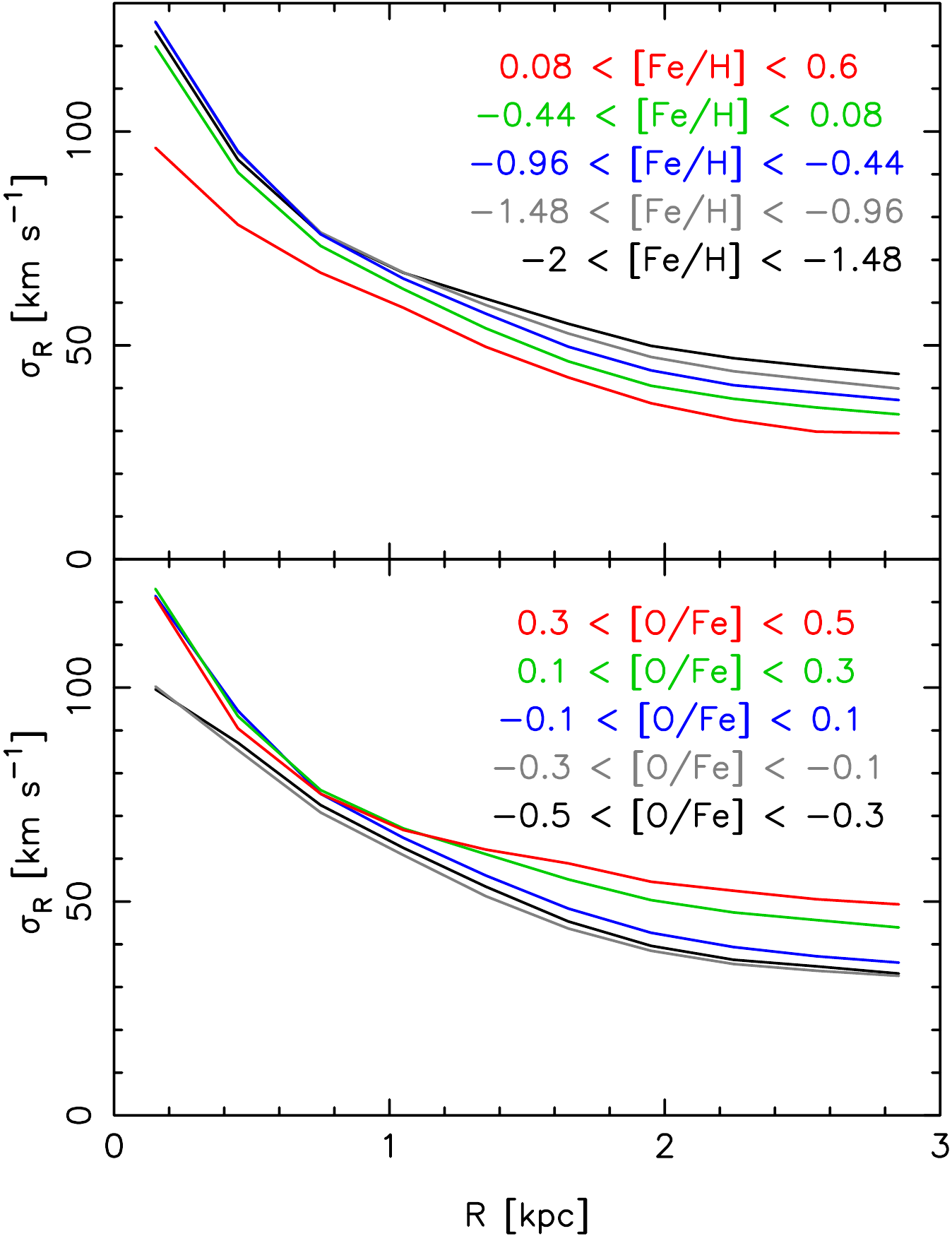}
}
\caption{The (anti)correlation between (\feh) \alfe\ and \sig{R} at
  $t=2\Gyr$, before the bar has formed in the star-forming
  simulation.  These relations arise because \sig{R}, \feh\ and
  \alfe\ all evolve with time.}
\label{fig:chemkin}
\end{figure}

Thus overall, at $2\Gyr$, before the bar has formed, the age of a
stellar population correlates with \avg{\sig{R}}, and \avg{\alfe}, and
anti-correlates with \avg{\feh}.  These dependencies on age imply that
\sig{R}\ correlates with chemistry, which we show explicitly in Fig.
\ref{fig:chemkin}.

Therefore, on the basis of the pure $N$-body simulations, we expect
that the bar will cause the different populations to separate
morphologically, forming a stronger bar in the younger, metal-rich
populations and a more ellipsoidal distribution in the older,
metal-poor populations.  As with \sig{R}, \sig{z}\ also correlates
with age, as expected since dispersions increase continuously at this
time.  Thus, \sig{z}\ will also correlate with chemistry, but the
driver of the separation of populations will be \sig{R}, not \sig{z},
as we showed in Section \ref{ssec:run742} above for the pure $N$-body
simulations.

\subsection{The final system}
\label{ssec:finalsystem}

\begin{figure}
\centerline{
\includegraphics[angle=-90.,width=\hsize]{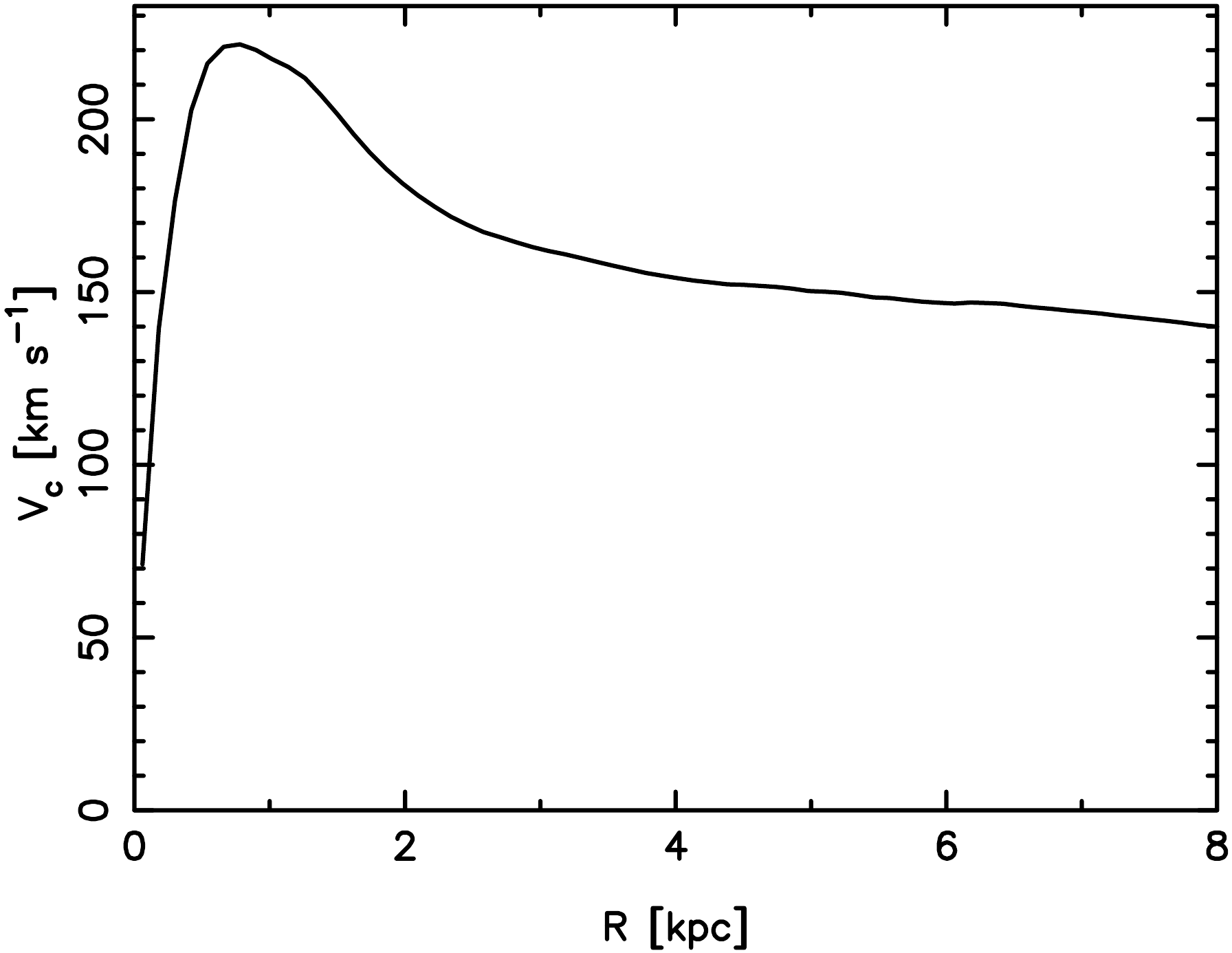}
}
\caption{The azimuthally averaged rotation curve of the star-forming
  simulation after $10\Gyr$.}
\label{fig:rotcurve708}
\end{figure}

The azimuthally averaged rotation curve of this model at $10\Gyr$ is
shown in Fig. \ref{fig:rotcurve708}.  This exhibits a central peak
which then drops off and flattens.  Because we are studying the bulge
of the model, our velocity scaling is determined by comparing the
bulge with that of the Milky Way, which has a rather flatter rotation
curve \citep[e.g.][]{bland-hawthorn_gerhard16, lizhi+16}.  Further
out, the rotation velocity of our model is lower than that of the
Milky Way.

\subsubsection{Velocity dispersions}
\label{ssec:vdisps}

Fig. \ref{fig:kinematics} shows the evolution of \avg{\sig{R}}\ and
\avg{\sig{z}}.  We average in the range $1.5 \leq R/\kpc \leq 3$ in
order to avoid the effects of the nuclear disc \citep{cole+14,
  debattistaMWND+15}.  The formation of the bar raises
\avg{\sig{R}}\ sharply, while \avg{\sig{z}}\ increases more slowly.
As in the pure $N$-body simulations, \avg{\sig{R}}\ of the different
populations evolve towards each other, while their
\avg{\sig{z}}\ remain separated, with continued late time heating
driven by the growing bar.  While we find no single violent buckling
event like in the pure $N$-body simulations, small-scale bends in the
disc can be detected after $2\Gyr$, which contribute to the vertical
heating of the inner disc.

\subsubsection{Vertical distribution}
\label{ssec:vdistribution}

\begin{figure}
\centerline{\includegraphics[angle=-90.,width=\hsize]{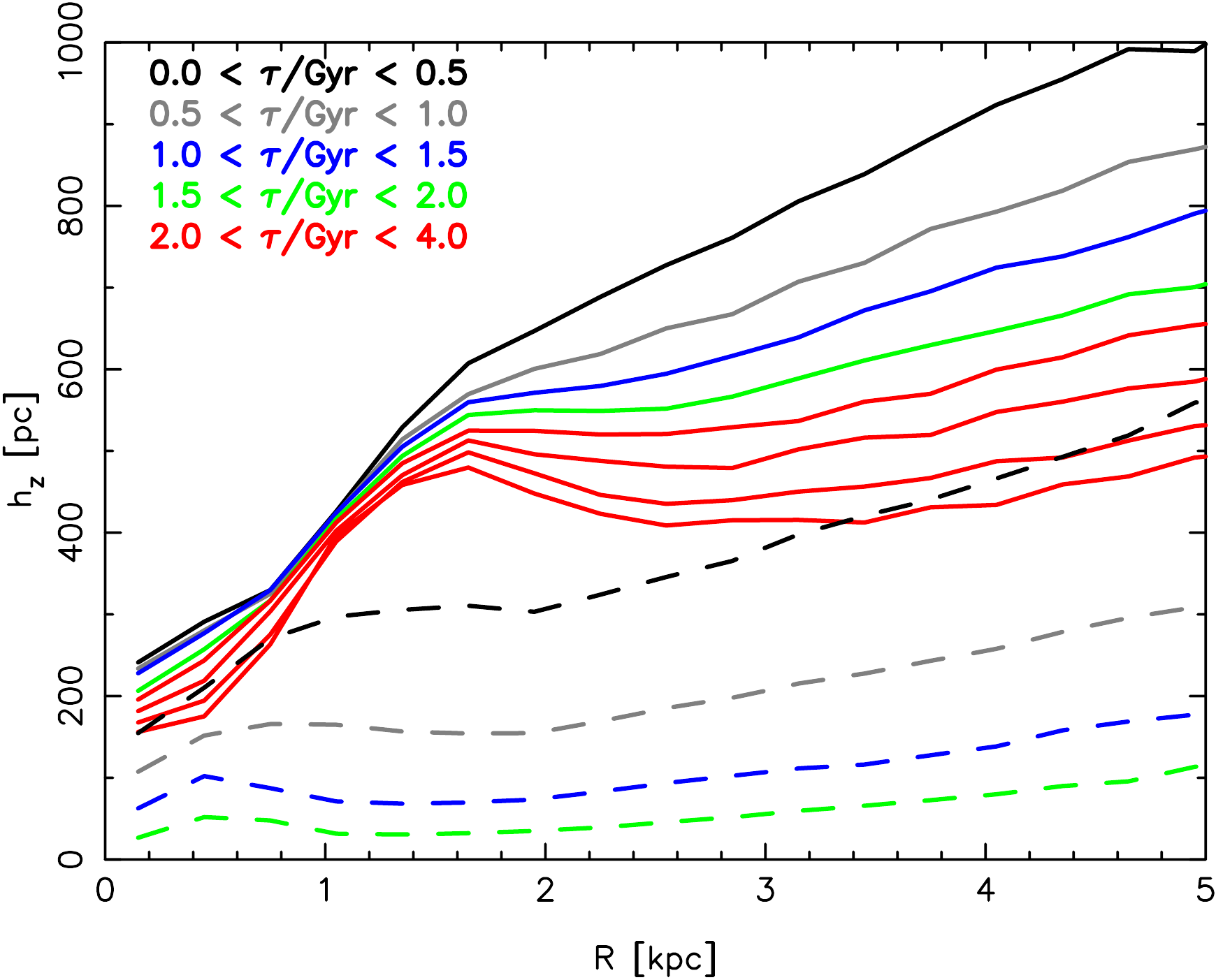}}
\centerline{\includegraphics[angle=-90.,width=\hsize]{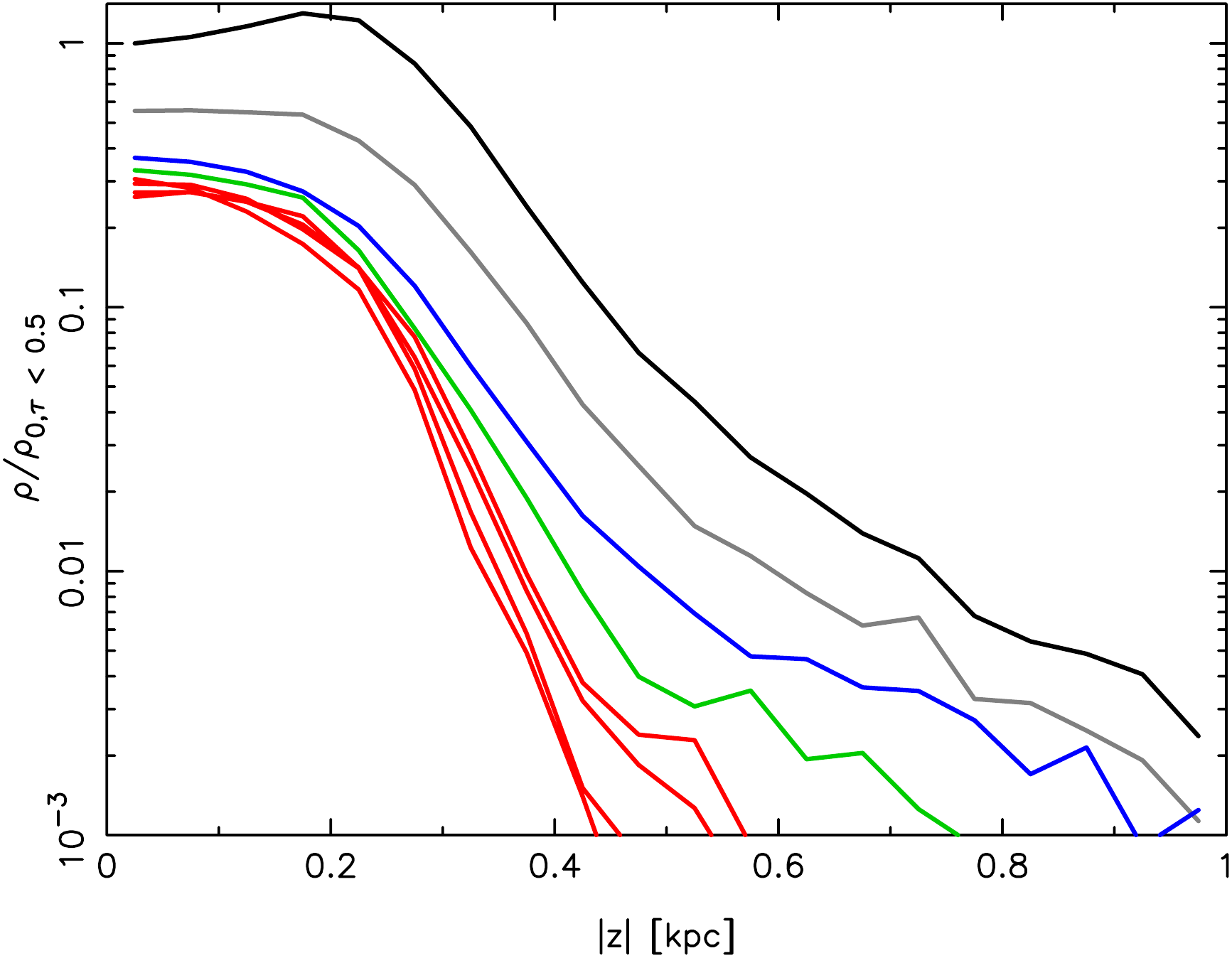}}
\caption{Top: vertical height of stars at $t=2\Gyr$ (dashed lines) and
  at $10\Gyr$ (solid lines) for the different \tf\ populations in the
  star-forming simulation.  The younger stars have a prominent local
  peak at $R\sim1.7\kpc$, coinciding with their stronger peanut shape.
  Bottom: vertical density profile along the minor axis of the disc at
  $10\Gyr$.  The profiles are normalized to the central density of the
  oldest population.  Colours are as in the top panel. The red lines
  show four bins, of equal width in \tf, of stars born at $2\leq
  \tf/\Gyr \leq 4$.  The younger populations extend to lower height
  than the older populations.}
\label{fig:heights}
\end{figure}

The top panel of Fig. \ref{fig:heights} plots profiles of $h_z$ for
the different \tf\ populations.  All populations thicken considerably
outside $R = 500 \pc$ between $2$ and $10\Gyr$, with the height of
the oldest population roughly doubling at $R=2\kpc$.  For stars with
$2 \leq \tf/\Gyr \leq 4$ the profile is locally peaked at $R \simeq
1.6 \kpc$ rather than monotonically increasing, indicating that these
younger populations are peanut-shaped.

The bottom panel of Fig. \ref{fig:heights} plots the vertical density
profiles along the minor axis of the disc at $10\Gyr$ for different
\tf\ populations, normalized by the mid-plane density of the oldest
population.  The oldest population is clearly the one that extends the
furthest in height and we have verified that still younger populations
reach even lower height.  The density of the oldest population peaks
at $z \simeq 200\pc$ rather than at the mid-plane.  This may be
related to the boxy-core structure reported by \citet{li_shen15}.

\subsubsection{Bar strength}
\label{ssec:barstrength}

\begin{figure*}
  \centerline{
    \includegraphics[angle=0.,width=0.33\hsize]{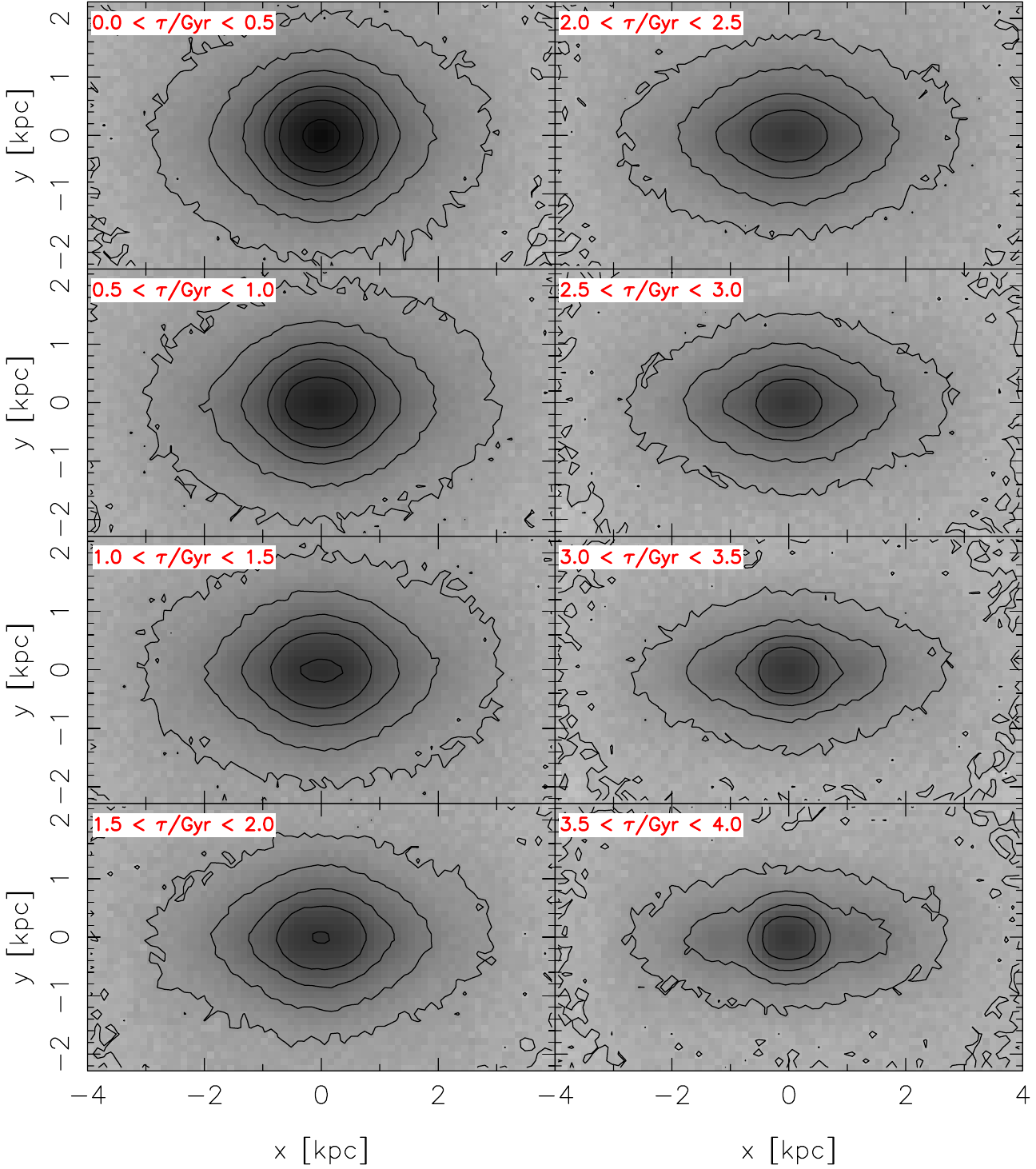}
    \includegraphics[angle=0.,width=0.33\hsize]{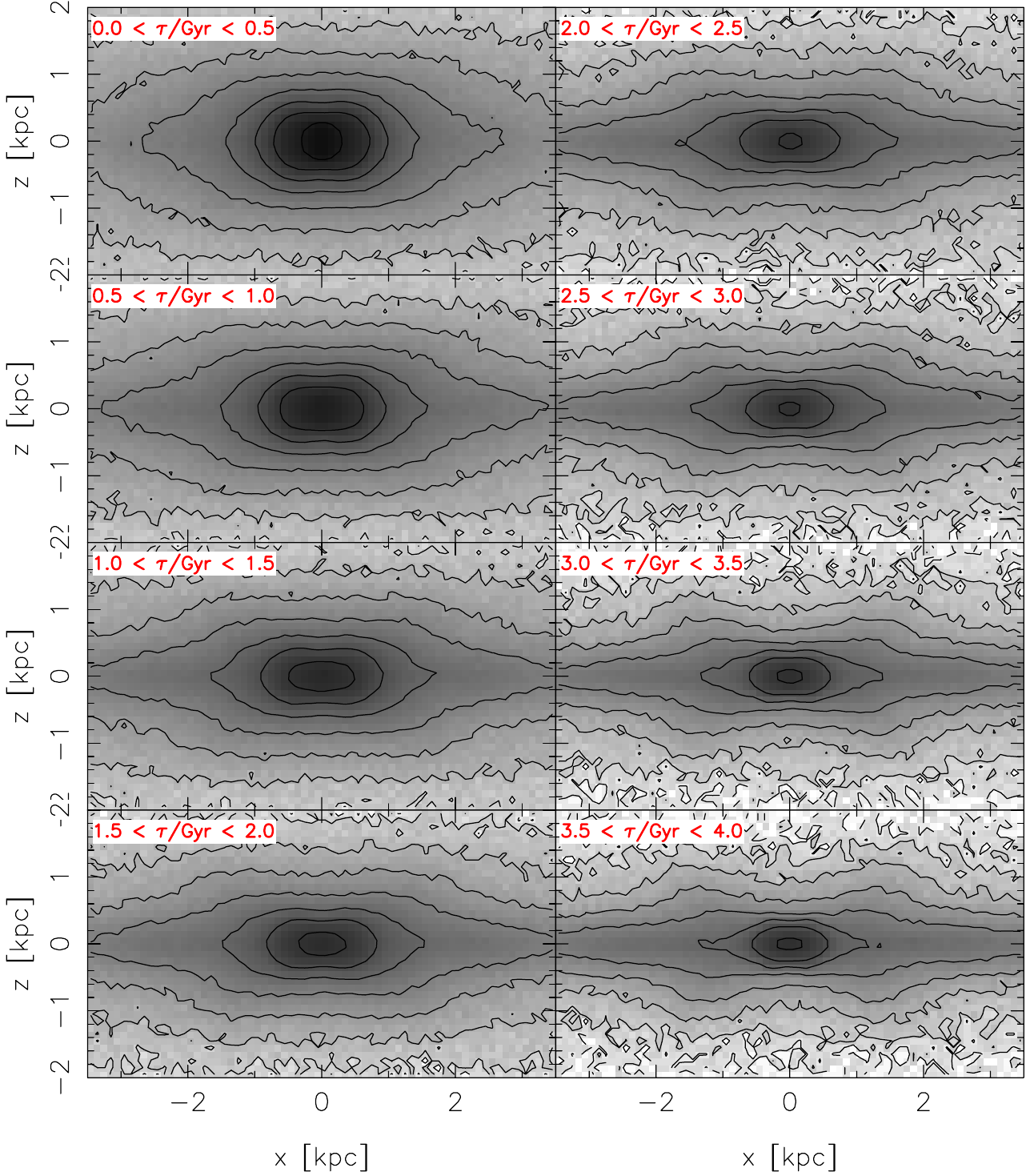}
    \includegraphics[angle=0.,width=0.33\hsize]{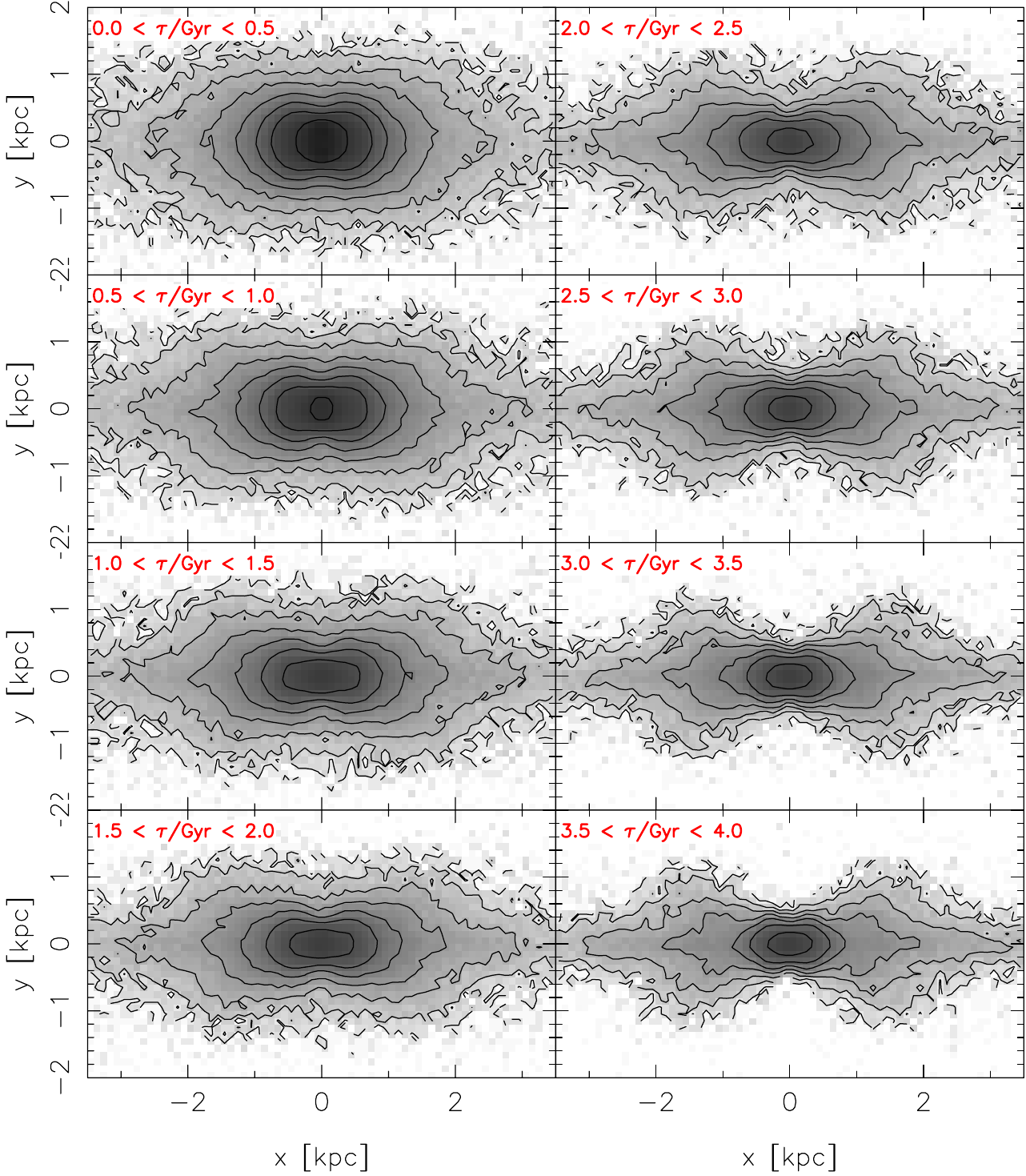}
  }
\caption{The face-on (left two columns) and edge-on (middle two
  columns) surface density distribution for different
  \tf\ populations, as indicated, in the star-forming simulation at
  $t=10\Gyr$.  The right two columns show a cross-section with $|y|
  \leq 0.2\kpc$ through the edge-on views of the middle panels.  In
  the edge-on panels, the bar is perpendicular to the line of sight
  (\ie\ it is viewed side-on).  Note that slightly different spatial
  scales are used for the face-on and the edge-on views.  The
  quadrupole moment and peanut shape both increase rapidly with \tf.}
\label{fig:feosnapshots}
\end{figure*}

\begin{figure}
\centerline{\includegraphics[angle=-90.,width=\hsize]{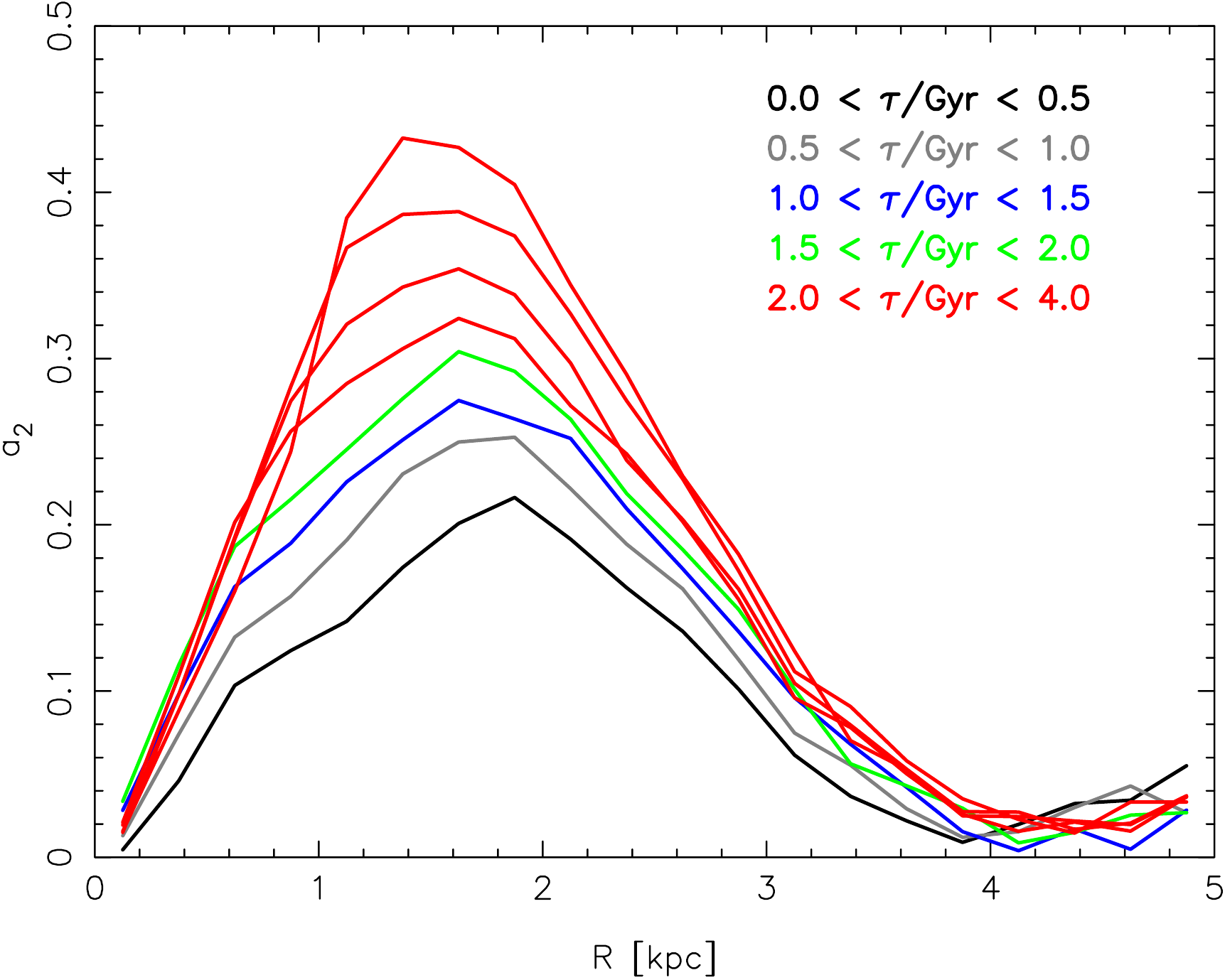}}
\centerline{\includegraphics[angle=-90.,width=\hsize]{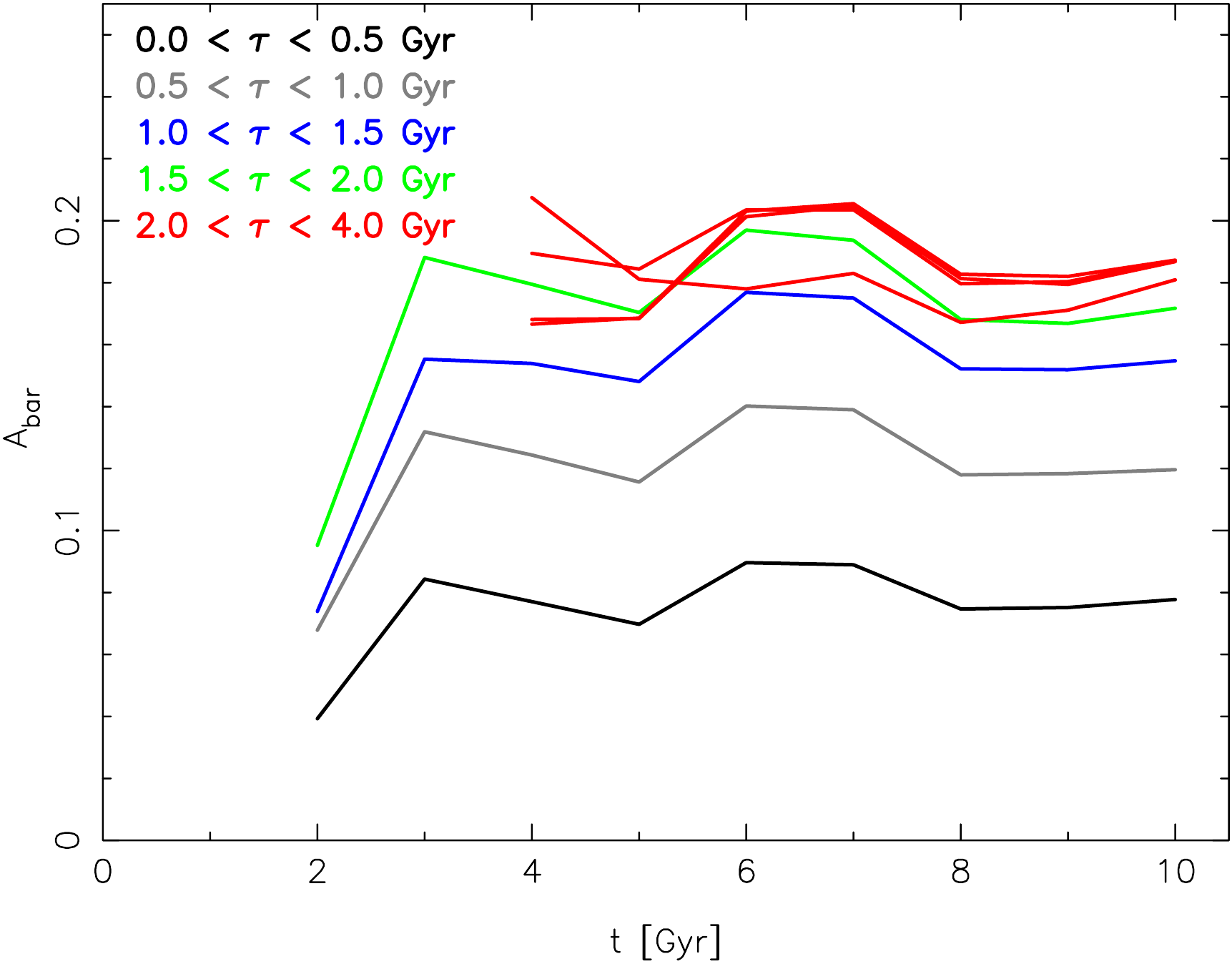}}
\caption{Top: final radial $m=2$ amplitude, $a_2$ ($m=2$ Fourier
  amplitude in radial bins), of the surface density of stars in
  different \tf\ populations in the star-forming simulation.  Bottom:
  the evolution of bar amplitude, \abar, for different
  \tf\ populations, as indicated. Globally and at each radius, the bar
  is weakest in the oldest stellar populations.  The red lines show
  four bins, of equal width in \tf, of stars born at $2\leq \tf/\Gyr
  \leq 4$.}
\label{fig:baramp}
\end{figure}

The left-hand panels of Fig. \ref{fig:feosnapshots} show the final
projected stellar surface density in different populations.  There is
a clear trend for the stellar distribution to become more strongly
barred the later the stars formed.  For stars formed before the bar
($\tf \leq 2\Gyr$), there is a strong dependence of bar strength on
age.  The bottom panel of Fig. \ref{fig:baramp} shows the evolution of
bar amplitude, \abar, as a function of \tf.  Populations with $\tf <
1.5 \Gyr$ end up in a weaker bar compared with populations that form
later.  The various populations with $2 \leq \tf/\Gyr \leq 4$ have
very similar global amplitude.  The bar amplitude of the old
populations evolves in parallel with that of the younger, cooler
populations without ever becoming as strong.

The top panel of Fig. \ref{fig:baramp} presents the radial profile of
the $m=2$ Fourier moment, $a_2$, showing that the peak amplitude
decreases with age.  Populations formed during $2 \leq \tf/\Gyr \leq
4$ also have different profiles, in spite of their equal overall bar
strength, with peak amplitude increasing with \tf.  In Fig.
\ref{fig:feosnapshots}, this is directly evident as an increasingly
elongated density distribution.  None the less the bar has the same
semi major axis size for all populations.

\subsubsection{Peanut shape}
\label{sssec:bpshape}

The middle and right-hand panels of Fig. \ref{fig:feosnapshots} show
the edge-on distribution of different \tf\ populations, with the bar
viewed side-on.  As \tf\ increases, the disc becomes thinner,
particularly at the centre and the shape changes from boxy to peanuty,
in agreement with the radial profiles of $h_z$ shown in the top panel
of Fig.  \ref{fig:heights}.  The very oldest populations have a thick,
boxy spheroidal shape.  The younger populations are thinner at the
centre, resulting in a peanut structure.  As with the bar strength,
this result is in agreement with the pure $N$-body simulations.

\subsubsection{Chemistry}
\label{sssec:chemistry}

The bottom row of Fig. \ref{fig:feh200and1000} shows maps of
\avg{\feh} at $t=10\Gyr$, which can be compared with the maps in the
top row at $t=2\Gyr$.  The central \avg{\feh}\ remains high, although
it is distributed over a slightly larger region than at $2\Gyr$.
Other \avg{\feh}\ gradients within each \tf\ population are largely
erased, evidence of the significant mixing over this region.

\begin{figure}
\centerline{\includegraphics[angle=0.,width=\hsize]{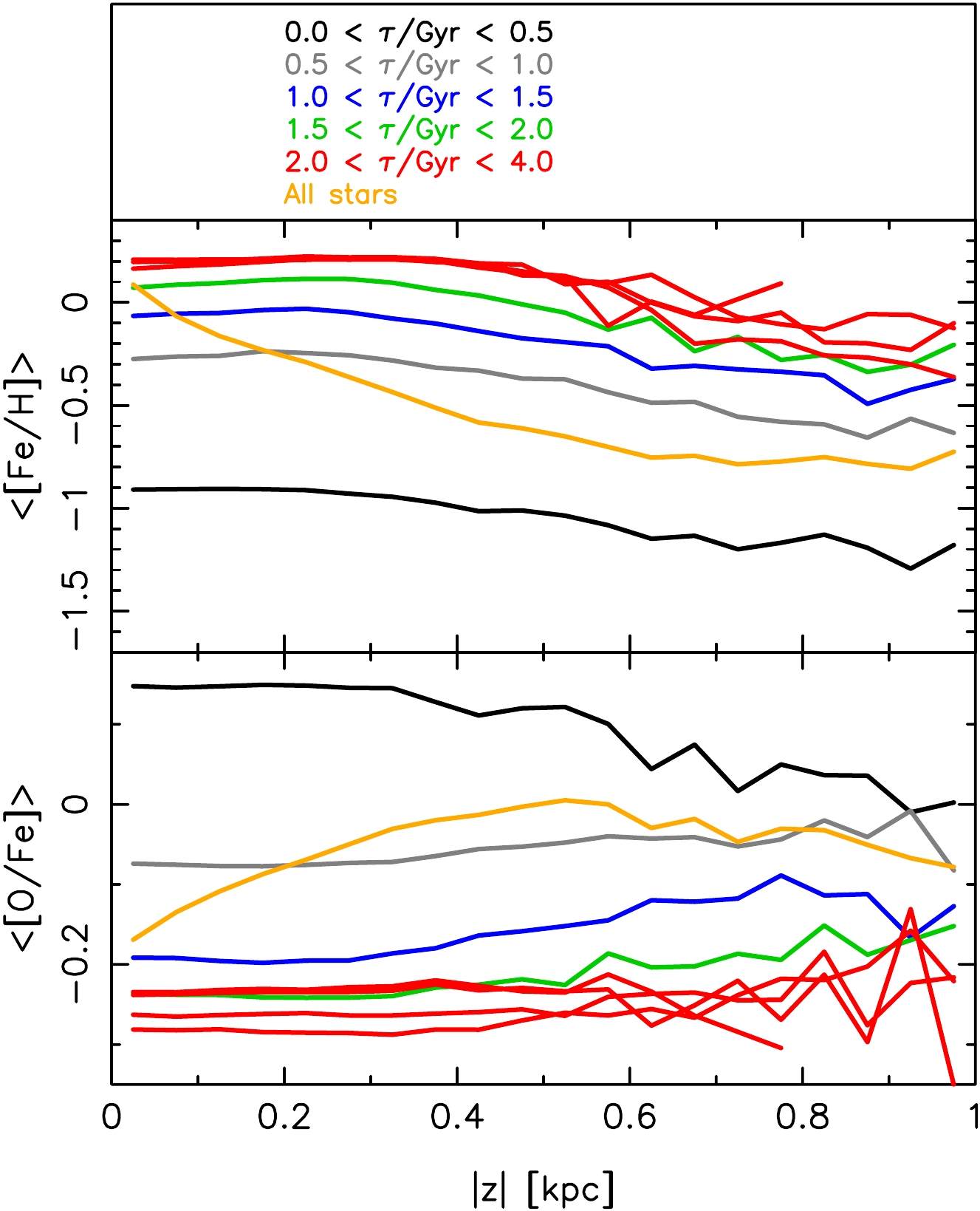}}
\caption{Vertical profiles of stellar chemistry on the minor axis of
  the bar (with $R < 0.3 \kpc$) at $10 \Gyr$ for different populations
  in the star-forming simulation.  The top panel shows \avg{\feh},
  while the bottom one shows \avg{\alfe}.  The total profiles are
  indicated by the orange lines.  The strong gradients are largely due
  to the different vertical extent of different age populations, not
  to internal gradients within each \tf\ population, which are quite
  shallow.  The red lines show four bins, of equal width in \tf, of
  stars born at $2\leq \tf/\Gyr \leq 4$.
  \label{fig:vertfehprof}}
\end{figure}

The top panel of Fig. \ref{fig:vertfehprof} shows the vertical
\avg{\feh}\ profiles on the bar's minor axis.  A non-zero gradient is
present which at large $|z|$ is dominated by the oldest population.
The vertical \avg{\feh}\ gradient for all stars varies with $|z|$,
becoming flatter with height.  The bottom panel shows similar profiles
for \avg{\alfe}.  The oldest population is very $\alpha$-enhanced,
dropping rapidly with \tf.  The \avg{\alfe}\ of all stars is low near
the mid-plane where stars are still forming but is quickly dominated
by the oldest stellar population by $\sim 400 \pc$.  In both panels,
the profile within each \tf\ bin is flatter than the overall profile,
indicating that the overall steep gradients in \avg{\feh}\ and
\avg{\alfe}\ are produced by variations in the relative densities of
different age populations.  Once the oldest population dominates the
vertical profile, the chemical gradients become much flatter.

\begin{figure}
\centerline{\includegraphics[angle=0.,width=\hsize]{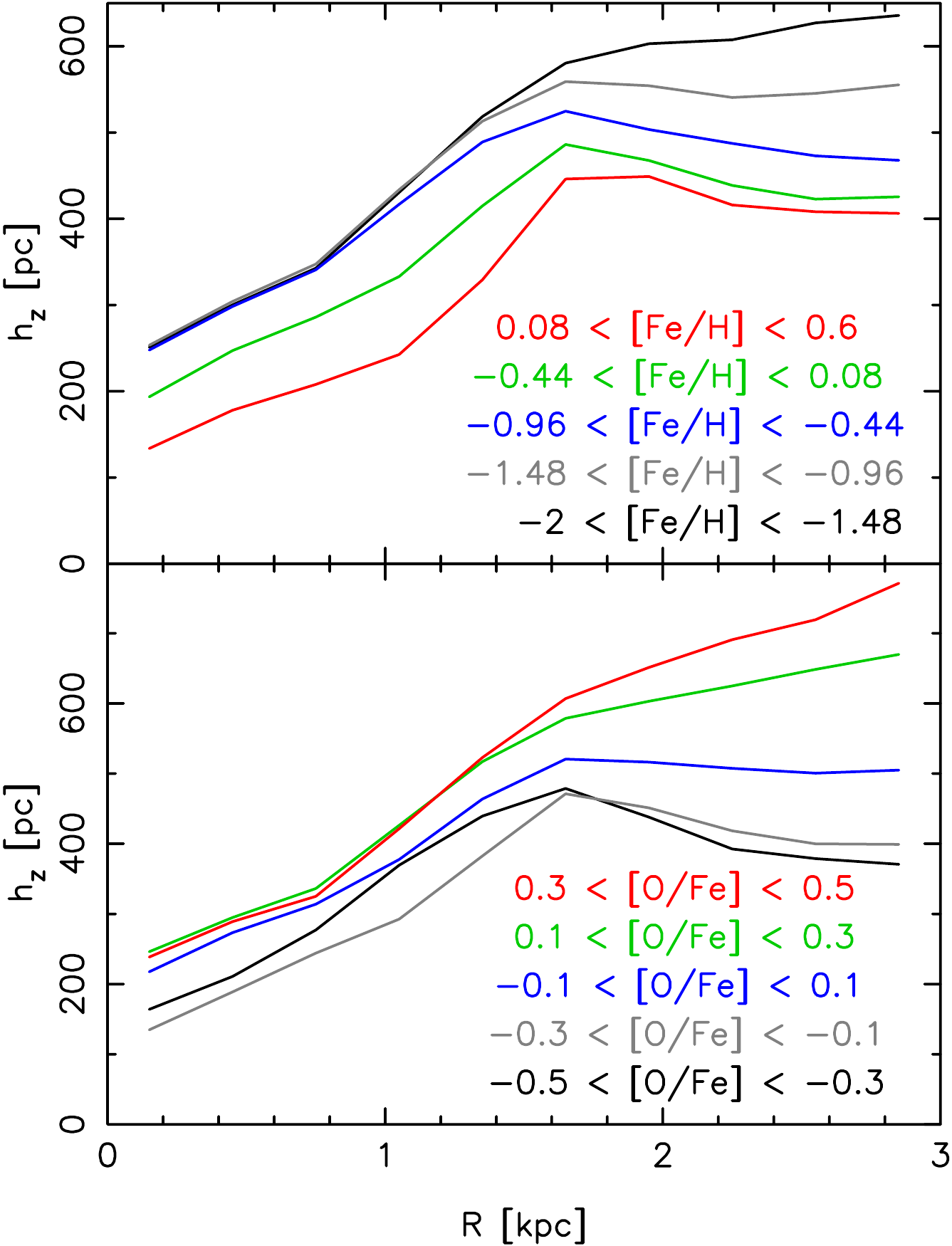}}
\caption{Profiles of the vertical height for different \feh\ (top) and
  \alfe\ bins (bottom) at $10\Gyr$ in the star-forming simulation.
  Peaks, associated with a peanut, rather than boxy, shape, are
  evident at high \feh\ and low \alfe.
  \label{fig:hzvsfehofe}}
\end{figure}

\subsubsection{Dependence of height on metallicity}

The top panel of Fig. \ref{fig:hzvsfehofe} plots the height profiles
of stars for different bins in \feh.  The metal-rich stars comprise a
thinner distribution than do the metal-poor ones.  A peak is present
at $R \simeq 1.6\kpc$ in the high metallicity $h_z$ profile, whereas
it rises monotonically for the more metal-poor stars.  The peak in the
metal-rich stars is a signature of the peanut-shaped distribution in
the younger populations, and the absence of a peak in the lowest
\feh\ stars is consistent with the absence of a peanut shape in old
stars.  The height profiles for different \alfe\ bins are
qualitatively similar, with the \alfe-rich population monotonically
increasing in height, whereas the \alfe-poor population exhibits a
local peak.  Bearing in mind the anti-correlation between age and
\feh, and the correlation with \alfe, Figs. \ref{fig:heights} and
\ref{fig:hzvsfehofe} tell a consistent story.

\subsection{Synthesis of the pure $N$-body and star-forming simulations}

We have shown that the evolution of both the kinematics and the
morphology of the star-forming simulation are consistent with those of
the pure $N$-body simulations: populations with smaller initial
in-plane velocity dispersions end up vertically thinner with a
pronounced peanut shape and a strong bar.  Conversely, the hotter
populations form a thicker distribution with a boxy, not peanut, shape
and with a bar that has a significantly weaker quadrupole moment.  The
pure $N$-body simulations have, by their nature, no chemistry but if
we reasonably assume that the high \sig{R}\ populations are older,
then we expect them to have lower \feh\ and higher \alfe, as in the
star-forming simulation.  Then, the pure $N$-body simulations would
predict a strong peanut shape in the metal-rich population and a weak
bar with a boxy bulge in the oldest, metal-poor population.  The
evolution of the bulge in the star-forming simulation can therefore be
understood simply as resulting from the separation induced by
differences in in-plane kinematics at bar formation.

%%%%%%%%%%%%%%%%%%%%%%%%%%%%%%%%%%%%%%%%%%%%%%%%%%%%%%%%%%%%%%%%%%%%%%%%%%%%%

\section{Application to the Milky Way}
\label{sec:milkyway}

Having understood how kinematic fractionation gives rise to
morphological and kinematic differences between different stellar
populations, we now explore whether the resulting trends can explain
observations of the Milky Way's bulge by comparing the star formation
model with the Milky Way.  We also take this opportunity to predict
trends that have not been observed before.  We scale the model in the
same way as above and view the model from the Sun's perspective, using
the standard Galactic coordinates $(l,b$).  We assume that the Sun is
8 \kpc\ from the Galactic Centre, placing the observer at $y=-8$ \kpc,
and orienting the bar at an angle of $27\degrees$ to the line of sight
\citep{wegg_gerhard13}.  We emphasize that in this section we are only
interested in understanding trends that arise in the Milky Way's
bulge, not in matching them in detail.

\subsection{Caveats about the model}

Before comparing the model with the Milky Way, it is important to
mention some limitations of the model.  The model is very useful for
interpretating global trends, but should not be construed as a
detailed model of the Milky Way, even after we rescale it.  It suffers
from a number of limitations which preclude efforts to test the model
in a detailed quantitative way.  One important difference is that the
X-shape is less prominent in the model than in the Milky Way, and
rises to smaller heights.  This is possibly due to the fact that the
bar has not grown quite as much as in the Milky Way.  The bar in the
model has a radius of $\sim 3\kpc$ \citep{cole+14}, while the bar in
the Milky Way extends to $\sim 4.5-5\kpc$ \citep{wegg+15}.  The
relatively high gas inflow rate to the centre of the model, with the
attendant angular momentum transport and high star formation rate, is
probably the source of this slow growth.  \citet{cole+14} showed that
this gas inflow produces a nuclear disc, which
\citet{debattistaMWND+15} proposed is the origin of the high-velocity
peaks seen in the mid-plane line-of-sight velocity distributions at $l
\simeq 6\degrees-8\degrees$ in the APOGEE data \citep{nidever+12}.
\citep[Alternatively][proposed that these high-velocity peaks are
  produced by young stars trapped by the bar from the
  disc.]{aumer_schoenrich15}

A second limitation of the model is in its chemical evolution.  The
simulation did not include the diffusion of metals between gas
particles, without which too many low-metallicity stars form at all
ages.  This has the effect of broadening the metallicity distribution
function and of weakening the trends between metallicity and age.
With metal diffusion, a tighter correlation between age and chemistry
would have resulted, making the chemical separation by the bar even
stronger.  Also, while the yields we use for the \feh\ enrichment have
been shown to lead to very good matches to the metallicity
distribution functions across the Milky Way \citep{loebman+16}, the
oxygen yields may have significant offsets \citep[e.g.][]{loebman+11}.

Lastly, and most obviously, the simulation was only run for 10 \Gyr.
However by the end of the simulation, the model has long been in a
phase of slow secular evolution, growing slowly while continuing to
form stars.

\subsection{X-shape}
\label{ssec:xshape}

\begin{figure*}
\centerline{\includegraphics[angle=0.,trim=2.5cm 0.0cm 3.0cm
   0.0cm, clip=true,width=\hsize]{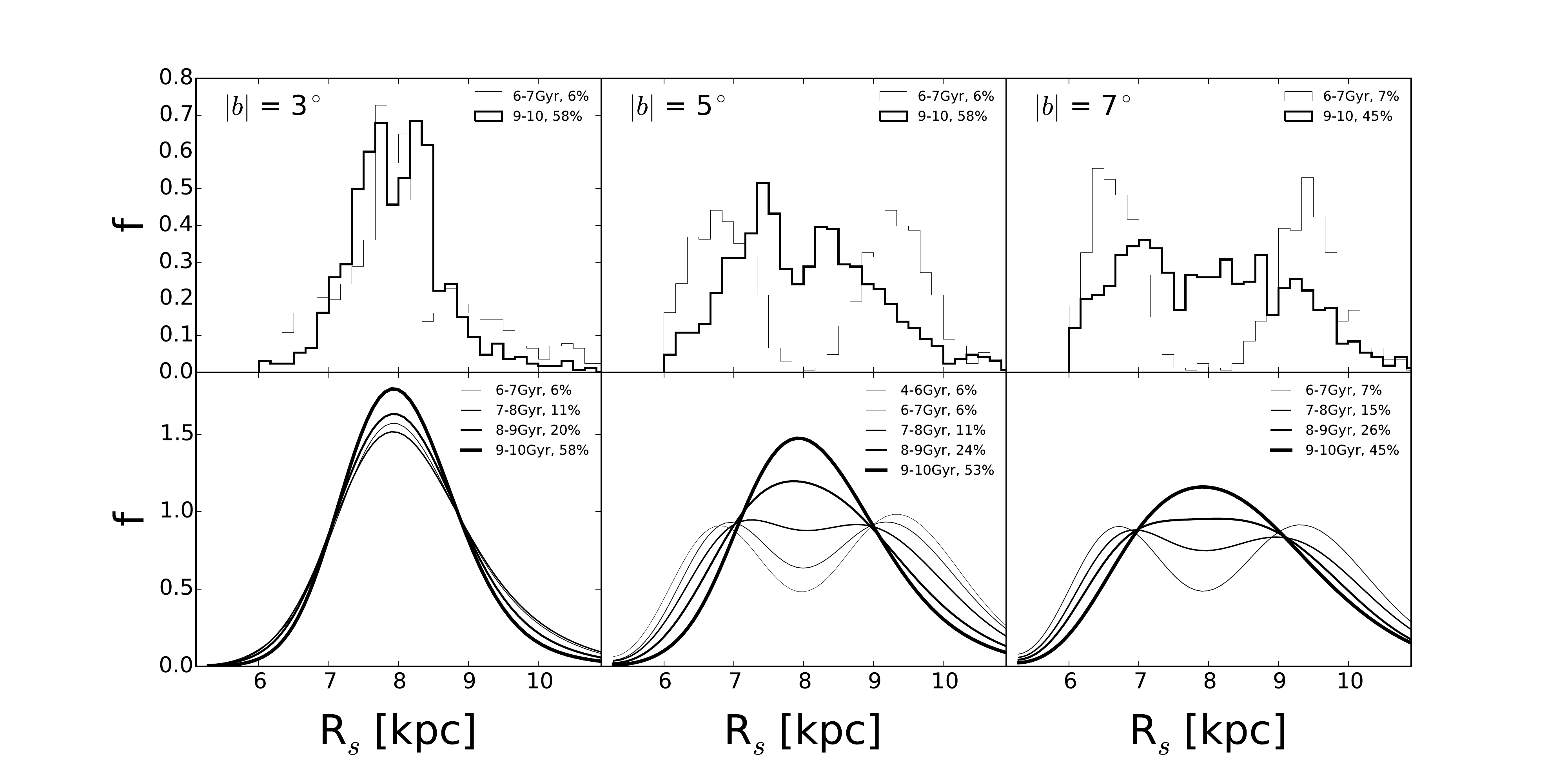}}
\caption{The distribution of red clump stars in the star-forming
  model, at distances from the Sun in the range $6 \leq \rs/\kpc \leq
  11$, at three different lines of sight on the minor axis, with the
  stars divided into age bins. Here, age is defined as $10\Gyr -
  \tau$. Top: raw data with no observational error added.  Bottom: the
  histograms convolved with the intrinsic scatter of the red clump
  magnitude of $\sig{} = 0.17$ dex.  Younger (metal-rich) red clump
  stars would appear distributed on an X-shape, while older
  (metal-poor) ones would not, given the intrinsic scatter.  More
  precise distance measurements however can reveal a weak bimodality
  also in metal-poor populations.}
\label{fig:xshape}
\end{figure*}

Several of the currently known morphological properties of the Milky
Way bulge are based on the magnitude measurement of red clump stars.
A number of studies have used bulge red clump stars from the Two
Micron All-Sky Survey and the VISTA Variables in the Via Lactea (VVV)
ESO public survey \citep{minniti+10} to derive the mean line-of-sight
distances across the bulge, thus mapping its 3D shape.  This has
resulted in the discovery of an X-shaped structure as a bimodal
distance distribution \citep{mcwilliam_zoccali10, nataf+10, saito+11,
  wegg_gerhard13, gonzalez+15}.  \citet{li_shen12} showed that a
peanut-shaped bulge viewed from the Solar perspective is responsible
for this X-shape. By separating red clump stars into metal-rich ($\feh
> -0.5$) and metal-poor ($\feh < -0.5$) populations, \citet{ness+12},
\citet{uttenthaler+12}, and \citet{rojas-arriagada+14} determined that
metal-poor stars do not follow the bimodal distance distribution and
therefore do not trace the X-shape of the bulge.

We now investigate the properties of the X-shape in the simulation.
\citet{gonzalez+15} have already compared the model, scaled and
oriented the same way as here, with the Milky Way.  They showed that
the model produced a split red clump at $b = -8.5\degrees$ for
$-2\degrees \leq l \leq +2\degrees$; at $l=+2\degrees$ the bright peak
is more prominent, while the faint peak is more prominent at
$l=-2\degrees$.  These are the same trends as observed in the VVV
survey and can be explained by an X-shaped overdensity of stars along
the line of sight.

In order to test if the model is able to reproduce the trends with
metallicity, we observe the star particles assuming they are red clump
stars.  Because young, metal-poor stars (caused by the absence of gas
metallicity mixing) contaminate the old, metal-poor bins when
selecting by metallicity, in this analysis we deconstruct the X-shape
by age, rather than by \feh.  The method to derive distances from red
clump stars is based on the construction of the luminosity function of
the bulge towards a given line of sight where the red clump can be
easily identified and its mean magnitude obtained \citep{stanek+94}.
We set the absolute magnitude of stars in the model to $M_K = -1.61$,
selecting equal numbers of stars in the magnitude bins $12.3 < K <
12.9$ and $12.9 < K < 13.58$ (chosen to sample the near and far arms
of the X-shape with equal numbers of stars, independent of the
detailed density distribution, reminiscent of the ARGOS selection
strategy). We convolve the magnitude distribution with a Gaussian
kernel of $\sigma$ = 0.17 mag. to simulate the intrinsic scatter of
the red clump \citep[e.g.][]{gerhard_martinez-valpuesta12}, and then
transform back to distance using the distance modulus appropriate for
$8 \kpc$.

The bottom row of Fig.~\ref{fig:xshape} shows the distribution of the
red clump stars in the model, at distances from the Sun $6 \leq
\rs/\kpc \leq 11$, for three different lines of sight on the
minor axis ($|l| < 2\degrees$), where the stars are divided into four
age bins.  At low latitudes ($b \sim 3\degrees$) the distance
distribution generally consists of a single peak for all stellar ages,
as expected since the arms of an X-shape only become sufficiently
separated at higher latitudes.  No split is present in the oldest
stars even at higher latitudes, but one is visible for younger stars.
The separation between the peaks increases and the depth of the minima
between peaks increases with decreasing stellar age. Both the
separation and the depth of the minima also become more prominent with
increasing latitude.  The prominence of the X-shape depends strongly
on stellar age, with the older stellar populations showing a weak (or
no) bi-modality and the younger populations exhibiting a strongly
bimodal distance distribution.

We infer that the age dependence in the distribution of stars tracing
the X-shape qualitatively explains the metallicity dependence of the
split clump observed in the Milky Way, where the most metal-rich stars
show the largest bimodality and stars of $\feh < -0.5$ show only a
single peak in their distance distribution. We identify the oldest
stars in the simulation, which trace the X-shape weakly, if at all,
with the metal-poor stars in the Galaxy.  Conversely, we identify the
more metal-rich stars in the Milky Way, which do trace the X-shape,
with the slightly younger stars in the simulation.

The bottom row of Fig.~\ref{fig:xshape} also shows that stars younger
than $7 \Gyr$ in the model also have a split red clump.  The
percentage of stars in each age bin at each latitude is indicated in
each panel of Fig.~\ref{fig:xshape}.  Stars younger than $7\Gyr$
contribute $<14\%$ of stars in this region, a number which reflects
details of the model's evolution.  In the Milky Way, this fraction
would be even lower, since observations based on turnoff studies show
that such stars contribute $<5\%$ of bulge stars
\citep[e.g.][]{ortolani+95, kuijken_rich02, zoccali+03, sahu+06,
  clarkson+08, clarkson+11, brown+10, valenti+13}.  The crucial point
is that an age difference of just $2 \Gyr$ is sufficient for the bulge
red clump distributions to change from a single peak to a bimodal
distribution.

The absence of a bimodal distribution in stars older than $8\Gyr$ is
not absolute.  We have convolved the magnitude distribution with an
intrinsic scatter of $\sig{} = 0.17$ mag.  If the observational
uncertainty were lower, then even the oldest population exhibits a
weakly bimodal distribution.  The top row of Fig. \ref{fig:xshape}
presents examples of distance distributions unconvolved with any
observational uncertainty.  In this case, even the oldest population
(age $>9\Gyr$) has a bimodal distribution at $|b| = 5\degrees$.  A
prediction that follows is that, with higher distance precision, a
weaker split at lower metallicities may also be observed.  A second
prediction from the unsmoothed data is that, for the younger
populations, the stars are distributed in two distinct distributions,
corresponding to the arms of the X-shape, with a near zero density of
stars in between.  We emphasize that this particular prediction is for
ages, not \feh\ since at any age the distribution of \feh\ is probably
not perfectly single-valued.

\subsection{Interpretation of the different distributions of 
Milky Way bulge red clump stars and RR~Lyrae}
\label{ssec:rrlyrae}

\begin{figure}
\includegraphics[width=9cm, angle=0,trim=0.0cm 0.0cm 0cm
  0.0cm,clip=true]{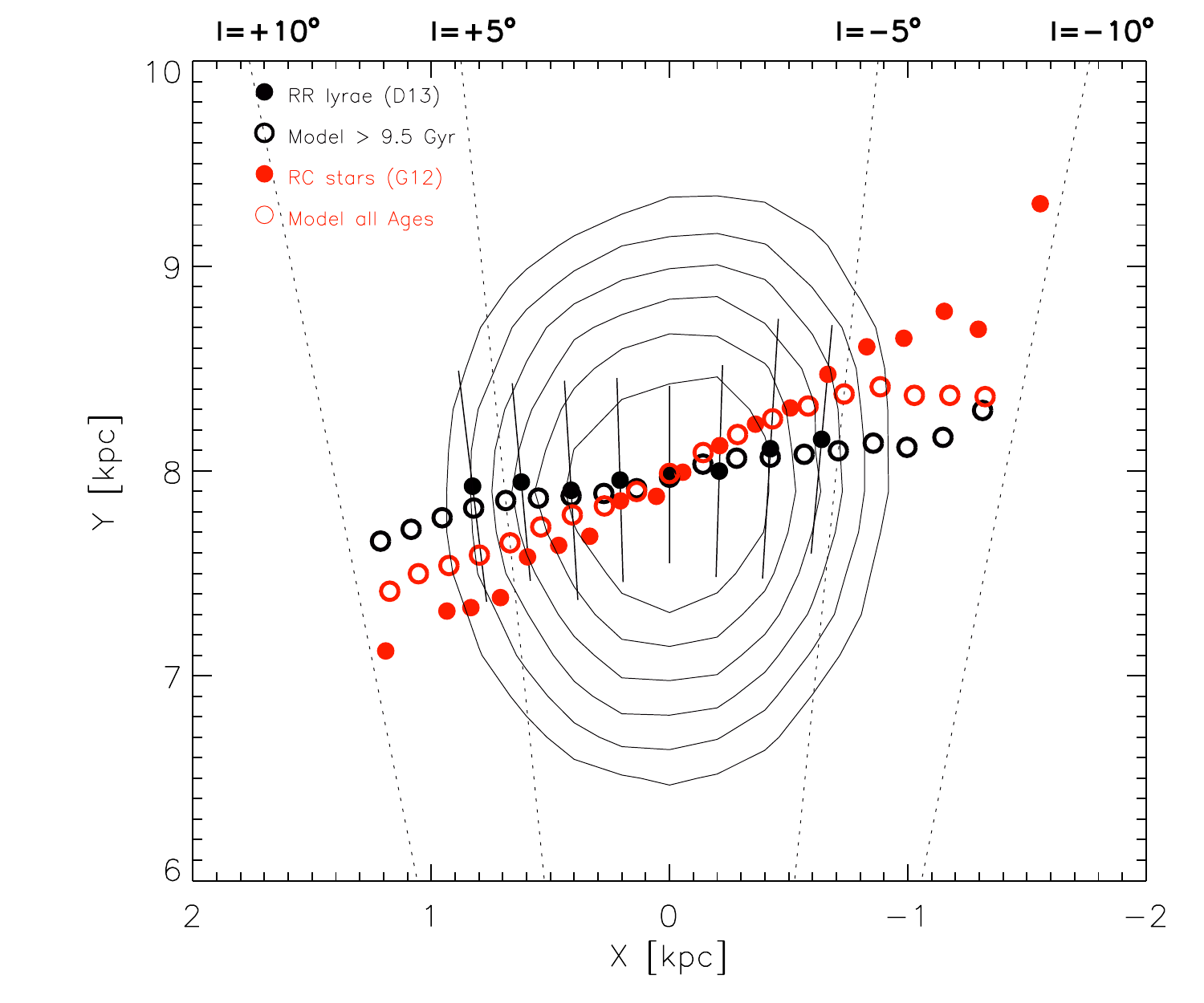}
\caption{Face-on projection of the simulated mean line-of-sight
  distances for star particles at $3.5\degrees < b < 4.5\degrees$ with
  ages $>$ 9.5 Gyr (black open circles) and for all the star particles
  (red open circles) in the star-forming simulation.  The projected
  mean line-of-sight distances of RR~Lyrae from \citet{dekany+13}
  (black filled circles) for latitudes $3.5\degrees \leq b \leq
  4.5\degrees$ and of red-clump stars in the same latitude range from
  \citet{gonzalez+12} (red filled circles) are also shown.  The
  surface density contours of the full RR Lyrae sample of
  \citet{dekany+13} are included in the plot. All distances, for
  simulated and observed stars, were calculated adopting an absolute
  magnitude of $M_K = -1.61$ mag in order to bring the model to the
  mean distance of the RR Lyrae.  The simulation matches the weaker
  barred distribution of RR~Lyrae and stronger bar in the red clump
  stars seen in the Milky Way's bulge.}
\label{fig:rrlyrae}
\end{figure}

The dependence of bar strength on age, shown in
Figs.~\ref{fig:feosnapshots} and \ref{fig:baramp}, is directly
relevant to interpreting the difference in the distribution of
RR~Lyrae and red clump stars noted by \citet{dekany+13}.  The adoption
of red clump stars as distance tracers includes the entire range of
stellar ages that is found in the Milky Way bulge. On the other hand,
RR~Lyrae are old stars \citep{Walker89}.  \citet{dekany+13}
investigated the spatial distribution of the oldest population of the
bulge by measuring the distances to RR~Lyrae based on their
near-infrared light curves from VVV photometry. They found a
distribution of RR~Lyrae in the Galactic bulge that is quite round and
exhibits little evidence of the bar morphology traced by the red clump
stars.  Fig. \ref{fig:rrlyrae} shows the projected mean distance to
red clump stars from \citet{gonzalez+12} compared to that from RR
Lyrae. We used the individual distances to the RR Lyrae from
\citet{dekany+13} to calculate a mean RR Lyrae distance using a
Gaussian fit to the distance distribution in each line of sight. The
projected distance is shown in Fig. \ref{fig:rrlyrae} as well as the
surface density contours using the individual distances of the entire
RR Lyrae sample from \citet{dekany+13}.  \citet{dekany+13} interpreted
the spheroidal distribution of this old population as evidence for a
dynamically distinct population that formed separate from the disc in
the bulge.

We assess the need for such a composite bulge scenario by measuring
the mean distances to the simulated stars, in a similar way as done
for the observations. We first select all the stars within the
distance range $\rm 4 \leq \rs/\kpc \leq 12$ to sample the integrated
population of the red clump stars observed across the extent of the
bulge. We then consider only the oldest stellar population with $\tf
\leq 0.5 \Gyr$ to obtain a sample representative of the oldest
population of the bulge, corresponding to the population traced by the
RR Lyrae. We then measure the mean distances to both samples in
$(l,b)$-space at a fixed latitude of $3.5\degrees \leq b \leq
4.5\degrees$ in $1\degrees$--spaced longitude bins between $-9\degrees
\leq l \leq +9\degrees$.  As we did in comparing with the observed
X-shape, in order to match the observational uncertainties, we first
convert the line of sight distance of every star particle to an
observed magnitude by adopting an absolute magnitude for the red clump
of $M_K$ = -1.61 mag \citep{alves00}.  The magnitude distribution of
stars towards each line of sight is then convolved with a Gaussian
with $\sig{} = 0.17$ mag. in order to account for the intrinsic
scatter of the bulge red clump population \citep{gonzalez+11,
  gerhard_martinez-valpuesta12}.  We then produce a Gaussian fit to
the magnitude distribution towards each line of sight and retrieve the
corresponding mean magnitude of the distribution. Finally, this mean
magnitude value is converted back to a distance using the same $M_K$
value. For the oldest stars, instead, we ignore the small distance
uncertainties.  Fig.~\ref{fig:rrlyrae} shows the mean positions
projected on to the $(x,y)$-plane for both cases: including all the
stars, and selecting only the oldest stars from the simulation.

Purely by measuring distances to the oldest stars, the distance
profile becomes flatter than when including the entire range of ages,
with a remarkable similarity to the RR~Lyrae measurements in the Milky
Way.  Furthermore, when all the stars of the simulation are included,
the distance profile becomes steeper and traces the position angle of
the bar. Thus this result explains the different morphological
signatures traced by red clump stars and RR~Lyrae within the Galactic
bulge, without the need for a second, dynamically distinct, component.
Instead, the continuum of bar quadrupole moments, from weak to strong
as a function of formation time, is able to explain the observed shape
difference.

A natural interpretation of the nearly spheroidal distribution of the
RR~Lyrae stars in the Galactic bulge is that these are older than the
bar itself.  By the time the bar formed, they constituted a radially
hotter stellar population, possibly as part of the stellar halo, and
therefore did not acquire a strong quadrupole moment.  A consequence
of this interpretation is that the RR~Lyrae should not exhibit a
strong peanut shape.

Recently, \citet{kunder+16} presented radial velocity measurements of
947 RR~Lyrae from the Bulge RR Lyrae Radial Velocity Assay (BRAVA-RR)
survey located in four bulge fields at Galactic latitude $-5\degrees <
b < -3\degrees$ and longitude $|l| < 4\degrees$. \citet{kunder+16}
concluded that the bulge RR~Lyrae are part of a separate component,
\ie\ a classical bulge or inner halo, based on the null rotation
observed in this population when compared to the RC stars. If this
result is confirmed after extending the rotation curve of the RR~Lyrae
to larger Galactic longitudes ($|l| < 10\degrees$), it could suggest
that the population traced by these stars is linked to the most
metal-poor ($\feh <-1.0$) population of red clump stars that shows a
similar kinematic behaviour as observed by the ARGOS survey
\citep{ness+13b}. Here, we showed that the oldest population found in
the model follows a weakly barred distribution.  However, we do not
find a velocity difference between RR~Lyrae and red clump stars as
large as that seen in the BRAVA-RR data.

A more continuous change in the orientation of the bar with age has
been measured in bulge Mira variables \citep{catchpole+16}. Mira
variables follow a period-age relationship, with short periods
corresponding to older stars \citep{wyatt_cahn83}, spanning a range of
ages from $\sim 3\Gyr$ to globular cluster ages.  \citet{catchpole+16}
find that the younger ($\la 5\Gyr$) Mira variables in the bulge at
$(l,b) = (\pm8\degrees,-7\degrees)$ follow a clear bar structure,
while the older ones are more spheroidal.  The angle of the major axis
of the distribution of Miras to the line of sight twists continuously
with their period, as would be expected for populations with
continuously varying quadrupole moments
\citep[e.g.][]{gerhard_martinez-valpuesta12}.

\subsection{Stellar kinematics}
\label{ssec:stellarkine}

\begin{figure*}
\centerline{\includegraphics[angle=0., width=\hsize]{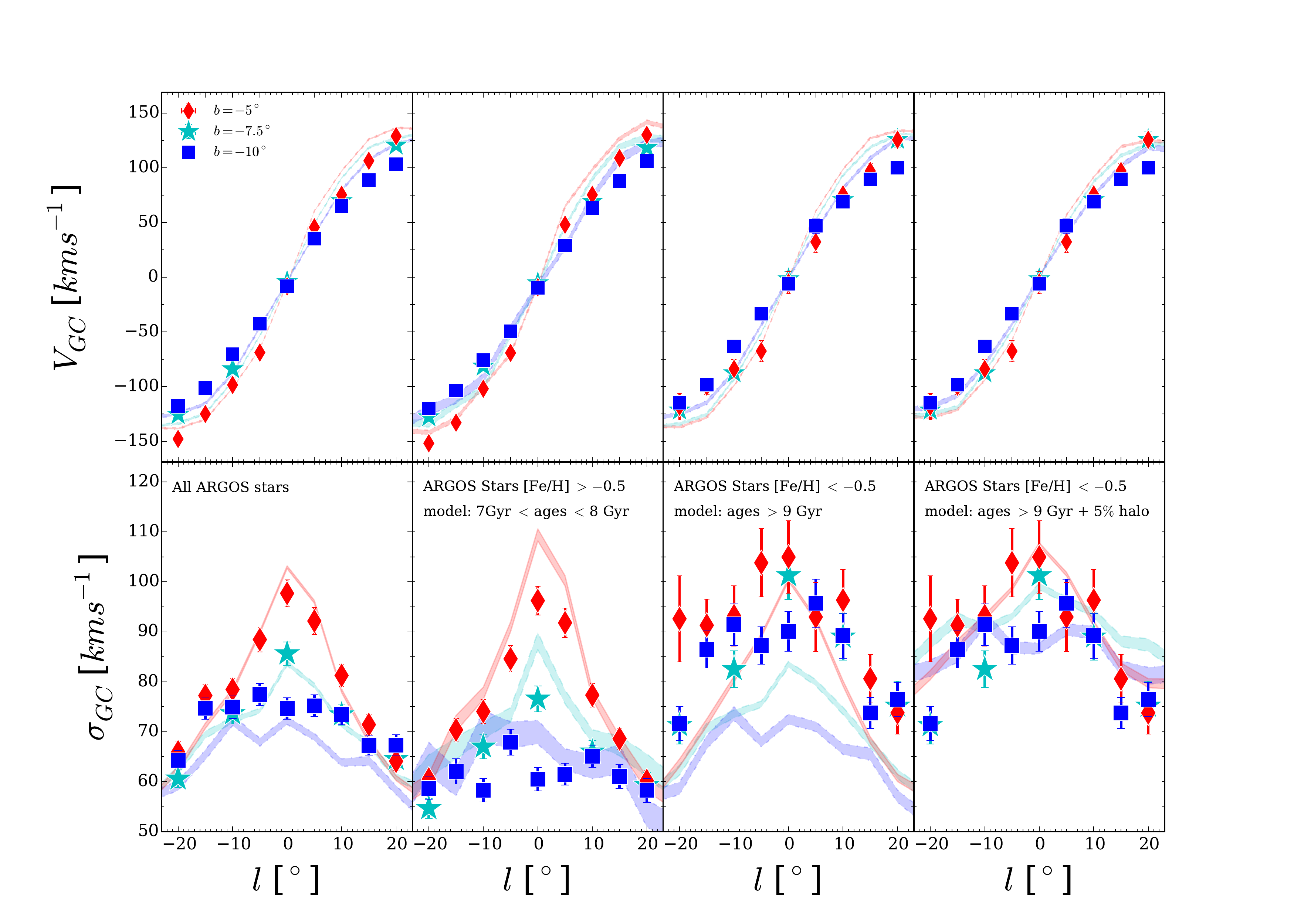}}
\caption{The ARGOS data (points with error bars) compared to the
  star-forming simulation (shaded region showing the sampling
  uncertainty; the shading uses the same colour scheme as the ARGOS
  data points to indicate latitude).  Left: all stars in the
  simulation along a given line of sight.  Second column: stars with
  $\feh > -0.5$ compared to model stars with ages between $7$ and
  $9 \Gyr$.  Third column: ARGOS stars with $\feh < -0.5$ compared
  with old model stars (age $> 9 \Gyr$).  Right: the same old stars in
  the model with a $15\%$ admixture of a hot population as described
  in the text.  This fraction corresponds to $5\%$ of the stellar mass
  of the bulge; this small addition leads to an excellent match to the
  observed kinematics of the metal-poor stars.}
\label{fig:fullkinematics}
\end{figure*}

We now compare the kinematics of the bulge as observed in ARGOS,
selecting stars in the simulation to be within $3.5 \kpc$ of the
Galactic Centre, as was done for red clump stars (with distance errors
$<1.5\kpc$) in ARGOS.  The results of this section are not too
sensitive to these distance cuts; removing the distance cuts changes
the mean velocities and the dispersions by no more than $10\%$.

The right column of Fig.~\ref{fig:fullkinematics} shows the rotation
and dispersion profiles in the simulation compared to the ARGOS data,
at latitudes of $b = 5\degrees$, $7.5\degrees$ and $10\degrees$.  The
rotation for both ARGOS and the simulation is cylindrical, with very
little difference in rotation speed at different latitudes.
Cylindrical rotation of the Galactic bulge has been observed by
several studies \citep[e.g.][]{howard+08, kunder+12, zoccali+14,
  ness+13b, ness+16b} and is a generic feature of $N$-body models of
boxy bulges \citep{combes+90, athanassoula_misiriotis02,
  saha_gerhard13}.  The dispersion profiles of the observations and
the simulation are also very similar, peaking in the centre and
decreasing with increasing longitude. The peak velocity dispersion is
observed near the centre ($b \sim 2\degrees$) on the minor axis, as
first reported by the GIBS survey \citep{zoccali+14}, and decreases
with latitude.  The model's velocity dispersion at $b = 2\degrees$ is
$\sim 1.5$ times the velocity dispersion at $5\degrees$, which is
comparable to the fractional difference reported in GIBS
\citep{zoccali+14} and ARGOS \citep{ness+16}. Outside the boxy bulge,
in the disc ($|l|>15\degrees$), the dispersion is similar at all
latitudes.

The ARGOS stars span a metallicity range of $-2 < \feh < 0.5$,
although $95\%$ of stars have $\feh > -1$. The velocity dispersions of
stars vary as a function of their metallicity \citep{babusiaux+10,
  ness+13b, ness+16b}.  We investigate these observed differences in
the kinematics using the stellar age in the simulation as a proxy for
\feh, as we did previously for the X-shape analysis.  The second
column of Fig.~\ref{fig:fullkinematics} shows the kinematics of the
metal-rich ($\feh > -0.5$) ARGOS stars, together with intermediate age
($7\Gyr < $ age $< 8 \Gyr$) stars in the model.  The model stars have
kinematics that are qualitatively similar to the observational trends
of the metal-rich stars, with a larger difference in velocity
dispersion between $5\degrees \leq |b| \leq 10 \degrees$, and a lower
and flatter dispersion profile at higher latitudes, compared to all
stars (left column).  The $\sim 10\%$ increase in the velocity
dispersion at $b = -5\degrees$ of the intermediate-age stars relative
to the full distribution is not observed in ARGOS for the more
metal-rich stars.  The origin of this discrepancy is unclear.

The third column of Fig.~\ref{fig:fullkinematics} compares the
metal-poor ($\feh < -0.5$) ARGOS stars with old (age $>9 \Gyr$) stars
in the model.  The trends observed for this population, including the
relatively high and flat velocity dispersion that changes slowly with
latitude, are not reproduced by the model.  We have explored a variety
of age cuts in the model, and all of them fail to match these
kinematic trends.  This suggests that the model is missing a component
that can produce these kinematics.  A similarly hot population at low
metallicity was found by \citet{babusiaux16}.  In order to explore
this population further, we note that ARGOS finds that stars with
$\feh < -1$ have a $\sim 50\%$ rotation velocity and an average
velocity dispersion $\sim 120\kms$ across $5\degrees \leq b \leq
10\degrees$ \citep[component D of][]{ness+13a}.  We add such a
population to the old stellar population in the model.  The right
column of Fig.~\ref{fig:fullkinematics} shows the outcome of adding
$15\%$ of stars with the properties of population D from ARGOS
(\ie\ rotation speed which is $50\%$ of the metal-rich stars and a
dispersion of $120\kms$) to the old population in the model.  The
qualitative match to the observations is now very good, with flat,
nearly constant velocity dispersion with longitude and a relatively
small drop with latitude.  Bearing in mind that the $\feh < -0.5$
population accounts for $30\%$ of all the ARGOS stars, and that the
model needed only $\sim15\%$ of such stars to match this behaviour, we
conclude that this component accounts for $\sim 5\%$ of the mass of
the central Milky Way.  This additional population is most likely the
stellar halo, which must also be present in the inner Milky Way.  This
low contribution of a hot component is significantly more stringent
than the estimates of \citet{jshen+10} for the presence of a hot,
unrotating component.

\subsection{Trends in the age distribution}
\label{ssec:agetrends}

\begin{figure}
\centerline{\includegraphics[angle=0.,width=\hsize]{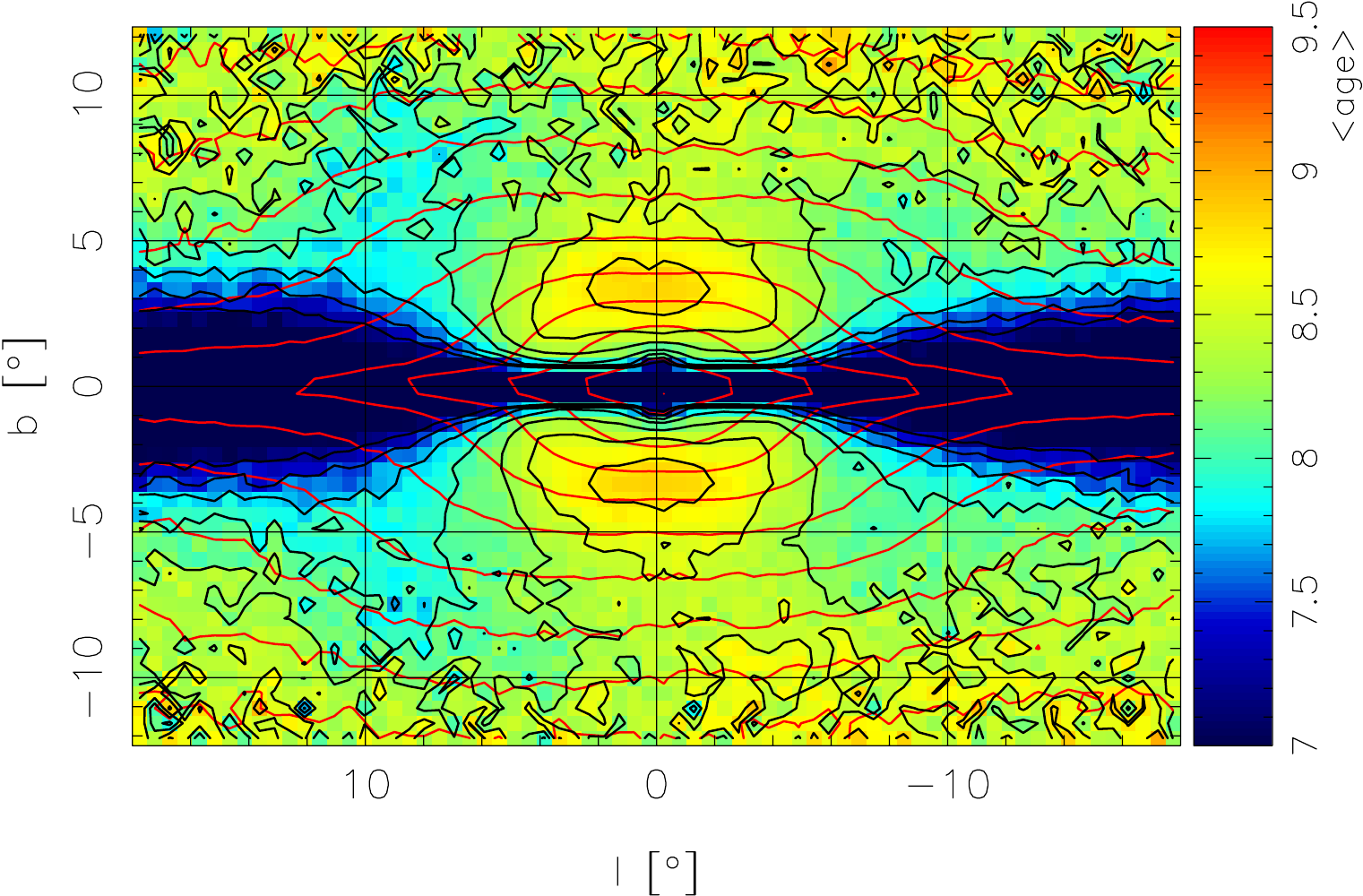}}
\centerline{\includegraphics[angle=0.,width=\hsize]{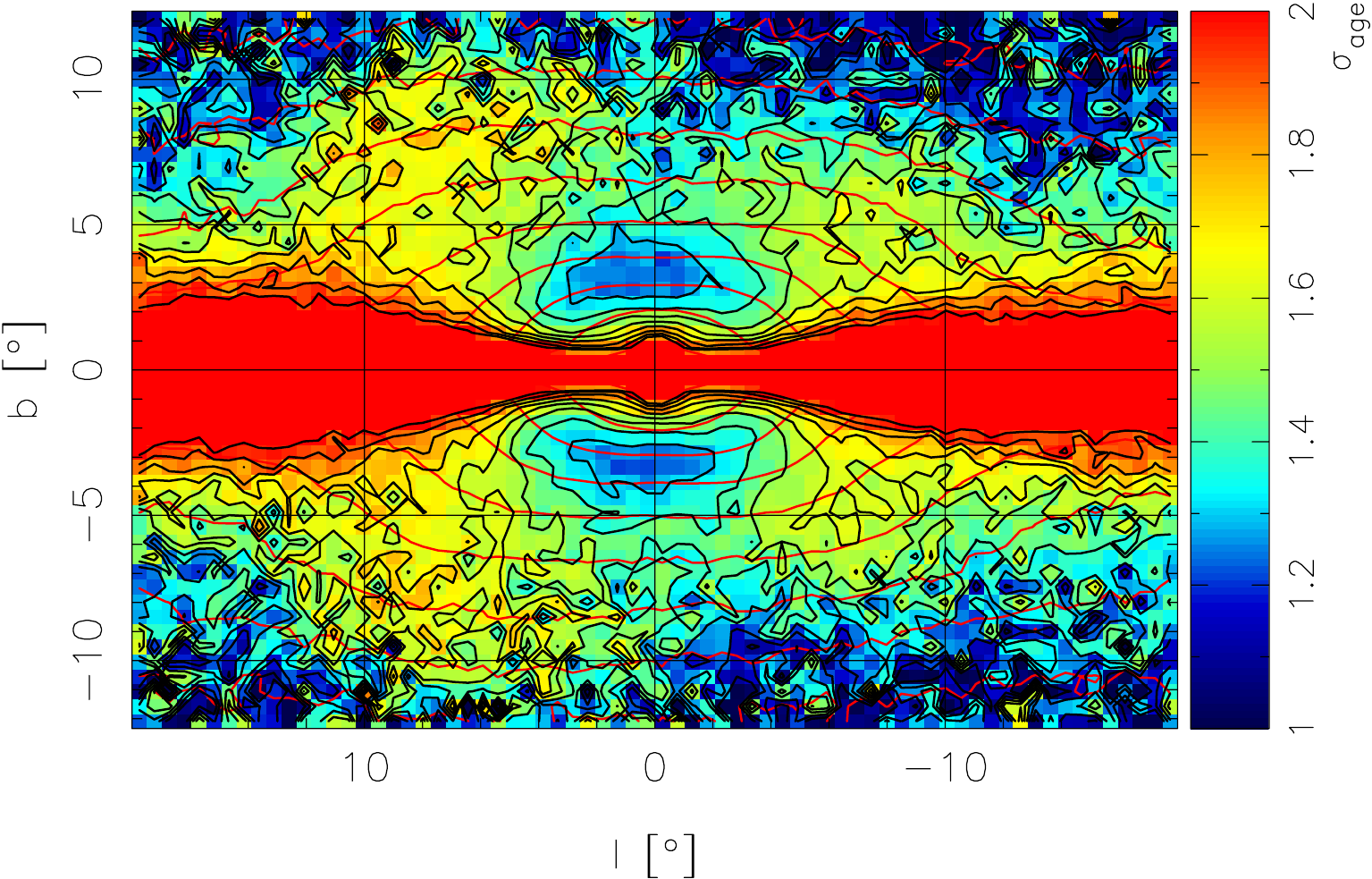}}
\centerline{\includegraphics[angle=0.,width=\hsize]{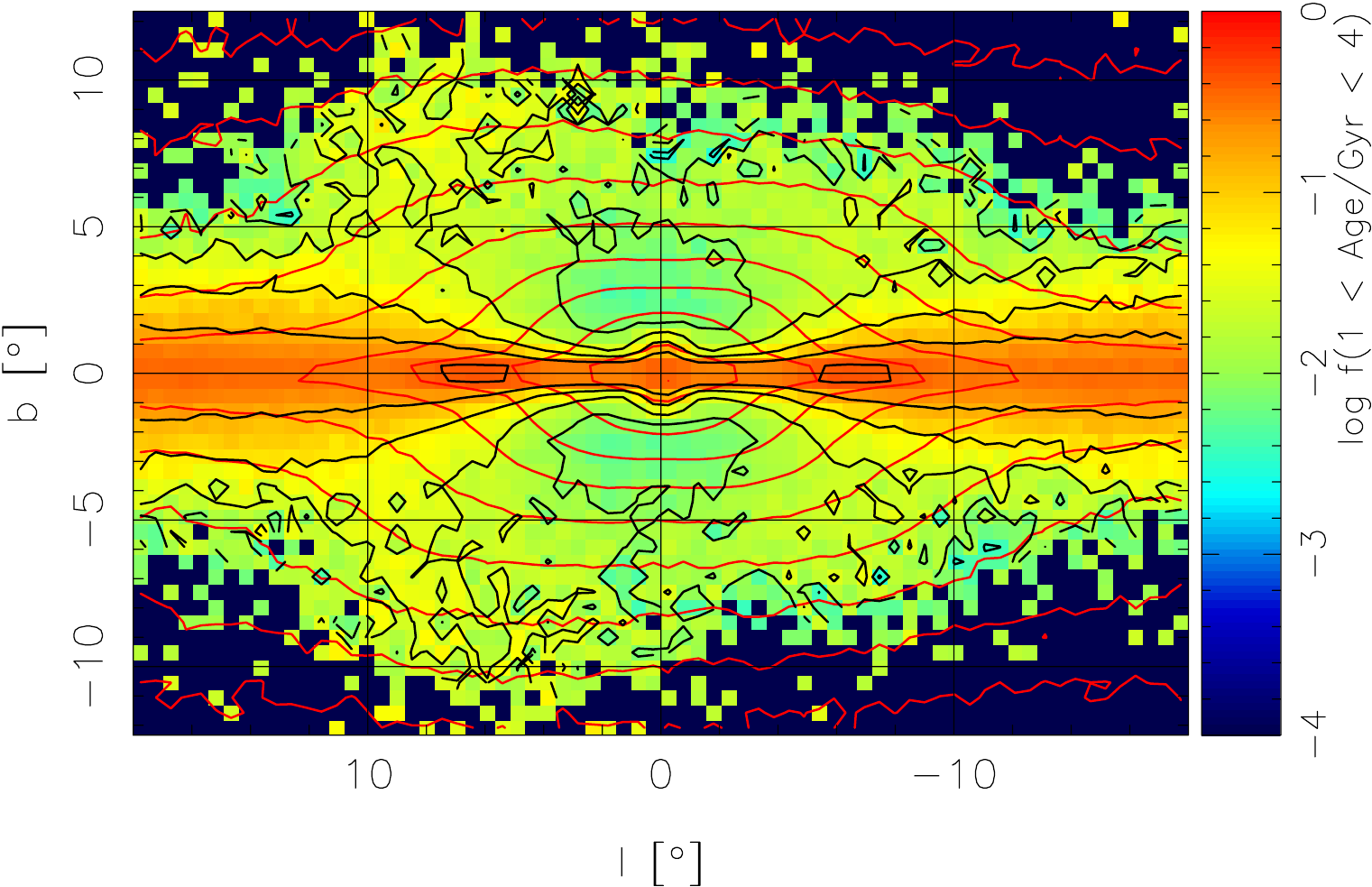}}
\caption{Maps of \avg{age}\ (top), \sig{age} (middle) and fraction of
  intermediate-age stars (bottom) in $(l,b)$-space as seen from the
  Sun for the star-forming simulation.  Only stars in the range $4
  \leq \rs/\kpc \leq 12$ are considered.  Black contours show constant
  plotted quantity, while red contours show constant surface density.
  Note the difference between the age distribution and the density, and
  the dominance of old stars on the minor axis.  }
\label{fig:ages}
\end{figure}

Fig. \ref{fig:ages} shows the average age, \avg{age}, (top panel) and
the standard deviation in the age, \sig{age}, (middle panel) in
$(l,b)$-space.  The peak of \avg{age}\ is on the minor axis, although
a small shift to positive $l$ due to the effect of perspective is
evident.  At the same location, \sig{age}\ is at a minimum.  The peak
in \avg{age}\ and the minimum in \sig{age}\ are due to the fact that
it is the oldest stars that ascend the furthest on the bar's minor
axis, giving rise to a vertical age gradient and a minimum in
\sig{age}\ as seen from the Sun.  Younger populations are instead more
peanut-shaped.  The dominance of the oldest stars above the mid-plane
is consistent with observations in the Milky Way
\citep[e.g.][]{ortolani+95, kuijken_rich02, zoccali+03, sahu+06,
  clarkson+08, clarkson+11, brown+10, valenti+13}.  \citet{ness+14}
showed that, none the less, younger stars are also mixed in with the old
stars, particularly when the most metal-rich populations are
considered, even at latitudes $|b| \ge 2\degrees$.  This qualitatively
agrees with the finding of younger microlensed dwarf stars
\citep{bensby+11, bensby+13}.  In the bottom panel of
Fig. \ref{fig:ages} we show the fraction, $f$, of all stars which have
intermediate ages, which we define as those with ages between $1$
and $4\Gyr$ ($9 \geq \tf/\Gyr \geq 6$).  Outside the disc, the
intermediate age population always accounts for less than $10\%$ of
the stars, in good agreement with Milky Way estimates from isochrone
fitting \citep[e.g.][]{clarkson+11}, but lower than estimated from
microlensing \citep[e.g.][]{bensby+13}.  Neglecting the disc, the
youngest \avg{age}\ is at $l \simeq \pm 10\degrees$.  As with
\avg{age}\ and \sig{age}, the minimum of $f$ is on the minor axis
displaced slightly to positive $l$.  There is a mismatch between the
contours of surface density and those of \avg{age}, \sig{age}\ and
$f$, with all these more pinched than the density.

\begin{figure*}
\centerline{
\includegraphics[angle=0.,width=0.5\hsize]{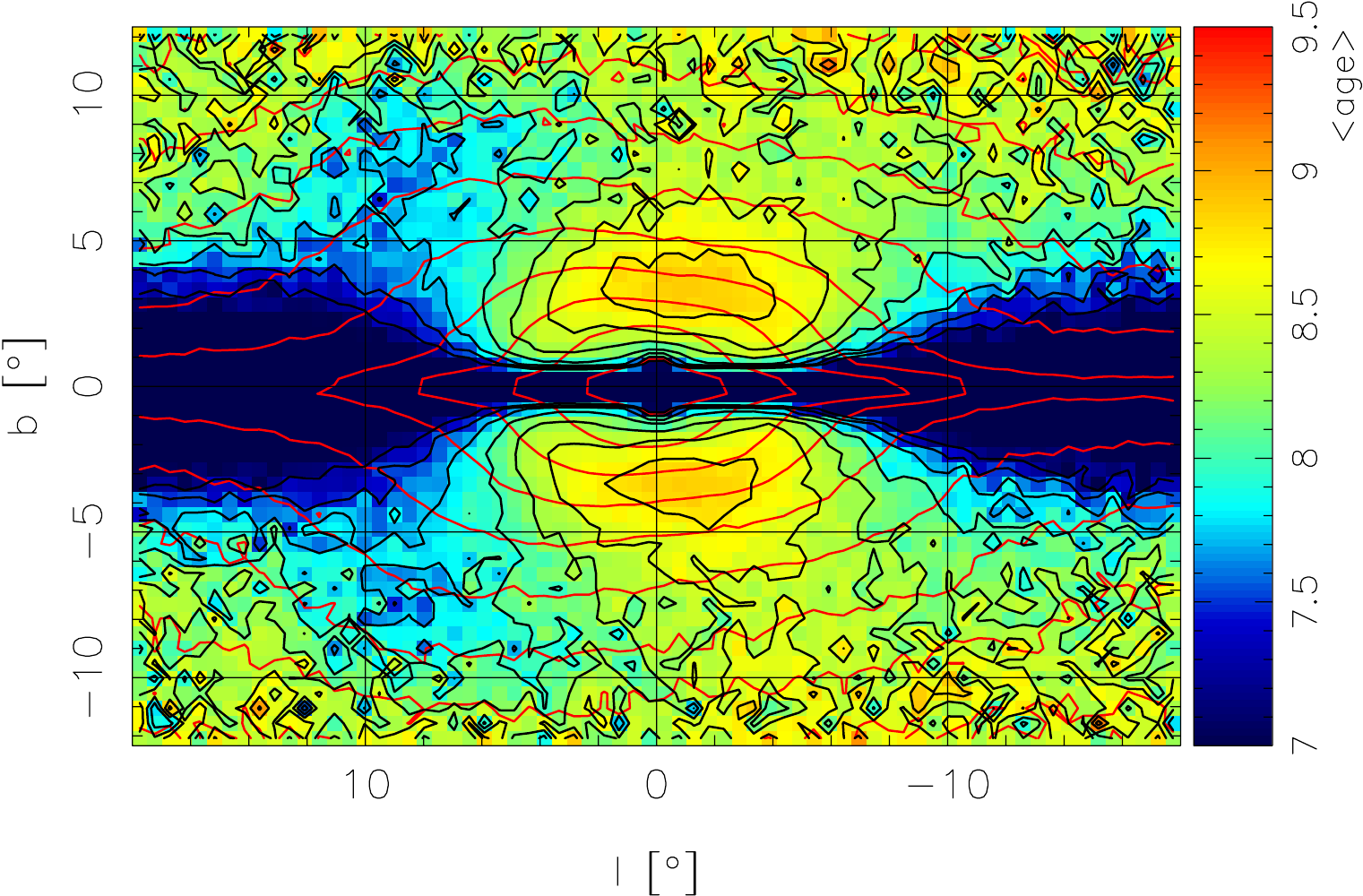}
\includegraphics[angle=0.,width=0.5\hsize]{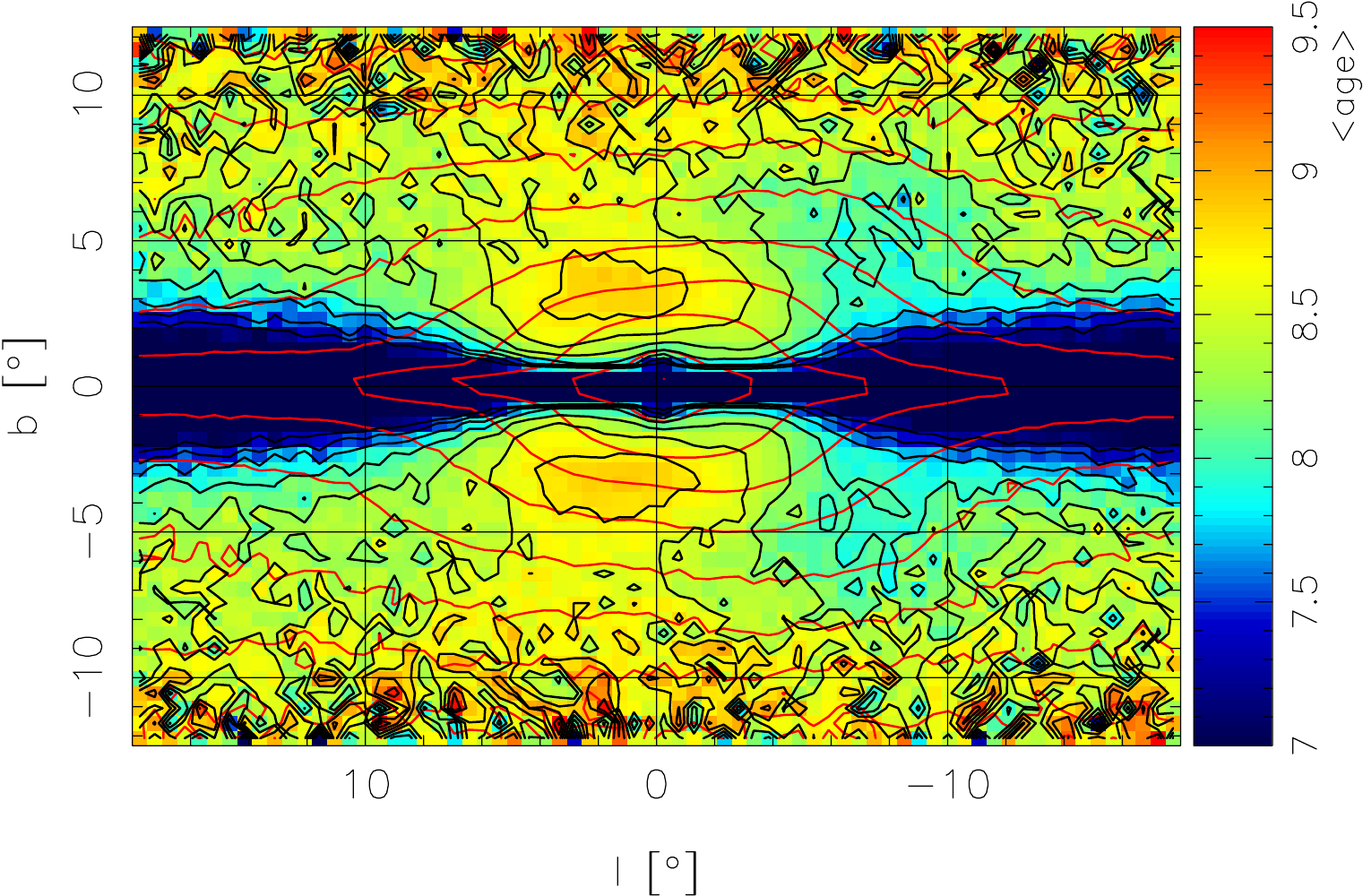}
}
\centerline{
\includegraphics[angle=0.,width=0.5\hsize]{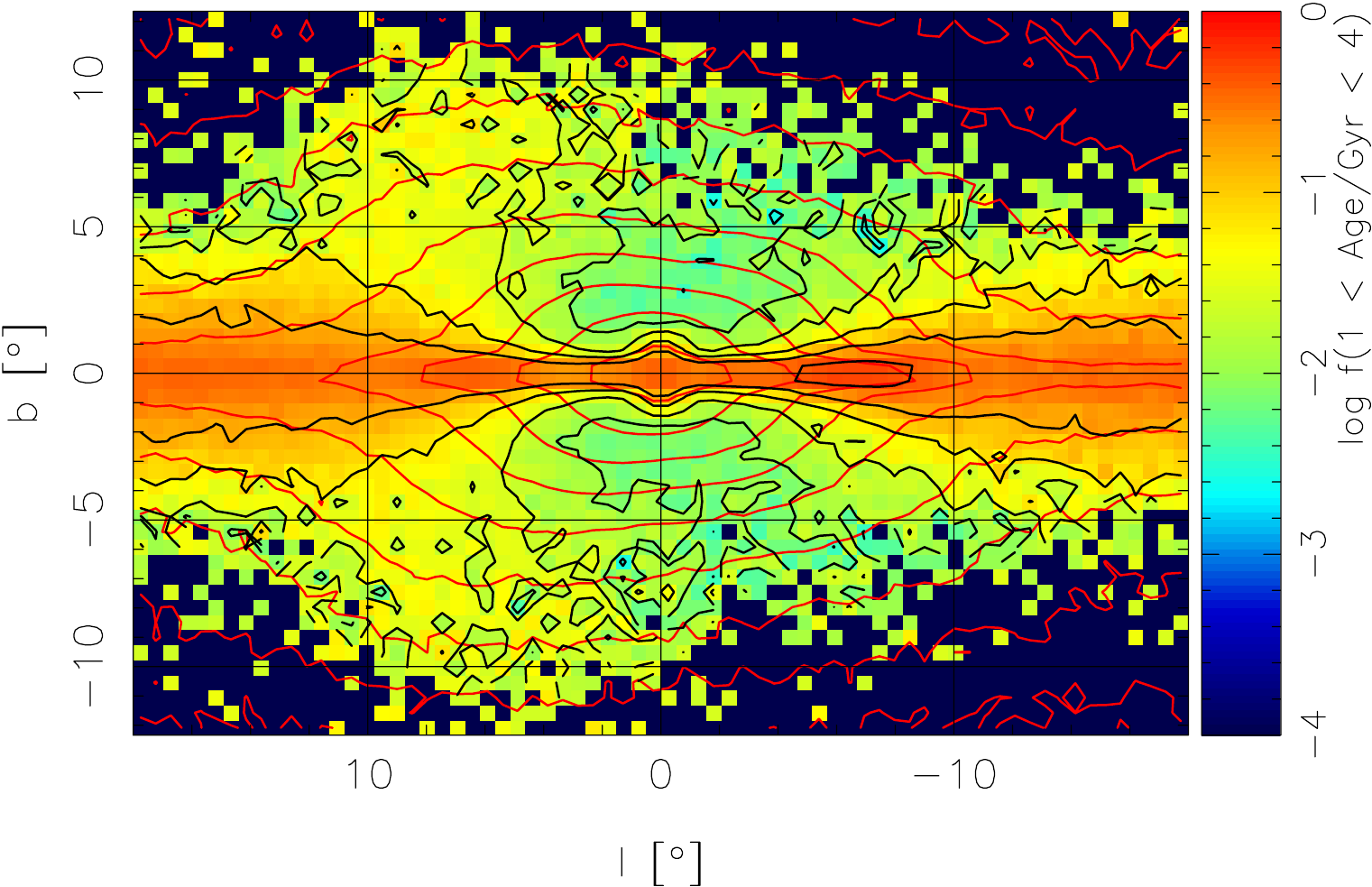}
\includegraphics[angle=0.,width=0.5\hsize]{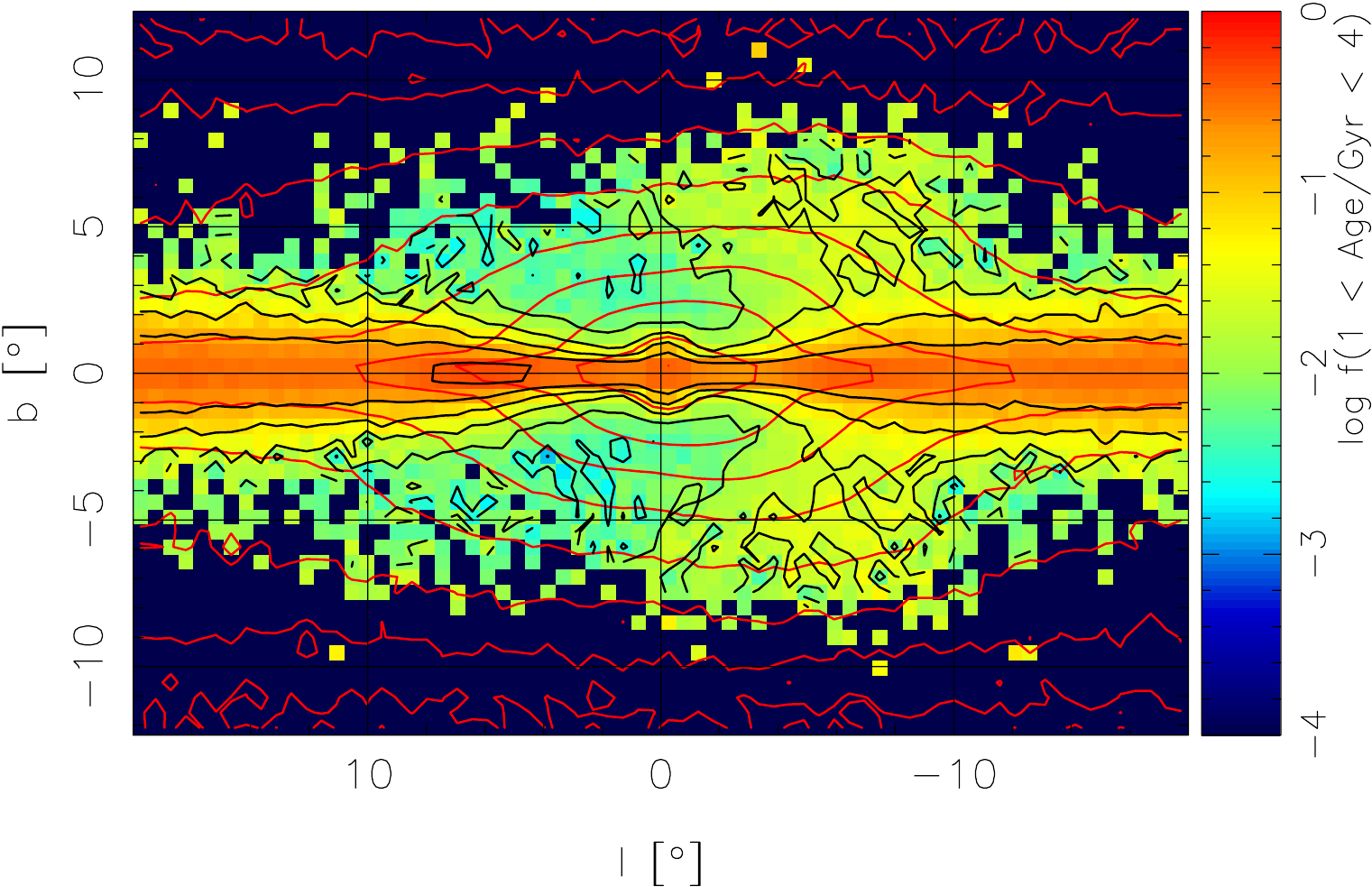}
}
\caption{Near- versus far-side stellar age distributions in the
  star-forming simulation. Top row: maps of \avg{age}.  Bottom row:
  maps of fraction of intermediate-age ($1\Gyr \leq$ age $\leq 4\Gyr$)
  stars.  The maps on the left are for near-side stars ($4 \leq
  \rs/\kpc \leq 8$), while those on the right are for the far-side
  stars ($8 \leq \rs/\kpc \leq 12$).  Black contours show constant
  plotted quantity while red contours show constant surface density of
  the population under consideration.  Away from the minor axis
  younger stars contribute most on the side where the arms of the
  X-shape are located.  Thus, the greater vertical extent of the
  intermediate-age stars is at $l>0\degrees$, a perspective effect.}
\label{fig:nearfarages}
\end{figure*}

Fig. \ref{fig:nearfarages} considers the difference in the
intermediate-age distribution for near- and far-side stars.  We define
the near side as $4 \leq \rs/\kpc \leq 8$ and the far side as $8 \leq
\rs/\kpc \leq 12$.  On the near side (left-hand panels of Fig.
\ref{fig:nearfarages}), the peak \avg{age}\ is at $l < 0\degrees$
while more intermediate-age stars are at $l > 0\degrees$ off the
mid-plane.  The situation is reversed on the far side.  The
interpretation of this result is that the X-shape contains more
intermediate-age stars ($\sim 6\%$ in the model), and this structure
exhibits a strong perspective effect.  This distribution has an
important consequence for microlensing.  If microlensing events are
dominated by sources on the far side of the Galaxy \citep{evans94,
  zhao+96, evans_belokurov02}, then the fraction of lensed
intermediate-age stars should be higher at $l < 0\degrees$.  We split
the microlensed star sample of \citet{bensby+13} at negative latitudes
by age into three bins: young stars (age $< 2\Gyr$), old stars (age $>
9\Gyr$), with the remainder as intermediate age stars.  Only three
microlensed dwarfs are in the young bin, and we ignore them, assuming
they are disc stars.  There are 38 stars in the intermediate-age bin
and 30 in the old bin.  Remarkably, the intermediate age stars
comprise $\sim 50\%$ of the total (24 of 49 stars) at $l>0\degrees$
and $\sim 74\%$ of the total (14 of 19 stars) at $l<0\degrees$.
Accounting for the uncertainties in the stellar ages, the fraction of
intermediate age lensed stars remains higher at negative longitudes
compared with positive longitudes.  Even if the sample of
\citet{bensby+13} is contaminated by foreground stars, as seems likely
given the relatively high fraction of intermediate-age stars, the
bulge remains the most likely source of an age asymmetry across
longitude.  While these results are suggestive, the sample remains too
small to claim a significant detection of an asymmetry across
longitude. Moreover, the microlensed stars for which spectra were
obtained were selected heterogeneously (for example in the left-right
asymmetry), making a detailed comparison with the model difficult.

\subsection{Trends in the chemical distribution}
\label{ssec:chemicaltrends}

\begin{figure*}
\centerline{
\includegraphics[angle=0.,width=0.5\hsize]{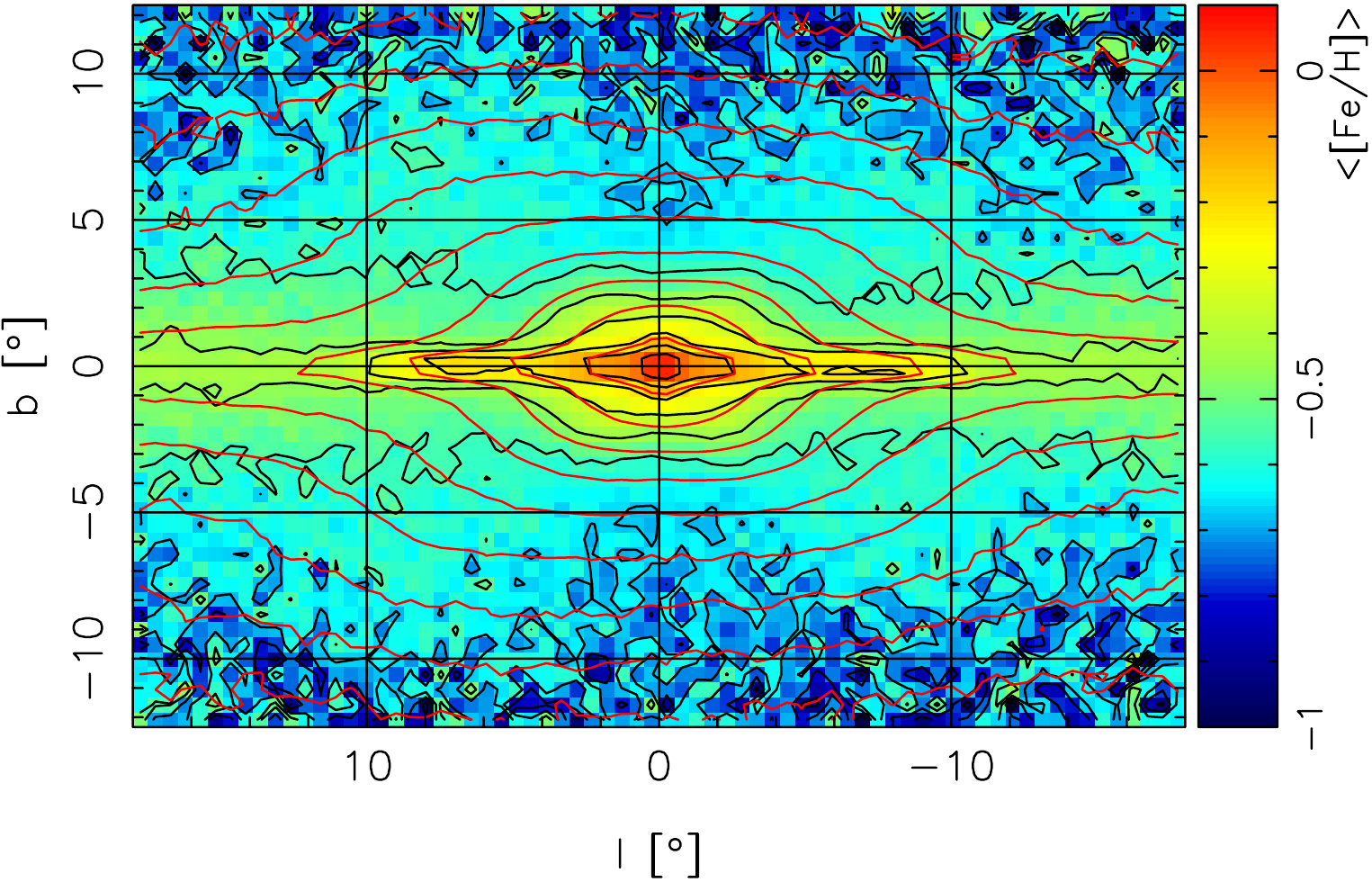}
\includegraphics[angle=0.,width=0.5\hsize]{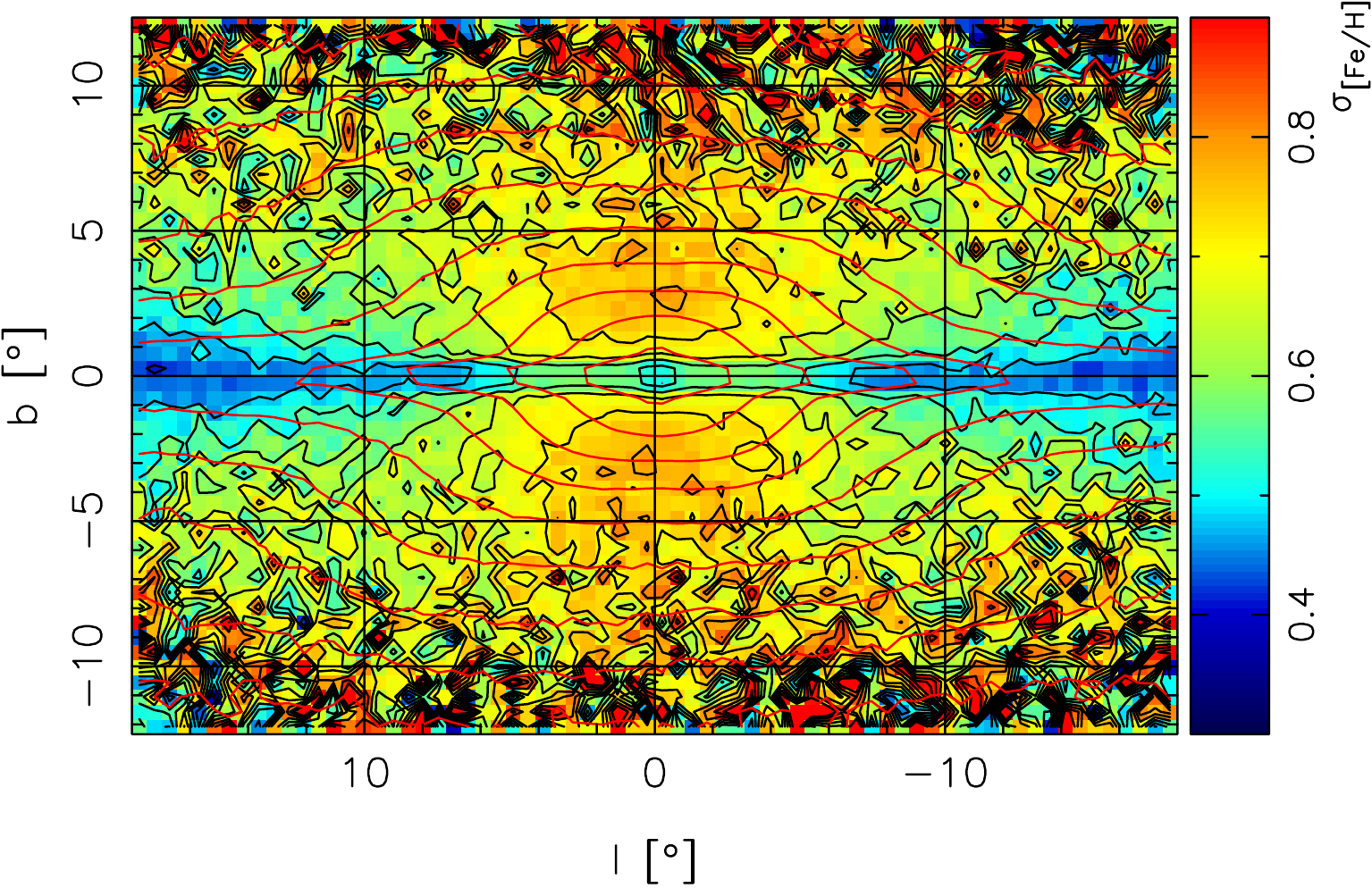}
}
\centerline{
\includegraphics[angle=0.,width=0.5\hsize]{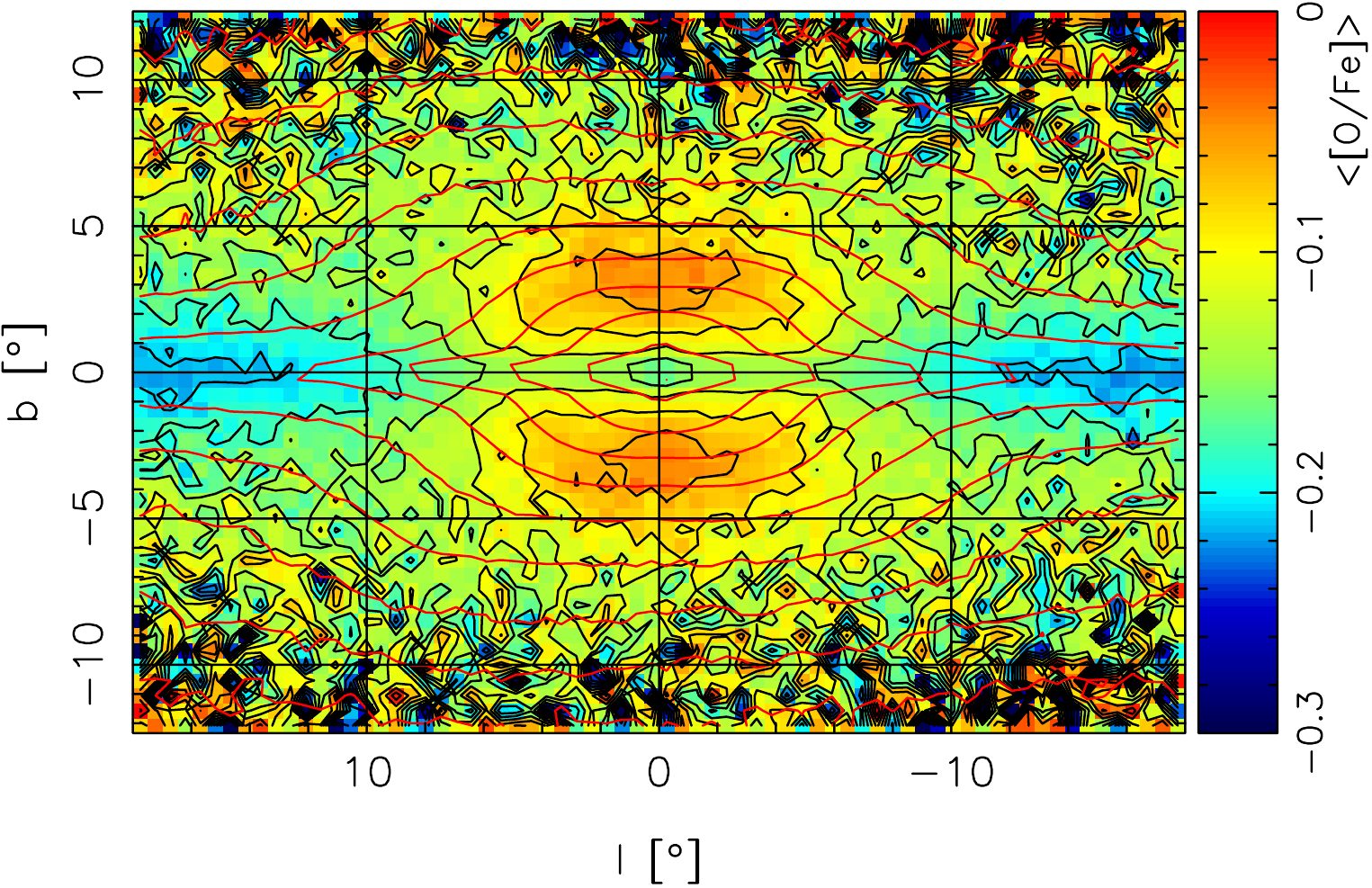}
\includegraphics[angle=0.,width=0.5\hsize]{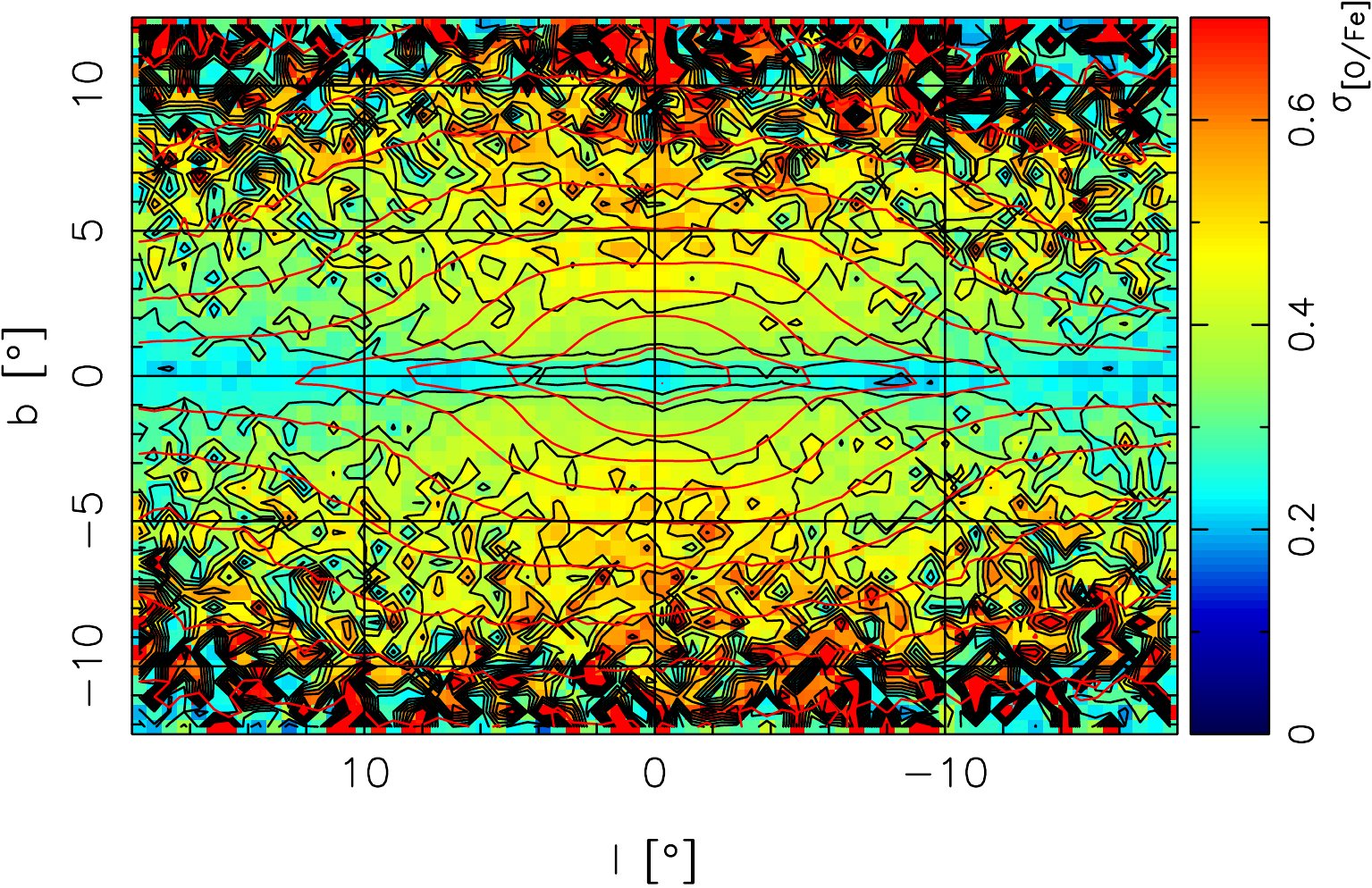}
}
\centerline{
\includegraphics[angle=0.,width=0.5\hsize]{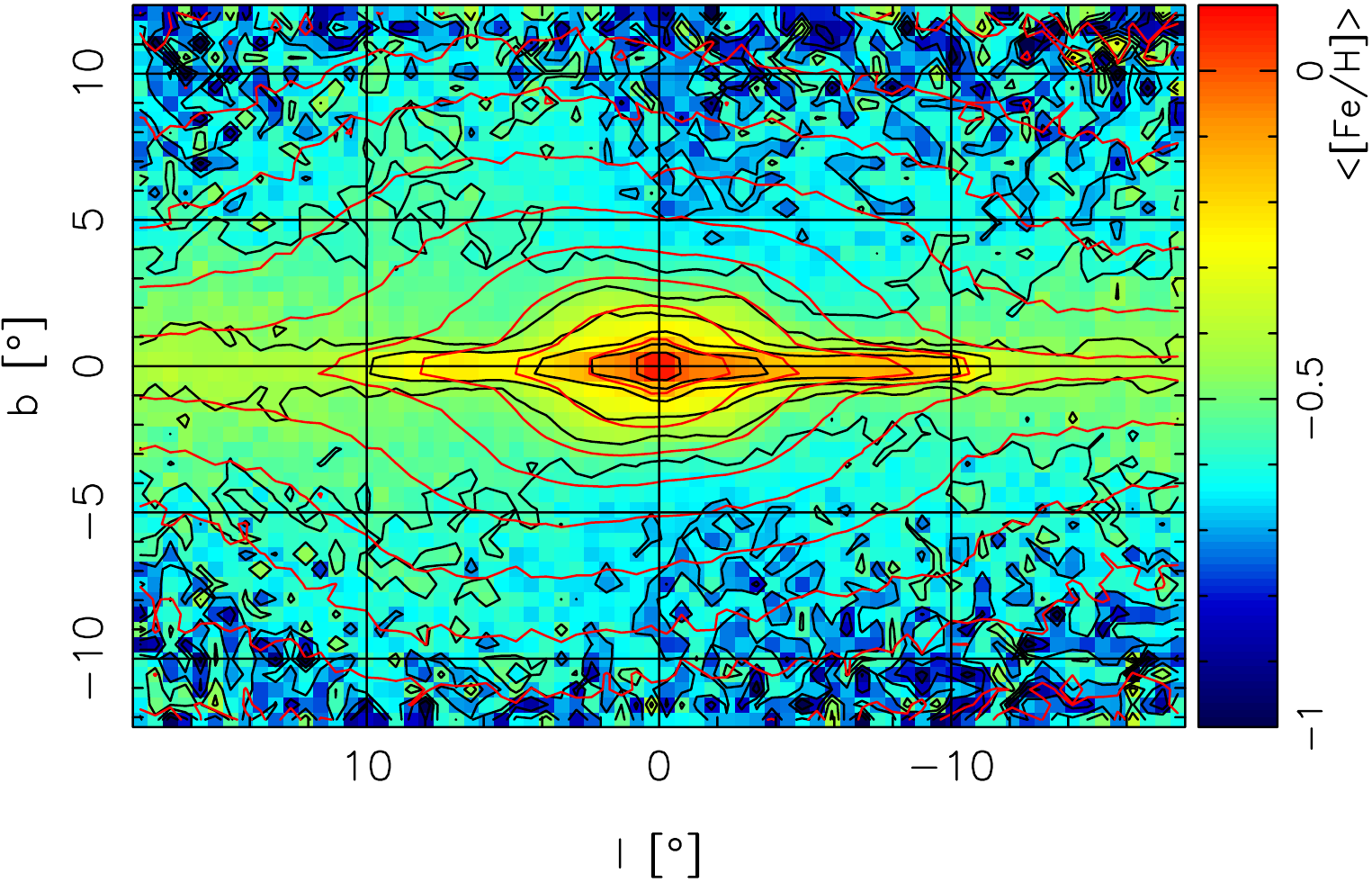}
\includegraphics[angle=0.,width=0.5\hsize]{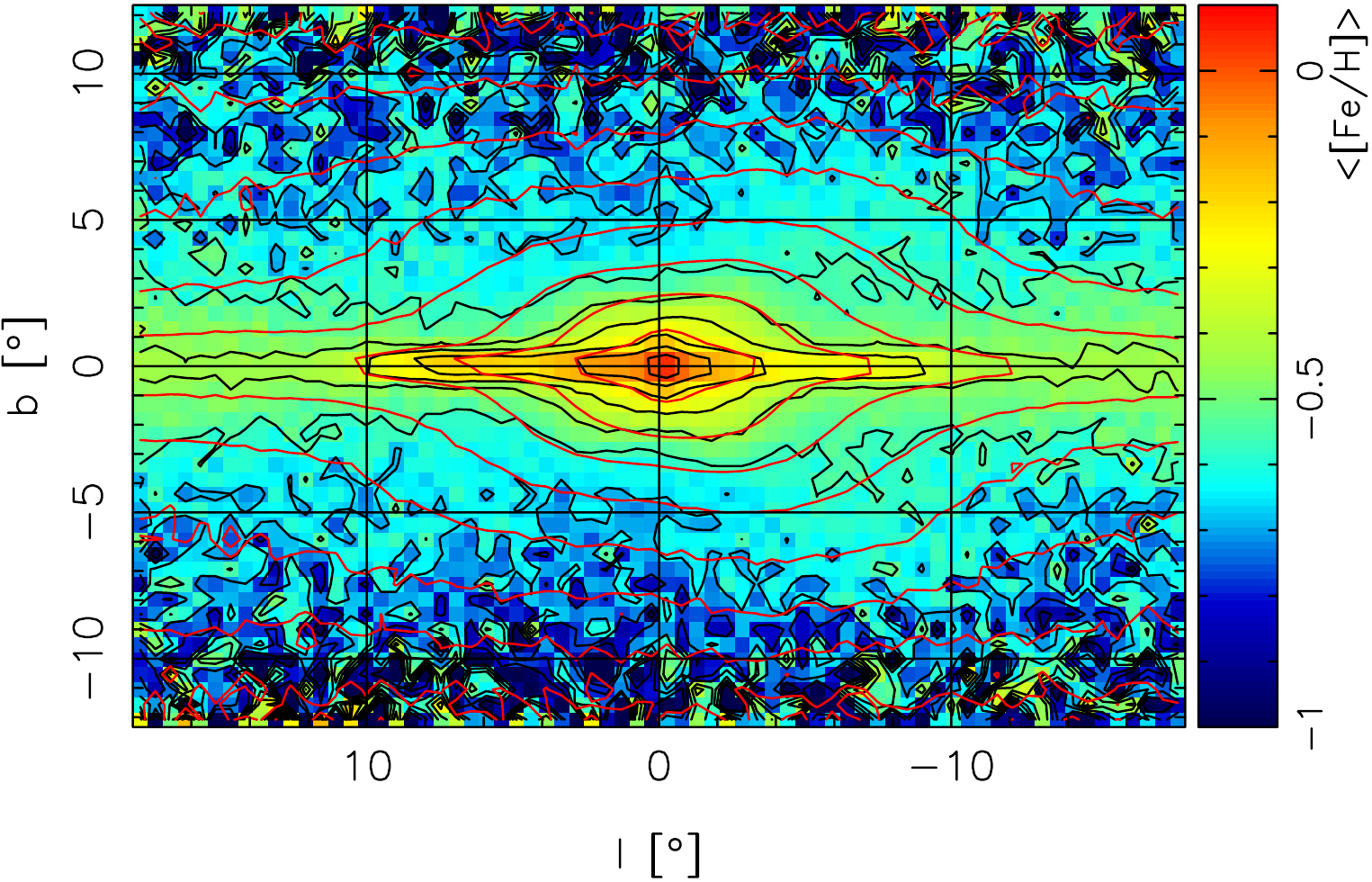}
}
\caption{The stellar chemistry in the star-forming simulation.  Top:
  maps of \avg{\feh}\ (left) and \sig{\feh} (right) in $(l,b)$-space
  as seen from the Sun.  Only stars in the range $4 \leq \rs/\kpc \leq
  12$ are considered.  Middle: same as top row for \alfe.  Bottom:
  maps of \avg{\feh}\ for near-side (left) and far-side (right) stars.
  Black contours show constant plotted quantity, while red contours
  show constant surface density of the population under consideration.
  Note the vertical gradients in \avg{\feh}\ and \avg{\alfe} and the
  mismatch between density and chemistry contours.}
\label{fig:fehlbmap}
\end{figure*}

Having considered the distributions of stars in the star-forming
simulation as a function of age, we now turn to the distribution of
stars by their metallicity, which is observationally more readily
measureable.  The absence of metallicity diffusion in this simulation
leads to somewhat lower \avg{\feh}\ and enhanced \sig{\feh}, which
should be kept in mind when considering the results in this section,
but the overall trends we focus on should remain valid.

The top-left panel of Fig. \ref{fig:fehlbmap} presents a
map of \avg{\feh}\ in Galactic coordinates.  Ignoring the nuclear disc
at $|b| < 1\degrees$, we find a metallicity gradient along the minor
axis of $\sim -0.06 \pm 0.01$ dex deg.$^{-1}$, in broad agreement with
\citet{gonzalez+11}.  A notable feature of this map is that it is more
pinched in \avg{\feh}\ than in the density.  In the model, this
difference is already apparent by $|b| \simeq 5\degrees$.  In our
model, it is the anti-correlation between in-plane dispersion and
\feh\ that sets up the {\it long-lived} vertical gradient.  The
top-right panel of Fig. \ref{fig:fehlbmap} shows \sig{\feh}; this is
also more pinched than the density distribution itself, although it
does not have the boxy part at the centre that the \avg{\feh}\ map
does.

The middle row of Fig. \ref{fig:fehlbmap} shows maps of \avg{\alfe}
and \sig{\alfe}\ at $10 \Gyr$.  Ignoring the nuclear disc ($|b| <
1\degrees$), a distinct \al-enhanced region is present at $|l| <
5\degrees$ and $1\degrees < |b| < 5\degrees$.  As with \feh, the
\avg{\alfe}\ and \sig{\alfe}\ maps are more pinched than is the
density map.

Fig. \ref{fig:nearfarages} showed that selecting stars on the near or
far side produces differences in the mean age distribution across
$l=0\degrees$.  Since age anti-correlates with \feh, metallicity maps
of near versus far sides should also be asymmetric across
$l=0\degrees$.  The bottom row of Fig. \ref{fig:fehlbmap} shows such
maps; as expected there is an asymmetry across the minor axis, with
intermediate-metallicity stars extending to larger latitude at
$l>0\degrees$ on the near side and at $l<0\degrees$ on the far side.
However, this difference is not as prominent as in the ages.  As with
ages, this asymmetry will manifest in microlensed stars.  Using the
microlensing data of \citet{bensby+13}, we find that there is a higher
average metallicity for lensed stars at $l <0\degrees$ than at $l
>0\degrees$, but again the number of stars is too small for this
result to be very significant.

\begin{figure}
\includegraphics[angle=0.,width=\hsize]{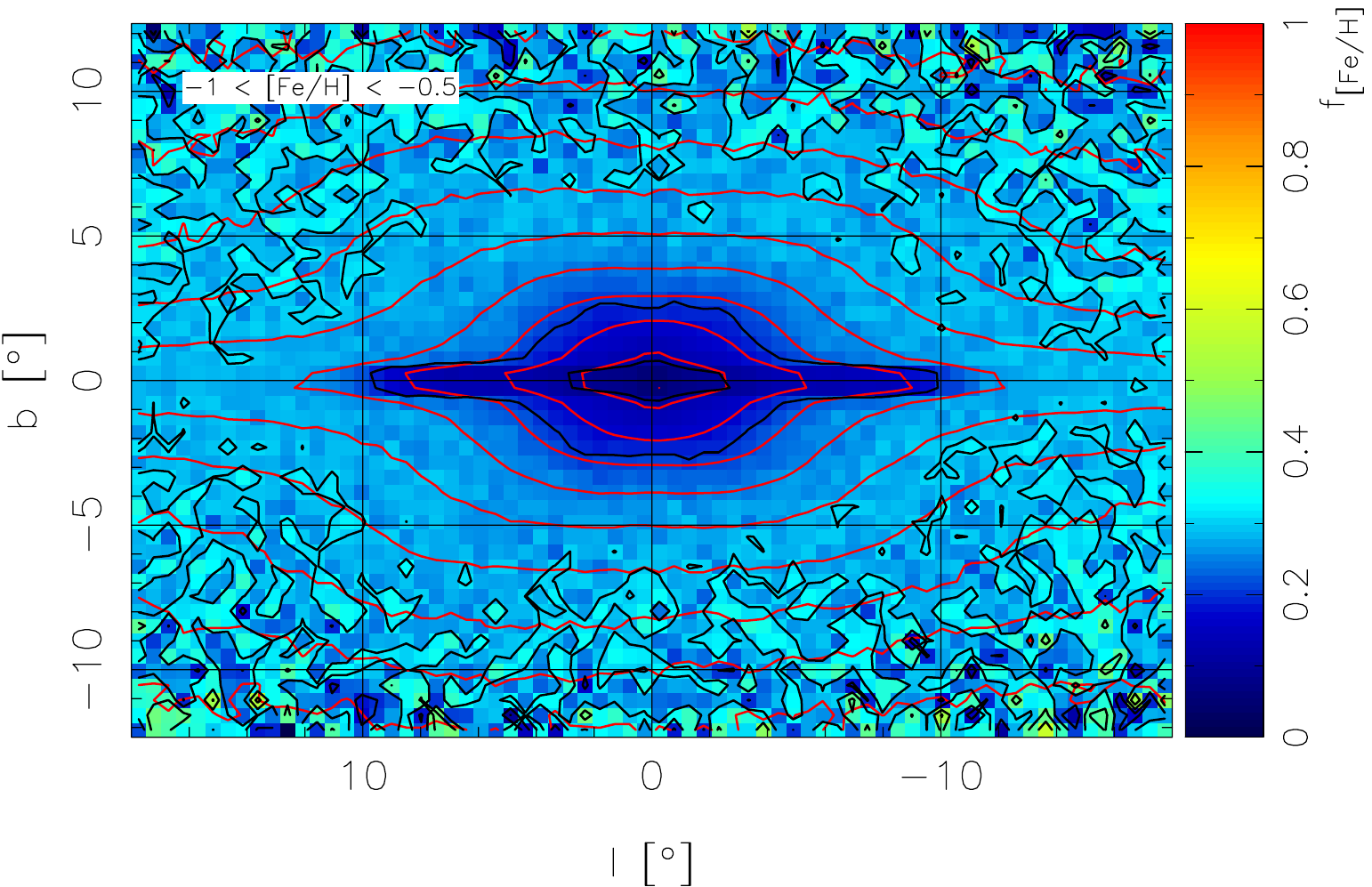}
\includegraphics[angle=0.,width=\hsize]{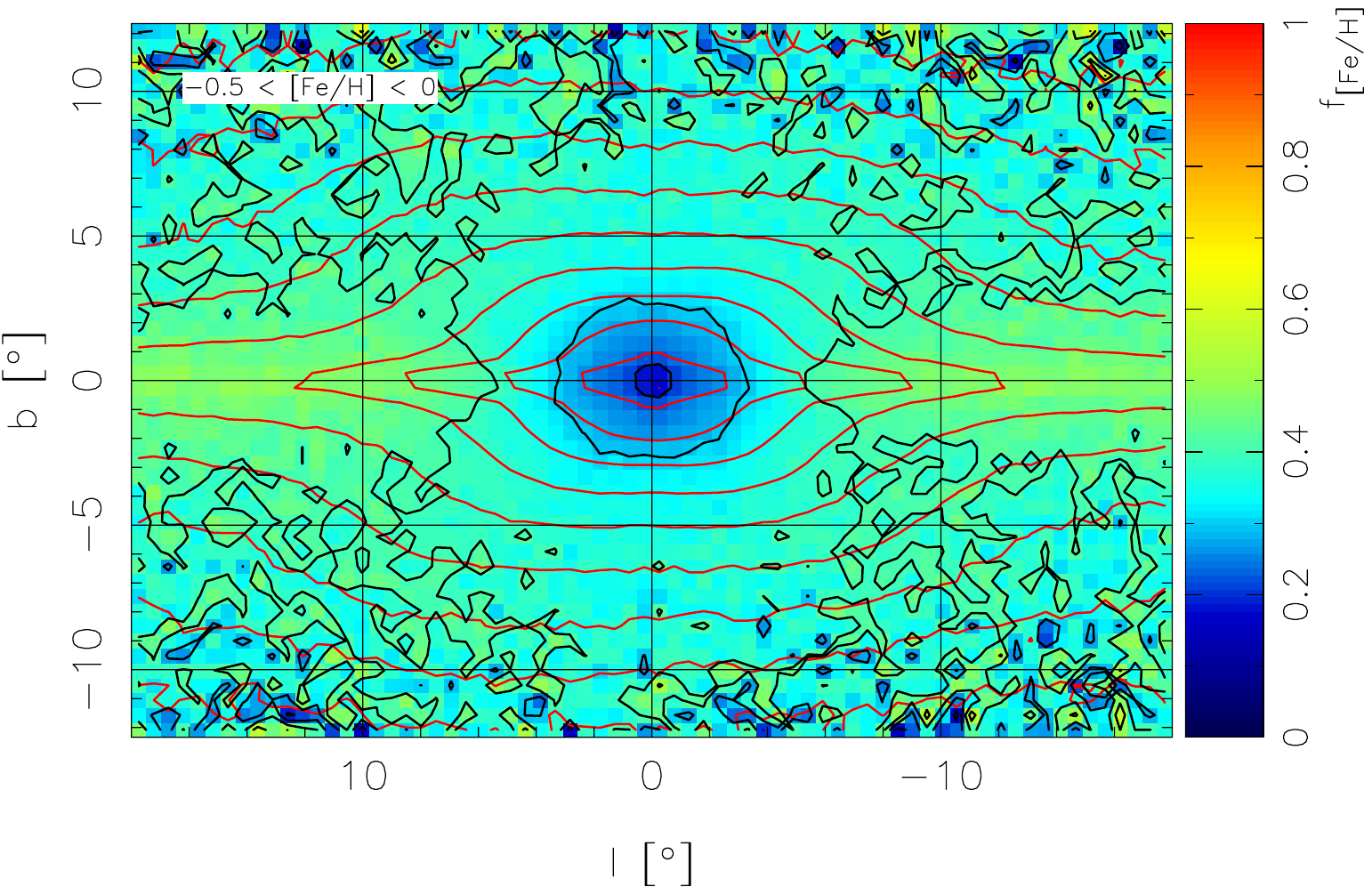}
\includegraphics[angle=0.,width=\hsize]{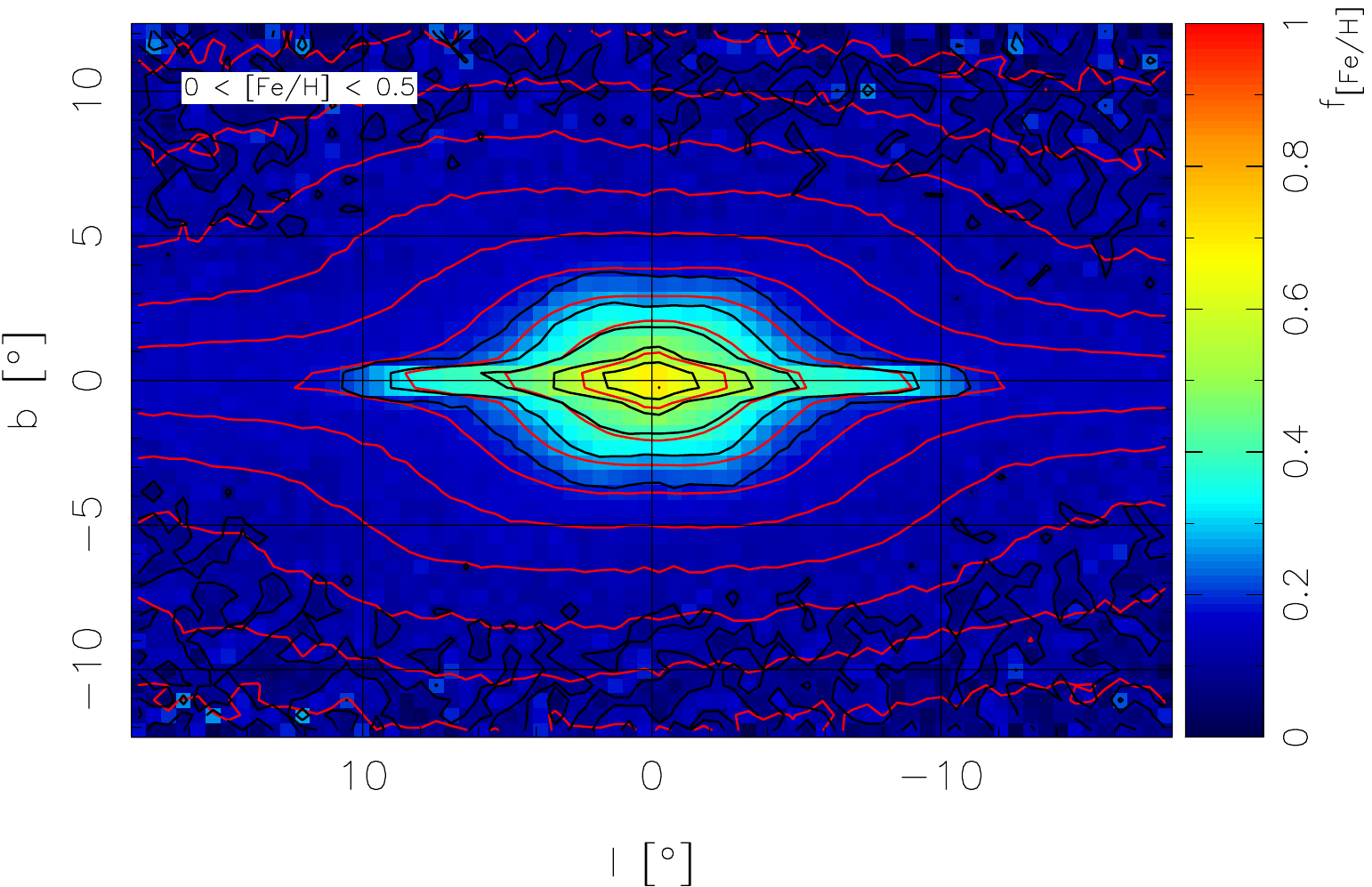}
\caption{The fraction of stars in the star-forming simulation in
  different metallicity bins (as indicated at top left) corresponding
  to the populations A (top), B (middle) and C (bottom) of
  \citet{ness+13a}.  Black contours show constant fractions, while red
  contours show constant overall surface density.  Metal-rich stars
  dominate at small $|b|$, while the intermediate-metallicity stars
  dominate everywhere else, as in ARGOS.}
\label{fig:reldensity}
\end{figure}

\citet{ness+13a} showed that, in the ARGOS survey, the fraction of
metal-rich (their component A) stars decreases with latitude.  In
contrast, the fraction of metal-poor stars (their component C)
increases relative to the intermediate-metallicity stars although they
remain less abundant.  Fig. \ref{fig:reldensity} shows maps of the
fraction of stars in the same three broad metallicity bins A, B and C
as in \citet{ness+13a} (we stress that these are only convenient bins,
not distinct structural components).  The metal-rich population A is
dominant at low latitudes, but drops off rapidly with increasingly
latitude, as found also by \citet{ness+13a}.  The metal-poor
population C instead follows the opposite trend, representing a
smaller population at low latitudes, but increases with latitude.  The
intermediate-metallicity population B dominates everywhere except near
the disc where population A dominates.  These are all trends that are
in agreement with the findings of \citet{ness+13a}.

Recently, \cite{zoccali+17} parametrized the metallicity distribution
in each of the 26 fields of the GIBS survey into a metal-poor and a
metal-rich component. They presented a density map for each of the two
components for $-10\degrees<l<10\degrees$ and
$-10\degrees<b<10\degrees$ showing that the metal-poor population is
more centrally concentrated than the metal-rich one, and has a more
axisymmetric spatial distribution. On the other hand, the metal-rich
population has a box-like distribution.  These trends are consistent
with the results of Fig. \ref{fig:reldensity}.

%%%%%%%%%%%%%%%%%%%%%%%%%%%%%%%%%%%%%%%%%%%%%%%%%%%%%%%%%%%%%%%%%%%%%%%%%%%%%

\section{Discussion}
\label{sec:discussion}

The separation by the bar of different stellar populations on the
basis of their in-plane kinematics presents a novel explanation for
the origin of many of the trends seen in the Milky Way's bulge, which
have seemed to require more than one formation mechanism.  Because
this mechanism has never been described before, we coin the term
`kinematic fractionation' to refer to it.

Kinematic fractionation circumvents the need for two or more
components to explain the trends observed in the bulge by employing a
continuum of stellar populations with different in-plane random
motions, a very natural outcome of internal evolution.  Heating can be
accomplished in a variety of ways, including star forming clumps,
spirals, lopsidedness, and other perturbations, supernova feedback,
gas turbulence, and even interactions, but there is no need to
generate a separate population (a thick disc or a classical accreted
bulge) in the process.  Moreover, early discs may form hotter directly
\citep[e.g.][]{bird+13}. As we showed in the star-forming simulation,
at early times even $2\Gyr$ is enough time to produce a significant
difference in the random motions between younger and older stars, and
in the Milky Way the bar may have required even longer to form.  Nor
do the trends observed in the Milky Way require that different spatial
regions of the disc contribute to different parts of the bulge.  For
instance, one of the key differences between our model and earlier
work is that the vertical metallicity gradient results not from a
radial gradient in the initial system, but from the correlation
between age (and metallicity) and kinematics at bar formation.  This
results in a long-lived vertical gradient.

The star-forming simulation we have compared the Milky Way to is not
a particularly good match to it.  Thus, the results presented here have
focused on trends, not on detailed matches to observations.  On the
other hand, the very fact that this quite generic model produces
trends that are also seen in the Milky Way argues very forcefully that
the model captures the essential physics that produces these
non-trivial trends, unless we are prepared to accept a quite
remarkable coincidence.

\subsection{Is a thick disc necessary?}

Implicit in our interpretation of the bulge of the Milky Way is that
there is a continuum of properties between old stars and those
slightly younger, most importantly in their in-plane kinematics.
Inevitably, a population that has hotter radial kinematics also has
hotter vertical kinematics, as is evident for the star-forming
simulation in Fig. \ref{fig:heights}.  Simulations
\citep[e.g.][]{brook+12, bird+13, stinson+13, grand+16} have suggested
that discs form `upside-down', i.e. thick initially slowly settling
into thinner structures.  This view is consistent with the observed
large gas velocity dispersions due to disordered motions in
high-redshift galaxies \citep[e.g.][]{law+09b, kassin+12, gnerucci+11,
  wisnioski+15}, while the lack of flaring in the thick disc of the
Milky Way \citep{bovy+16} is also suggestive of such formation.  Thus,
it may be that a thick disc was present already by the time of bar
formation.  An ab initio thick disc will still experience vertical
heating by the bar and the extent of this will be determined by the
in-plane motion.  Moreover, the in-plane motion of such a thick disc
will determine whether its stars form part of the X-shape (see Section
\ref{ssec:run742}).  A pre-existing thick disc will therefore produce
the trends observed in the Milky Way's bulge \citep{bekki_tsujimoto11,
  pdimatteo+15}, via kinematic fractionation.
  
However, the oldest population in this simulation does not constitute
an especially thick disc at $2\Gyr$, before the bar has formed.
Therefore, we argue that a pre-existing thick disc at the time of bar
formation, whether as part of a continuum with the thin disc
\citep{brook+04TD, schoenrich_binney09b, loebman+11, bird+13,
  bovy+12a}, or as the separate distinct structure envisaged by
\citet{bekki_tsujimoto11}, is {\it not necessary} to produce the
chemical trends observed in the Milky Way bulge.  However, it is not
inconceivable that the oldest population evolves into what,
structurally, is consistent with the present-day thick disc.  Indeed
in the star-forming simulation, the oldest population evolves to a
scaleheight of $1\kpc$ at a radius of $5\kpc$.  Thus, while the
observed bulge trends are consistent with a formation with a thick
disc already present \citep[e.g.][]{bekki_tsujimoto11}, a thick disc
is not necessary to explain them.

\subsection{Consequences for external galaxies}
\label{ssec:edgeons}

\begin{figure}
\includegraphics[angle=0.,width=\hsize]{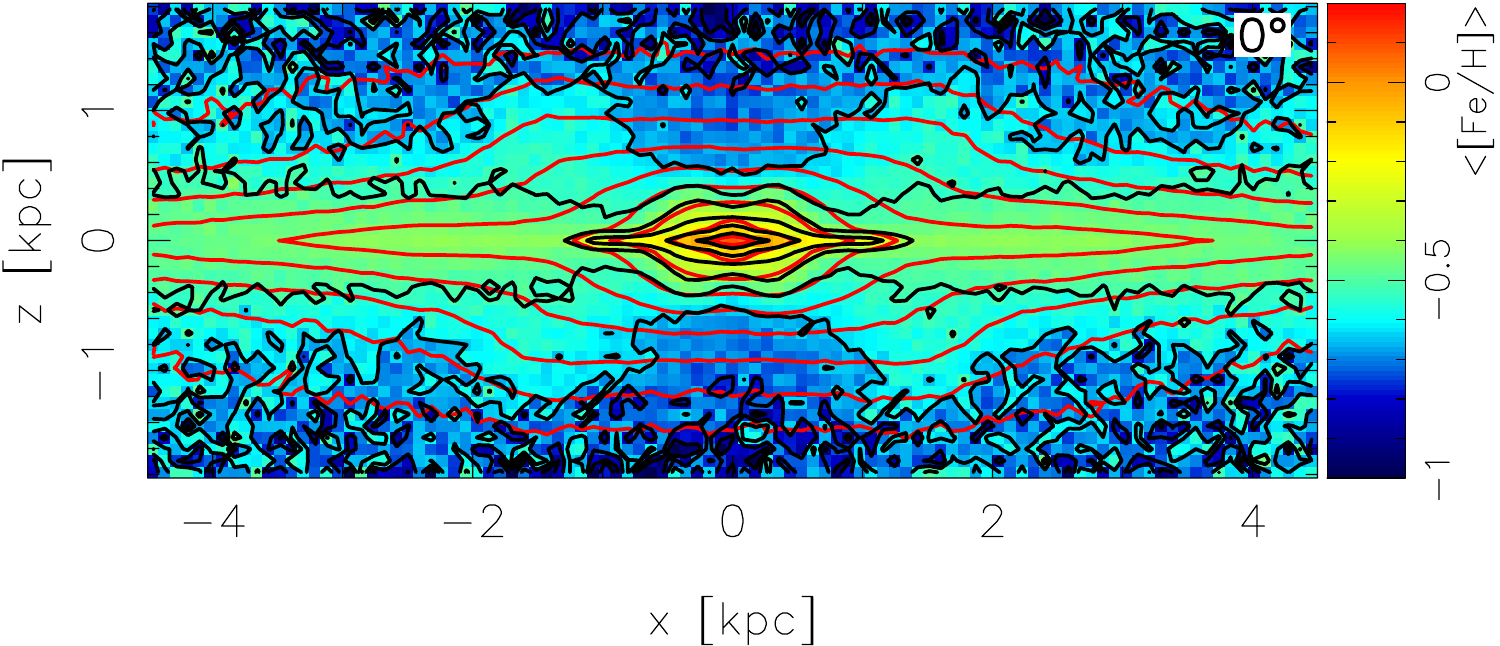}
\includegraphics[angle=0.,width=\hsize]{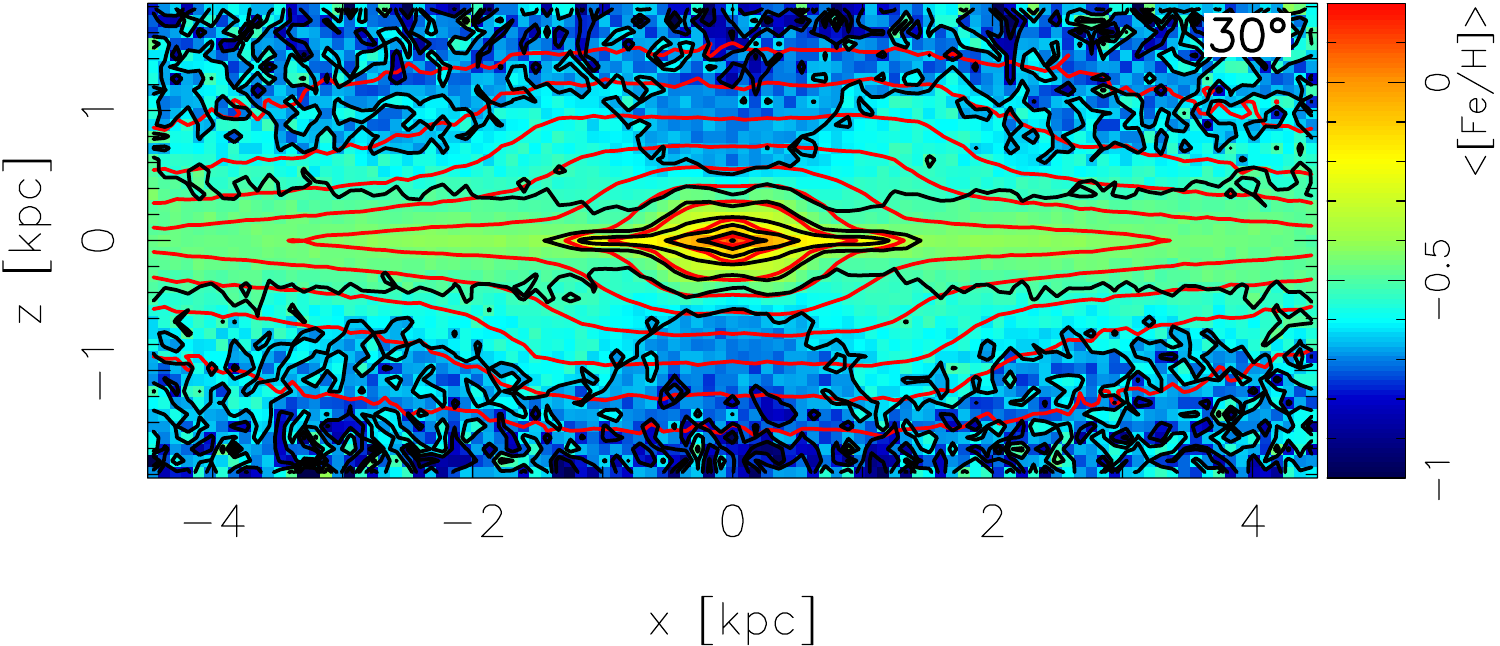}
\includegraphics[angle=0.,width=\hsize]{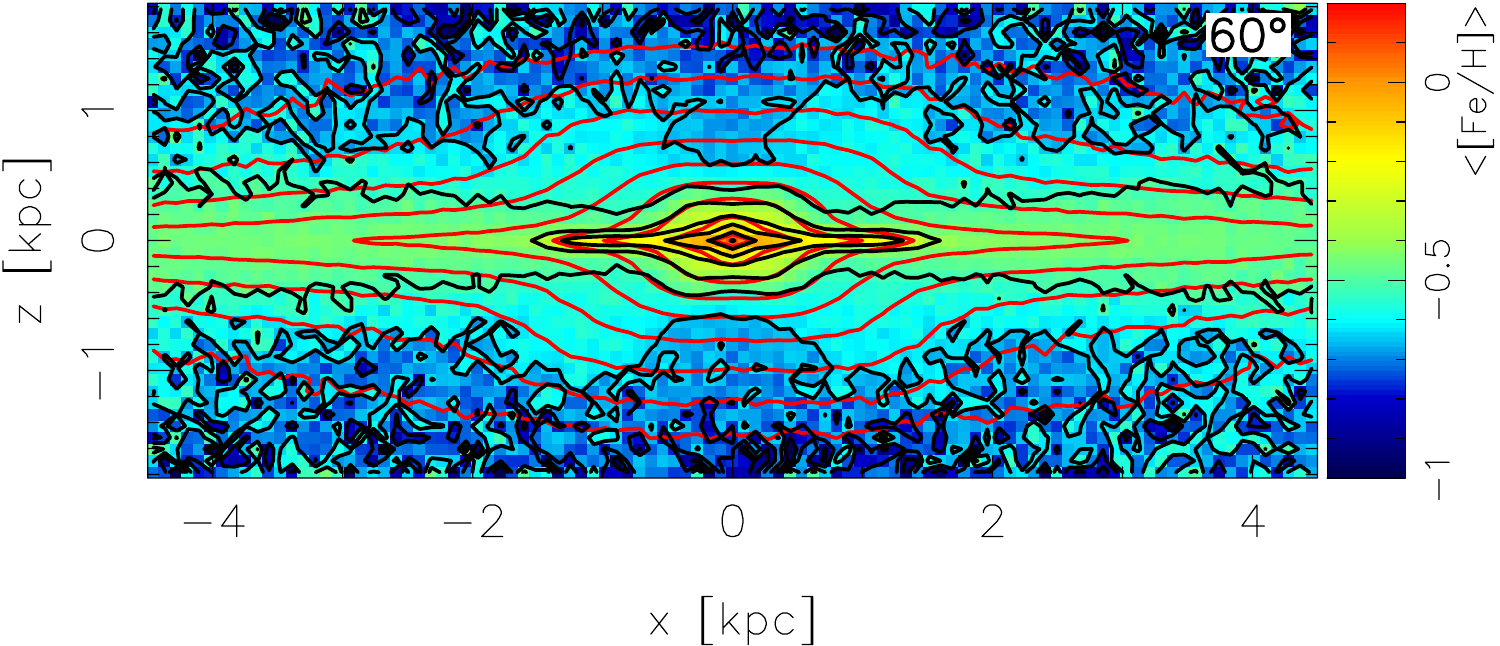}
\includegraphics[angle=0.,width=\hsize]{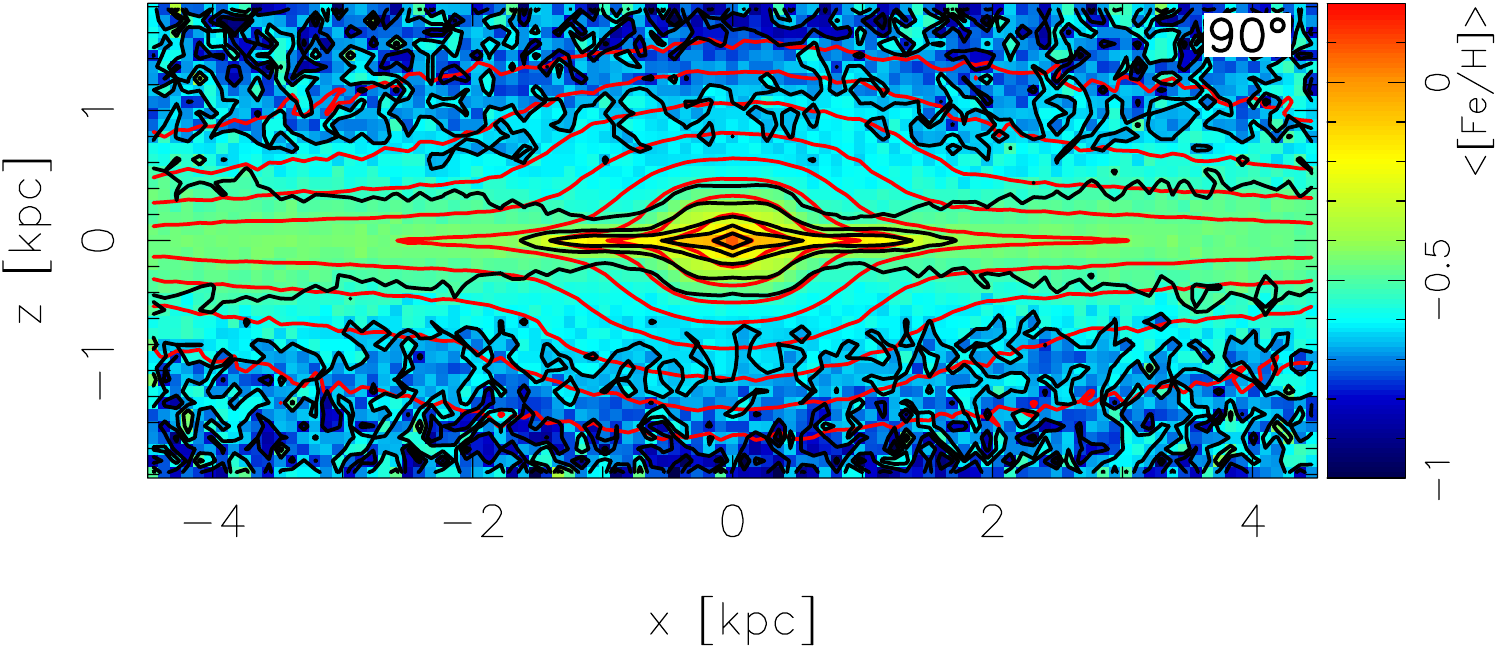}
\caption{Maps of \avg{\feh}\ (black contours and colours) and surface
  density (red contours) in the star-forming simulation seen edge-on
  with the bar side-on (top), end-on (bottom) and two intermediate
  orientations (middle two panels).  The bar's orientation is
  indicated in the top right of each panel.  These maps predict the
  trends in the metallicity distribution of external B/P-bulge
  galaxies.}
\label{fig:externalfehmaps}
\end{figure}

The evolution we have outlined is very general and should occur in
external galaxies that have evolved in isolation.  Additional support
for kinematic fractionation having played a dominant role in shaping
bulges can be obtained by testing for similar trends in external
galaxies.  A straightforward test comes from the metallicity
distribution compared with the density distribution in edge-on
galaxies.  As with the Milky Way, a map of mass-weighted
\avg{\feh}\ will appear more pinched-/peanut-shaped than the density
distribution off the mid-plane.  Fig. \ref{fig:externalfehmaps}
illustrates this behaviour and shows that the difference remains
detectable to a bar angle of $60\degrees$ to the line of sight, which
is also roughly the limit at which a peanut shape can be detected.

\begin{figure}
\includegraphics[angle=0.,width=\hsize]{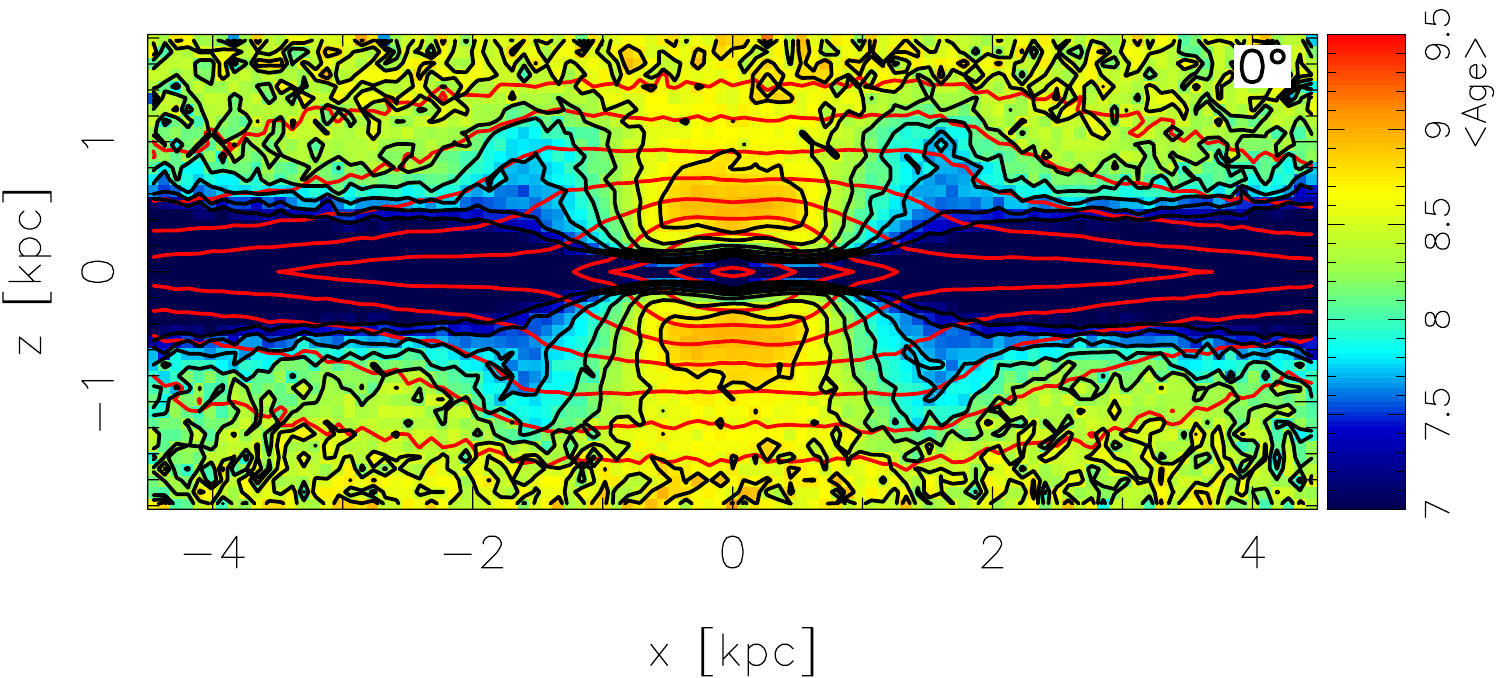}
\caption{Map of \avg{age}\ (black contours and colours) and surface
  density (red contours) in the star-forming simulation seen edge-on
  with the bar side-on.}
\label{fig:externalagemaps}
\end{figure}

Fig. \ref{fig:externalagemaps} presents an age map for the model.
This exhibits a more prominent peanut-shape than in the density
distribution.  However, the dynamic range outside the disc is
relatively small, $\sim 2\Gyr$, so may be difficult to measure
directly.

\subsection{Future prospects}

A number of future spectroscopic and photometric surveys will offer
critical comparisons to test our model in more detail. These include
APOGEE-2 \citep{apogee}, 4MOST \citep{4most}, MOONs \citep{moons} and
LSST \citep{lsst}.  APOGEE-2 (2016-2020), observes at high resolution
in the near-IR so that it can efficiently observe the bulge at low
latitudes, past the obscuring dust which limits optical
observations. MOONS (2019) will later offer the possibility to map the
inner Galaxy at a similar resolution and wavelength but it will make
use of the 8 m aperture of the VLT to reach deeper, to study fainter
stars. The highly complementary parameter space mapped by these
surveys, measuring radial velocities and abundances of more than 15
elements for thousands of bulge stars, will be supplemented by the
astrometric information from the near-IR photometric mapping of the
VVV survey. The kinematics, chemical abundances, proper motions and
even spectroscopic age measurements of large number of stars
\citep{martig+15, ness+16}, in addition to those of other Galactic
components provided by optical surveys such as {\it Gaia}, 4MOST,
WEAVE and LSST, will provide the critical constraints on our model,
helping us understand the detailed formation history of the Galaxy. In
this context, our model also presents the possibility of defining
optimal strategies during the planning stages of forthcoming surveys.

Our predictions are general and not specific to the Milky Way and so
comparisons to boxy bulges in other spiral galaxies are also very
important.  Modern IFU spectrographs and their dedicated surveys offer
a great opportunity for understanding the bulge of the Milky Way in
context \citep{gonzalez_gadotti16}, by measuring the morphological
aspects of other spiral galaxies and prevalence of mergers compared to
secular evolution at high redshift \citep[e.g.][]{gadotti+15}.

Our results also point the way forward in terms of future efforts to
simulate the Milky Way in detail.  Since simulations with full gas
physics are subject to much more stochasticity, and are
computationally expensive, pure $N$-body simulations will continue to
play an important role in trying to understand the bulge of the Milky
Way.  Past pure $N$-body simulations of the Milky Way
\citep[e.g.][]{fux97, jshen+10, martinez-valpuesta_gerhard11} have
produced good matches to the density distribution and kinematics of
the bulge.  Studying the age and metallicity distributions of such
models requires assigning to particles a population tag
(age/chemistry) in some way.  \citet{martinez-valpuesta_gerhard13}
used such a technique to match the observed vertical \feh\ gradient;
in their approach the metallicity of a particle was a function of its
initial orbital radius.  We propose that a more useful approach to
assigning a population tag to particles is based on their initial
orbital circularities, rather than on their initial position.

\subsection{Summary}

Our main conclusions can be summarized as follows
\begin{itemize}
\item Bars can efficiently separate disc stellar populations that
  differ by their initial in-plane kinematics (see Section
  \ref{sec:nbodysims}).  Hot populations form bars with weak
  quadrupoles, become vertically thick, particularly at the centre,
  and are box-shaped, while the cooler populations host stronger bars,
  are thinner at the centre and thus produce a peanut shape.  We refer
  to this separation of populations as {\it kinematic fractionation}.
\item The early rapid evolution of discs naturally imprints
  correlations (or anti-correlations) between age, \sig{R}, \feh\ and
  \alfe, which leads to relations between \sig{R} and chemistry (see
  Section \ref{ssec:prebar}).  After bar formation, kinematic
  fractionation leads to different populations having different bar
  strengths, vertical distributions and B/P-bulge strengths, naturally
  without the need for a separate origin for different populations.
  Instead the properties of the bulge vary continuously with age.  The
  underlying reason for these differences is primarily the difference
  in the initial \sig{R}\ and its strong dependence on age in young
  galaxies (see Section \ref{ssec:finalsystem}).
\item Kinematic fractionation also leads to the X-shape that is
  stronger in relatively younger (metal-rich) stars than in the older
  (metal-poor) ones.  In our star-forming model, a difference of just
  $2 \Gyr$ in age is sufficient to alter the final distribution of
  stars from a single peak to a bimodal distribution.  This is also
  consistent with observations in the Milky Way \citep{ness+12,
    uttenthaler+12, rojas-arriagada+14}.  Moreover we predict that,
  for tracers with small distance uncertainties, old populations will
  still appear bimodal, while younger populations will be well
  separated (see Section \ref{ssec:xshape}).
\item Kinematic fractionation of a hot, older population and a cooler,
  younger population produces a weaker old bar population and a
  stronger young bar population, even if the difference in age is only
  $\sim 2\Gyr$.  This is consistent with what is observed in the Milky
  Way, where the RR~Lyrae trace a weak bar, while red clump stars
  trace a stronger bar \citep{dekany+13}.  In addition, bulge Mira
  variables show a smooth variation from strongly barred to more
  spheroidal distribution as a function of their age
  \citep{catchpole+16} (see Section \ref{ssec:rrlyrae}).
\item Comparing with the ARGOS data \citep{ness+13b}, the model is
  able to match the overall kinematics and those of the metal-rich
  population.  Matching to the kinematics of the metal-poor
  population, with its relatively flat velocity dispersion which
  varies slowly with latitude, requires the addition of $15\%$ of
  stars in a hot, slowly rotating component.  This slowly rotating
  population amounts to $\sim 5\%$ of all bulge stars.  Assuming a
  bulge-to-total stellar mass ratio of $\sim 25\%$
  \citep{bland-hawthorn_gerhard16}, this amounts to $1.25\%$ of the
  stellar mass of the Milky Way.  We attribute these stars to a
  metal-poor halo, leaving no room for a classical bulge.  This
  conclusion is in agreement with the independent result derived by
  Kormendy et al. (in progress and private communication) that the
  Milky Way has no classical bulge component: aside from the boxy
  bulge, near the centre they find only a nuclear star cluster and a
  small discy pseudobulge.  \cite{bland-hawthorn_gerhard16} review
  some of the trends which have been used to argue for a bulge
  component built through mergers (a `classical bulge' component);
  we have shown here that these trends can be explained without such
  an accreted bulge (see Section \ref{ssec:stellarkine}).
\item A vertically declining \feh\ gradient, and vertically rising
  \alfe\ gradient, are established largely as a result of the
  different vertical extents of different ages.  A declining
  \feh\ gradient has been found in the Milky Way
  \citep[e.g.][]{zoccali+08, gonzalez+11, johnson+13}.  Separating
  populations into broad metallicity bins, as used in ARGOS
  \citep{ness+13a}, we showed that at low latitudes the metal-rich
  population dominates with the low-metallicity and
  intermediate-metallicity bins providing a smaller contribution.  The
  contribution of the metal-rich population drops rapidly at larger
  latitudes, with that of the low- and intermediate-metallicity bins
  rising.  The intermediate-metallicity bin dominates at large
  latitudes.  These trends are in agreement with observations in the
  Milky Way \citep{ness+13a} (see Section \ref{ssec:chemicaltrends}).
\item Because the minor axis is dominated by old, metal-poor stars, a
  prediction of the kinematic fractionation model is that the shape of
  the \avg{\feh}\ (or \avg{\alfe}\ or \avg{age}) map in the
  $(l,b)$-plane will appear more pinched/peanut-shaped than the
  density distribution.  These trends will need to be verified in
  future surveys (see Sections \ref{ssec:agetrends} and
  \ref{ssec:chemicaltrends}).
\item We predict that microlensed dwarf stars on the far ($\rs >
  8\kpc$) side of the Galaxy will contain more intermediate-age,
  metal-rich stars at negative longitudes compared with positive
  longitudes.  The data of \citet{bensby+13} suggest such trends, but
  the number of spectroscopically measured microlensed dwarfs is still
  too small to confirm this result (see Sections \ref{ssec:agetrends}
  and \ref{ssec:chemicaltrends}).
\item A simple test of the kinematic fractionation model is provided
  by external edge-on barred galaxies with B/P bulges, for which maps
  of \avg{\feh}\ will be more strongly pinched/peanut-shaped than the
  density distribution itself (see Section \ref{ssec:edgeons}).
\end{itemize}

\bigskip
\noindent
{\bf Acknowledgements.} 

\noindent
VPD is supported by STFC Consolidated grant no. ST/M000877/1.  MN is
funded by the European Research Council under the European Union's
Seventh Framework Programme (FP 7) ERC Grant Agreement no. 321035.  MZ
and DM acknowledge support by the Ministry of Economy, Development,
and Tourism’s Millennium Science Initiative through grant IC120009,
awarded to The Millennium Institute of Astrophysics (MAS), by the
BASAL-CATA Center for Astrophysics and Associated Technologies PFB-06,
Fondecyt Regular Proyects 1130196 and 1150345, and (MZ) by CONICYT's
PCI grant DPI20140066.  We acknowledge support from the ESF Exchange
Grant (number 4650) within the framework of the ESF Activity entitled
`Gaia Research for European Astronomy Training'.  The star-forming
simulation used in this paper was run at the High Performance
Computing Facility of the University of Central Lancashire.  The pure
$N$-body simulations were run at the DiRAC Shared Memory Processing
system at the University of Cambridge, operated by the COSMOS Project
at the Department of Applied Mathematics and Theoretical Physics on
behalf of the STFC DiRAC HPC Facility (www.dirac.ac.uk). This
equipment was funded by BIS National E-infrastructure capital grant
ST/J005673/1, STFC capital grant ST/H008586/1, and STFC DiRAC
Operations grant ST/K00333X/1. DiRAC is part of the National
E-Infrastructure.  We are deeply grateful to Larry Widrow for his
support with using {\sc GalactICS}, including providing us with a
development version that allows for two discs of different thickness.
VPD acknowledges with great pleasure the support of the Pauli Center
for Theoretical Studies, which is supported by the Swiss National
Science Foundation (SNF), the University of Z\"urich and ETH
Z\"urich.  He particularly thanks George Lake for arranging for his
sabbatical visit during which time this paper was completed.  We thank
Sandro Tacchella, Reynier Peletier, Peter Erwin and Marc Balcells for
useful discussions.  We thank the anonymous referee for useful
comments that have helped improve the paper.

%%%%%%%%%%%%%%%%%%%%%%%%%%%%%%%%%%%%%%%%%%%%%%%%%%%%%%%%%%%%%%%%%%%%%%%%%%%%%

\bigskip 
\noindent

\bibliographystyle{aj}
\bibliography{ms.bbl}
%\bibliography{allrefs}

%%%%%%%%%%%%%%%%%%%%%%%%%%%%%%%%%%%%%%%%%%%%%%%%%%%%%%%%%%%%%%%%%%%%%%%%%%%%%

\appendix
\section{Height dependence of orbital frequencies}
\label{app:orbits}

Here we demonstrate that the vertical frequency, $\nu$, of a star
declines more rapidly with increasing maximum height, $z_{max}$, than
does the orbital frequency, $\Omega$.  We do this directly by
computing orbits of stars in the pure $N$-body model with disc D5
described in Section \ref{sec:nbodysims}.  We select all stars in a
narrow annulus of guiding radius $3.95 \leq R_{gui}/\kpc \leq 4.05$,
where $R_{gui} = J_z/V_c$, $J_z$ is the angular momentum of the
particle about the $z$-axis, and $V_c$ is the circular velocity at
this radius.  We choose this annulus because the peanut shape is very
distinct at the end of the simulation in this region; however the
vertical distribution of the stars is much narrower than the final
peanut shape.  We integrate orbits for $\sim 9000$ of these particles,
using a timestep of $10^4$ yr, saving the phase space coordinates
of each star every timestep over a $1\Gyr$ period.  From these orbits,
we then compute the eccentricity of the orbit
\begin{equation}
\epsilon = \frac{R_{max}-R_{min}}{R_{max}+R_{min}}
\end{equation}
where $R_{max}$ and $R_{min}$ are the maximum and minimum radius
reached; $\epsilon=0$ corresponds to a circular orbit.  We also
compute $z_{max}$ as well as the tangential and vertical frequencies,
$\Omega$ and $\nu$, respectively.

\begin{figure}
\includegraphics[angle=0.,width=\hsize]{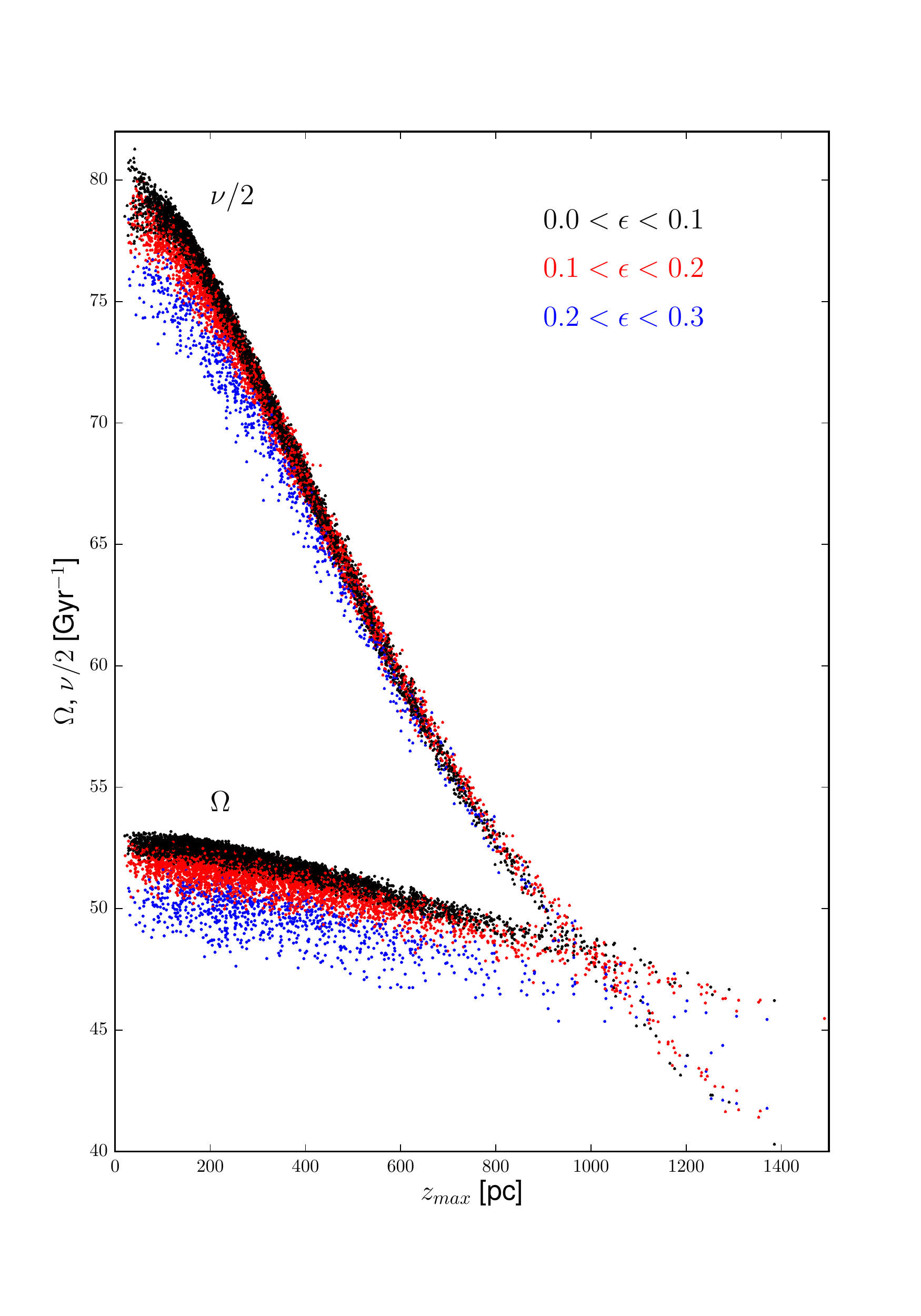}
\caption{Filled triangles show $\nu/2$ while filled circles show
  $\Omega$ for individual star particles.   Black, red and blue points
  show increasing eccentricity as indicated.  All orbits have guiding
  radius $3.95 \leq R_{gui}/\kpc \leq 4.05$.  }
\label{fig:frequencies}
\end{figure}

Figure \ref{fig:frequencies} shows the result of this calculation.  It
is very clear that $\nu/2$ declines much more rapidly with $z_{max}$
than does $\Omega$, which only changes by $\sim 3\Gyr^{-1}$ over this
vertical range.  The figure also shows the variation of $\Omega$ and
$\nu/2$ as a function of $\epsilon$.  The angular frequency $\Omega$
decreases with increasing $\epsilon$, while the change in $\nu/2$ with
$\epsilon$ is more modest.

\label{lastpage}

\end{document}